\documentclass[english,11pt]{article}

\usepackage{dsfont}

\usepackage{lmodern}        
\usepackage[T1]{fontenc} 
\usepackage{caption} 
\usepackage{subfigure}
\usepackage{float}
\usepackage[utf8]{inputenc}
\usepackage{fullpage}
\usepackage{float}
\usepackage{amsmath}
\usepackage{stmaryrd}
\usepackage{amsthm}
\usepackage{amssymb}
\usepackage{cancel}
\usepackage{graphicx,xcolor}
\usepackage{esint}
\usepackage[toc,page]{appendix}
\usepackage{chngcntr}
\usepackage{apptools}
\usepackage{cases}
\newcommand{\R}{{\mathbb R}}

\newcommand{\E}{{\mathbb E}}\makeatletter
\newcommand{\N}{{\mathbb N}}

\newcommand{\Swap}{{c_{swap}}}
\newcommand{\RC}{{CR}}

\newcommand{\RD}{{DSR}}

\newcommand{\Q}{\mathbb{Q}}
\newcommand{\cF}{\mathcal{F}}

\newcommand{\CompRate}{{r^{comp}_t}}

\floatstyle{ruled}
\newfloat{algorithm}{tbp}{loa}
\providecommand{\algorithmname}{Algorithm}
\floatname{algorithm}{\protect\algorithmname}
\theoremstyle{definition}
\newtheorem*{condition*}{\protect\conditionname}
\theoremstyle{plain}
\newtheorem{thm}{\protect\theoremname}[section]
\theoremstyle{plain}

\theoremstyle{definition}

\theoremstyle{plain}
\newtheorem{lem}[thm]{\protect\lemmaname}
\theoremstyle{plain}

\theoremstyle{plain}
\newtheorem{remark}[thm]{\protect\remarkname}
\theoremstyle{plain}

\theoremstyle{plain}
\newtheorem*{assumption*}{\protect\assumptionname}
\usepackage{algorithm,algpseudocode}
\numberwithin{equation}{section}
\makeatother
\usepackage{babel}
\providecommand{\assumptionname}{Assumption}
\providecommand{\conditionname}{Condition}
\providecommand{\corollaryname}{Corollary}
\providecommand{\definitionname}{Definition}
\providecommand{\lemmaname}{Lemma}
\providecommand{\propositionname}{Proposition}
\providecommand{\theoremname}{Theorem}
\providecommand{\remarkname}{Remark}
\providecommand{\examplename}{Example}
\providecommand{\keywords}[1]{\textbf{\textbf{Keywords:}} #1}

\begin{document}
\renewcommand{\contentsname}{Contents}
\renewcommand{\refname}{References} 
\renewcommand{\abstractname}{Abstract} 
\title{A full and synthetic model for Asset-Liability Management in life insurance, and analysis of the SCR with the standard formula.}
\author{Aur\'elien Alfonsi\thanks{Universit\'e Paris-Est, Cermics (ENPC), INRIA, F-77455 Marne-la-Vall\'ee, France. E-mail: {\tt aurelien.alfonsi@enpc.fr}}\and  Adel Cherchali\thanks{Universit\'e Paris-Est, Cermics (ENPC), INRIA, F-77455 Marne-la-Vall\'ee, France. E-mail: {\tt adel.cherchali@enpc.fr}}\and   Jose Arturo Infante Acevedo\thanks{ GIE AXA, 25 avenue Matignon 75008 Paris, France {\tt josearturo.infanteacevedo@axa.com}. } }
\maketitle
 
\begin{abstract}
The aim of this paper is to introduce a synthetic ALM model that catches the main specificity of life insurance contracts.  First, it keeps track of both market and book values to apply the regulatory profit sharing rule.  Second, it introduces a determination of the crediting rate to policyholders that is close to the practice and is a trade-off between the  regulatory rate, a competitor rate and the available profits. Third, it considers an investment in bonds that enables to match a part of the cash outflow due to surrenders, while avoiding to store the trading history. We use this model to evaluate the Solvency Capital Requirement (SCR) with the standard formula, and show that the choice of the interest rate model is important to get a meaningful model after the regulatory shocks on the interest rate.  We discuss the different values of the SCR modules first in a framework with moderate interest rates using the shocks of the present legislation, and then we consider a low interest framework with the latest recommandation of the EIOPA on the shocks. In both cases, we illustrate the importance of matching cash-flows and its impact on the SCR. 
\end{abstract} 

\section{Introduction}

Life insurance contracts are very popular in the world and involve very large portfolios. In 2017, the life insurer assets were about $7.5$ trillions of euros in Europe (source: Insurance Europe) and $7.2$ trillions of dollars in the United States (source: American Council of Life Insurers). To manage these large portfolios on a long run, insurance companies perform what is called an Asset and Liability Management (ALM). We refer to the recent paper~\cite{Albrecher_etal} for an overview of the current topics and issues of ALM.  Basically, insurance companies invest the deposit of policyholders in different asset classes (equity, sovereign bonds, corporate bonds, real estate, ...), while respecting a performance warranty with a profit sharing mechanism for the policyholders. Thus, insurance companies have to determine an appropriate allocation between the different types of asset. This allocation should be a good trade off between risk and returns, but also with the capital requirement imposed by the regulator to handle the portfolio.  To determine a suitable allocation strategy, it is worth to rely on an ALM model that takes into account the main specifies of the life insurance business.  

Many works in the literature have dealt with the fair valuation of insurance liabilities, see e.g. Briys and de Varenne~\cite{BrDV}, Bacinello~\cite{Bacinello} or more recently Delong et al.~\cite{DeDhBa}. However, when handling large portfolios of life insurance liabilities over on a long run, the fair valuation of the contracts under a risk neutral setting is not the only issue. The insurance company is also interested in investing the policyholders deposit optimally, which has to be made under the real world probability, like in the pioneering work of Merton~\cite{Merton}. Besides, the insurance company may also want to minimize or at least impose an upper bound on the Solvency Capital Requirement (SCR) related to this portfolio of liabilities. To address these questions, it is necessary to have a reliable ALM model that describes properly the life insurance business. One important specificity of these models is to keep track of both market and book values in order to determine the realized gains and losses that enter in the profit sharing mechanism. Up to our knowledge, one of the first model of this kind has been proposed by Gerstner et al.~\cite{GGHGH}. They consider an insurer that invests in bonds with a buy and hold strategy on zero-coupon bonds with a fixed maturity. The insurer keeps a constant allocation proportion in bonds, which leads sometimes to short sell bonds. The crediting rate to policyholders is the same as the one proposed by Grosen and J{\o}rgensen~\cite{GrJo}: it is basically the maximum of the guaranteed interest rate and the profit sharing rate. However, the book value is approximated without neither taking into account the history of trading nor the difference between buy and sell orders for updating of book values. Berdin and Grundl~\cite{BeGr} fill this gap and calculate the book values according to the German GAAP (General Accepted Accounting Principles). They also consider the investment in different asset classes and across different bond maturities, which is then more precisely described by Berdin et al.~\cite{BeKoPa}.

In this paper, we present a new ALM model that incorporates the main features of life insurance business and handles both market and book values. For simplicity of exposition, we consider only two asset classes: equity and riskless bonds, that we consider as a good approximation of top-rated sovereign bonds. The first original feature of our model brings on the determination of the crediting rate to policyholders. To determine this rate, we take into account the gains and losses made during the reallocation and the corresponding profit sharing rate. We consider also a competitor rate that model the rate given by competing insurance companies to their policyholders. Then, the insurance company drive its latent gain and losses and the profit sharing reserve in order to reach the targeted crediting rate, if it is possible. Otherwise, it tries to give the best rate possible while keeping a part of the profit sharing reserve, but in any case the crediting rate is above the minimal regulatory rate. Interestingly, the four cases that we distinguish to determine the crediting rate form a good indicator to monitor the ALM business. A second original feature of our model is that it takes into account dynamic surrenders (or lapses): their proportion is modeled as a function of the difference between the crediting and the competitor rates. Such a dynamic surrender rate is also considered in the model proposed by Floryszczak et al.~\cite{FlLCMa}. The third original point of our model is to consider an investment in an equally weighted portfolio of bonds with maturity going from $1$~year to $n$~years. The dynamic of the coupon rates of the different bonds is also precisely described. A similar but different idea is considered in~\cite{BeKoPa}.  The nominal value of the $1$-year bonds enables essentially to match the cash-flow of the surrenders. This is very important to hedge a part of the risk related to interest rates. Our ALM model is written with the French GAAP, but it could quite easily be adapted to other local GAAP.

Our original motivation to design such an ALM model is to evaluate the Solvency Capital Ratio (SCR) by using the Solvency II standard formula, which is part of the regulation of the European Union~\cite{DelReg}. Other papers have recently dealt with the standard formula: Gatzert and Martin~\cite{GaMa} compare the standard formula with internal models that basically use a Value-at-Risk of the basic own funds with a level of 99.5\%, Boonen~\cite{Boonen} compares the Value-at-Risk and the Expected shortfall risk measures by looking at the stress factor in the standard formula that would calibrate these measures.   In our numerical experiments, we evaluate the different SCR modules in our model with a constant allocation between bonds and stocks. We first examine a case with moderate interest rates around 2\% and then case with low interest rates where we use the latest recommendation of EIOPA~\cite{EIOPA2} for the shocks. We interpret the different cases corresponding to the shocks on equity and bonds. Interestingly, we find that interest rate models like Hull and White model that mean reverts toward a parametric curve are not really well suited for the standard formula. They are able to fit the shocks but then the calibrated curve oscillates too much and the model is meaningless. Our  numerical study also points some weaknesses of the standard formula. Last, we illustrate the importance of matching cash-flows in ALM, and discuss how to do it optimally in our model for minimizing the SCR requirement with the standard formula. 

The paper is organized as follows. Section~\ref{sec_ALM} introduce the main notation, presents the ALM model and the mechanism that determine the crediting rate to policyholders. This part is self-contained and does not rely on the model of the different assets, that is presented in Section~\ref{sec_asset_model}. We discuss in particular the choice of the interest rate model in view of the application of the standard formula. Last, Section~\ref{sec_num} presents and discusses the different numerical simulations.

\section{The ALM model}\label{sec_ALM}

We consider an insurance company that has sold life insurance contracts to many policyholders. To be precise, we consider here General Account (GA) guaranteed with profit contracts. We do not consider Unit-Link (UL) type of contracts where policyholders bear the risk due to market
variations, which is clearly simpler for insurance company to handle. GA contracts are mainly described by two drivers: the minimal guaranteed rate $r^G$ that triggers the minimal earnings, and the participation rate $\pi_{pr}\in[0,1]$ that forces the insurer to redistribute this proportion of gain on equity assets. The French legislation imposes that $\pi_{pr}\ge 0.85$ (see~\cite{BMDGL} p.~5). Policyholders do not receive intermediary payments: they are paid only when they exit the life insurance contract.

The insurance company then has to choose a strategic asset allocation that will enable to face up to the liabilities and provide some earnings. In particular, it is interested to assess the Solvency Capital Requirement (SCR) needed to run its strategic asset allocation. Typically, the insurance company invests in different asset classes such as equity, sovereign bonds, corporate bonds, real estate. Here, we will for simplicity consider two type of assets: equity and riskless bonds, the latter being a good approximation of top-rated sovereign bonds. We consider a time horizon $T\in \N^*$, usually greater than thirty years in practice. We will assume that the insurance company only make reallocations at times $t\in\N \cap[0,T)$ in order to reach a portfolio with respective weights $w^s_t\in [0,1]$ and $w^b_t=1-w^s_t$ in equity and bonds. The portfolio is assumed to be static on $(t,t+1)$, and at time~$T$, the portfolio is liquidated. The time unit can be in practice one year or one semester: in this paper, we take a one year unit for our numerical investigations.

We denote by $(S_t)_{t\ge 0}$ the equity asset that can be thought as a stock market index to reflect that the insurance company invests in many different stocks. Concerning interest rate products, we assume that there exist riskless zero-coupon bonds and bonds of any maturity that can be bought at par. Following the notation of Brigo and Mercurio~\cite{BrMe}, we denote by $P(t,t')$ with $t\le t'$ the price of a zero-coupon bond at time~$t$ with maturity $t'$. For simplicity, we assume that the different bonds pay coupon at the same frequency as the portfolio reallocation: thus at time~$t\in \N$, the value of a bond with maturity $t+n$, constant coupon~$c$ and a unit nominal value is given by
\begin{equation}\label{bond_value}
  B(t,n,c)=\sum_{i=1}^n cP(t,t+i) + P(t,t+n),
\end{equation}
and the swap rate given by 
\begin{align}
\Swap(t,n)=\frac{1-P(t,t+n)}{\sum_{i=1}^{n}P(t,t+i)}
\end{align}
is the value of the coupon leading to a unit value of the bond. For $t\in \N$, $n\in \N^*$ and ${\bf c}_t:=(c^i_t)_{i\in\{1,\dots,n\}}$, we consider a portfolio containing for any $i$, $1/n$ bond with maturity $t+i$ and coupon $c^i_t$. We denote
\begin{equation}\label{def_Bbar}\bar{B}(t,n,{\bf c}_t)=\frac{1}{n}\sum_{i=1}^n B(t,i,c^i_t)
\end{equation}
the value of this combination of bonds at time~$t$.

Before to model the mechanism of the ALM management, we also have to specify at which rate policyholders enter or exit. Since our purpose is to evaluate the Solvency Capital Requirement, we will only consider (as recommended in Solvency II) the case where policyholder contracts run off and exclude the arrival of new contracts, even though it could be obviously added to the model. We assume that the proportion of policyholders that exit on the period $(t,t+1)$ for $t\in \N$ is given by $p^e_t \in (0,1)$, and that policyholders exit uniformly on $(t,t+1)$. This corresponds to the case of infinitely many policyholders that exit at a continuous rate $\lambda^e_t=-\log(1-p^e_t)$ on $(t,t+1)$. So, our model assume that there is a large number of policyholders that exit independently, conditionally on the information they have at time~$t$. We will assume that
\begin{equation}\label{def_StructSurr}
  \forall t \in \N, \ p^e_t \ge \underline{p}>0.
\end{equation}
Thus, $\underline{p}$ quantifies the structural surrenders while $p^e_t-\underline{p}$ is the proportion of surrenders that evolves along the time (typically the dynamic surrenders and the mortality variations), that will be modeled afterwards in Section~\ref{sec_asset_model}.

\subsection{Main variables and portfolio initialization at time $t=0$}

The liability of the firm is divided into different reserves that must comply with the local accounting standards. Even if the main principles of the Solvency II directive are followed by most of the countries, there are some specific features from one country to another. In our model, we will take into account the French regulation rules, and we refer to~\cite{BMDGL} for a recent study of the French legal prudential reserve.

The Mathematical Reserve, denoted by the process $(MR_t)_{t \in \{0,\dots,T\}}$, is the main reserve in life insurance. It corresponds to the insurer's debt towards its policyholders. For sake of simplicity, we assume that the initial premium $MR_0$ is paid once and for all (single premium) by policyholders without fees. Thus, the initial value $MR_0$ of this reserve is given by the initial deposit of the policyholders. At the end of each year, the mathematical reserve is reevaluated by annual benefits (the crediting rate) paid by the insurer to the insured party.

The Capitalization Reserve, denoted by the process $(CR_t)_{t \in \{0,\dots,T\}}$, is imposed by the French legislation to buffer the capital gains obtained when selling bonds. The purpose of this reserve is twofold. First, it dissuades insurance companies from using interest rate movements to make profits on bonds, since it may impact negatively its policyholders on the long-run (typically, capital gains on bonds result in lower coupons). Second, it acts as a cushion against interest rate movements as the fund stored in the reserve can be used later on in order to absorb capital losses coming from selling bonds.
The capitalization reserve is a part of the equity capital of the insurance company.

Last, the Profit-Sharing reserve denoted by the process $(PSR_t)_{t \in \{0,\dots,T\}}$ is a legal provision used as a capital buffer against stock movements in order to smooth the crediting rate. A fraction of the capital gains obtained from selling equity are stored in this reserve and is distributed the next years. This reserve belongs to the policyholders: the French legislation imposes a maximum of 8 years to redistribute the accumulated profit to policyholders.

The capital gain of the insurance company is determined by the difference between the book value (i.e. the purchase price) and the market value of the sold assets. Thus, we denote respectively by $BV^s_t$, $BV^b_t$ and $BV_t=BV^s_t+BV^b_t$ the book values of the equity assets, the bonds and of the whole portfolio at time~$t$. We similarly denote  $MV^s_t$, $MV^b_t$ and $MV_t=MV^s_t+MV^b_t$ the respective market values. It is clear how to evaluate market values, and we will explain later on how book
values are calculated in our model. 

All these quantities $MR_t$, $CR_t$, $PSR_t$, $BV^s_t$, $BV^b_t$, $MV^b_t$ and $MV^s_t$ remain nonnegative for all $t\in \{0,\dots, T \}$ in our ALM model. 

At inception ($t=0$),  the insurance company receives all the policyholders deposit $MR_0$ and invests this amount in a reference portfolio according to target proportions $w^s_0$ in stocks and $w^b_0$ in bonds. We furthermore make the following hypothesis: the company has no existing back book of contracts sold in the past. Thus, there are no capitalization and profit-sharing reserves, and book values and market values coincides:
$$CR_0=PSR_0=0,\ BV^b_0=MV^b_0=w^b_0 MR_0 \text{ and } BV^s_0=MV^s_0=w^s_0 MR_0.$$
However, it would be easy at this stage to consider existing back book of contracts by initializing accordingly these values. 
We now specify the quantity of assets in each class. Let us note that during all the time, the quantities hold by the insurer are nonnegative. The initial holding $\phi^s_0$ in the stock asset is clearly given by
$$\phi^s_0=\frac{w^s_0MR_0}{S_0}. $$
Concerning the bonds, we assume that all bonds at bought at par during the whole strategy. We assume that the amount $w^b_0MR_0$ is invested in an asset which is an equally weighted basket (with weights $1/n$) of riskless coupon bearing bonds from maturity $1$ to $n$ with unitary face value. We thus set
$$c^i_0=c_{swap}(0,i), \ i=1,\dots,n.$$
From the definition of the swap rate, we have $\bar{B}(0,n,{\bf c}_0)=1$, with ${\bf c}_0=(c^i_0)_{i\in \{1,\dots,n\}}$ and $\bar{B}$ defined by~\eqref{def_Bbar}. Thus, the quantity of asset is simply given by
$$\phi^b_0= \frac{w^b_0MR_0}{\bar{B}(0,n,{\bf c}_0)}= w^b_0MR_0. $$
Here, we stress that we consider the investment in a basket of bonds with different maturities instead of only one bond. Thus, the insurance company is able to match a part of the cash outflows: the nominal value of the one-year bond is mainly used to pay back the policyholders that exit during the first year. This cash-flow matching is commonly used to hedge against interest-rate risk and is known as a so-called immunization technique. In order to hedge the minimal rate of surrenders $\underline{p}$, a natural choice is to take $n=\left\lceil \frac 1 {\underline{p}} \right\rceil$. If one wants to take into accounts  the additional market surrenders $p^e_t-\underline{p}$, it can be relevant to take more generally $n\le \left\lceil \frac 1 {\underline{p}} \right\rceil$, $n$ being roughly speaking an average of $1/p^e_t$. The choice of $n$ and the influence of this parameter will be discussed in the numerical section.

\subsection{Reallocation, claim payment and margin at time~$t\in\{1,\dots,T-1\}$}

This subsection presents the different steps of the portfolio reallocation at time $t\in\{1,\dots,T-1\}$. In particular, it describes the composition of cash-inflows and outflows, the legal profit-sharing mechanism, a way to determine the crediting rate for policyholders and the accounting margin for shareholders. The goal of the reallocation is to end with an asset side that is allocated according to the weights $w^s_t$ in equity and $w^b_t$ in an equally weighted portfolio of bonds with weights $1/n$ as described in~\eqref{def_Bbar}.  This amounts to have the quantities
$$\phi^s_t=\frac{w^s_tMV_t}{S_t} \text{ and } \phi^b_t=\frac{w^b_tMV_t}{\bar{B}(t,n,{\bf c}_t)} $$
at the end of the the reallocation procedure, where $MV_t$ is the market value of the assets and  ${\bf c}_t=(c^i_t)_{i\in \{1,\dots,n\}}$ are coupon values that will be precised afterwards. Concerning book values, since the capitalization reserve is managed with a separate accounting, the goal is to have at the end of the reallocation: $BV_t=MR_t+PSR_t$, i.e. that the liabilities exactly match the portfolio book value. The corresponding balance sheet is given in Table~\ref{BS_time_t}. For all balance sheets, the sum of the asset book values is equal to the sum of the liabilities. 
\begin{table}[htp]
\centering
\begin{tabular}{|l|l|} \hline
Assets & Liabilities\\ \hline\hline
$BV^s_{t}$ &$MR_{t}$\\ 
$BV^b_{t}$ &$PSR_{t}$\\ \hline
\end{tabular}
\caption{Book value balance sheet after the reallocation at time~$t$.}
\label{BS_time_t}
\end{table}

We present the whole reallocation procedure in five steps. For convenience, we set $t_1=t_2=\ldots=t_5=t$ the different steps of the ALM management procedure at time~$t$, the step $i+1$ being immediately executed after the step $i$. Some quantities are updated only once during the five steps, and we use then the index~$t$ for these quantities. Other quantities, such as the book value are updated at different steps, and we note $BV_{t_i}$ the book value after step~$i$ and $BV_t$ the book value after the last update in the whole procedure. Note that these quantities (except market values) are then kept constant on $(t,t+1)$ until the next reallocation. 

\subsubsection{Step 1: cash inflows}\label{paragraphe_Step1}
We recall that we assume that portfolio is static on $(t-1,t)$ and therefore the quantity of equity (resp. bond) assets at the beginning of the reallocation is $\phi^s_{t-1}$ (resp. $\phi^b_{t-1}$). More precisely, for all $i\in \{1,\dots,n\}$, the insurance company holds $\phi^b_{t-1}/n$ bonds with maturity $t-1+i$ and coupon $c^i_{t-1}$. The financial income corresponding to the coupon payments from each bond is thus given by
\begin{align}
FI_{t}=\phi^b_{t-1}\left(\frac{1}{n}\sum_{i=1}^{n}c^{i}_{t-1}\right).
\end{align}
Besides, the nominal value coming from matured bond is given by $\frac{\phi^b_{t-1}}{n} $. Thus, the insurer's overall cash inflow $CIF$ is obtained by aggregating terms: 
\begin{align}\label{cash_inflow}
CIF_{t}=FI_{t}+\frac{\phi^b_{t-1}}{n}
\end{align}
The book value in bond assets has to be updated. Following standard accounting procedures, the nominal value of the matured bonds have to be removed from the book value. We thus set
\begin{align*}
BV^b_{t_1}=BV^b_{t-1}-\frac{\phi^b_{t-1}}{n}
\end{align*}
In order to satisfy the bookkeeping condition, the insurer must redistribute the income $FI_t$ on the liability side. Table 2 sums up the insurer balance-sheet after step 1.
\begin{table}[htp]
\centering
\begin{tabular}{|l|l|} \hline
Assets & Liabilities\\ \hline\hline
$BV^s_{t-1}$ &$MR_{t-1}$\\ 
$BV^b_{t_1}$ &$PSR_{t-1}$\\ \hline
$CIF_{t}$ &$FI_{t}$\\ \hline
\end{tabular}
\caption{Book value balance sheet after step~1.}
\label{table_BV_t1}
\end{table}
\subsubsection{Step 2: claim payment}\label{paragraphe_Step2}
Cash outflows occur when policyholders exit their contract. We recall that the proportion of policyholders that exit on $(t-1,t)$ is given by $p^e_{t-1}$. We assume that these policyholders are paid with the minimum guaranteed rate $r^G$, pro rata the time elapsed between $t-1$ and the exit. Since we 
we assume that they exit uniformly on $(t-1,t)$, this amounts to the cash outflow
\begin{align}
COF_{t}=p^e_{t-1} MR_{t-1}\left(1+\frac{r^G}{2}\right)
\end{align}
On the liability side, the liabilities corresponding to remaining policyholders is then given by
\begin{align*}
MR_{t_2}=(1-p^e_{t-1})MR_{t-1}
\end{align*}
On the asset side, the difference between cash inflows $CIF_{t}$ and cash outflows $COF_{t}$ is called the liquidity gap $G_{t}$:
\begin{align*}
G_{t}=CIF_{t}-COF_{t}=CIF_{t}-p^e_{t-1} MR_{t-1}-\frac{r^G}{2}p^e_{t-1} MR_{t-1}
\end{align*}
A positive gap $G_{t}>0$ means that assets inflows are sufficient to cover claims. A negative gap $G_{t}<0$ means that additional liquidity is necessary to pay claim-holders. To fill the funding gap, the insurer must sale assets in this situation. Table~\ref{BS_t2} depicts the insurer balance after claim payment.
\begin{table}[htp]
\centering
\begin{tabular}{|l|l|} \hline
Assets & Liabilities\\ \hline\hline
$BV^s_{t-1}$ &$MR_{t_2}$\\ 
$BV^b_{t_1}$ &$PSR_{t-1}$\\ \hline
$G_{t}$ &$FI_{t}-\frac{r^G}{2}p^e_{t-1} MR_{t-1}$\\ \hline
\end{tabular}
\caption{Book value balance sheet after step~2.}
\label{BS_t2}
\end{table}
We thus set
\begin{equation}\label{def_FItilde}
  \widetilde{FI}_t=FI_{t}-\frac{r^G}{2}p^e_{t-1} MR_{t-1},
\end{equation}
that represents the coupon income corrected with the part of these earnings that are distributed to the surrendering policyholders.

\subsubsection{Step 3: reallocation}\label{paragraphe_Step3}
We assume that the insurer follows a static investment strategy  on $(t,t+1)$ and allocates its capital according to the portfolio weights $w^s_t$ between stocks and bonds $w^b_t$ at time~$t$. The available capital, which is also the market value of the portfolio $MV_{t}$, is given by the sum of the liquidity gap $G_{t}$ and the market value of each asset classes:
\begin{align}\label{def_MVt}
MV_{t}=G_{t}+\phi^s_{t-1}S_t+\frac{\phi^b_{t-1}}{n} \sum_{i=1}^{n-1}B(t,t+i,c^{i+1}_{t-1}),
\end{align}
where the function $B$ is defined by~\eqref{bond_value}. The term $\phi^s_{t-1}S_t$ is the market value of the equity and $\phi^b_{t-1}\left(\frac{1}{n}\sum_{i=1}^{n-1}B(t,t+i,c^{i+1}_{t-1})\right)$ is the market-value of the bonds that have not  reached maturity.  Note that before the reallocation, the bond portfolio is made with bonds with maturity $i\in\{1,\dots,n-1\}$ and coupon $c^{i+1}_{t-1}$.

In what follows, we assume that $MV_t>0$ and calculate the new quantities invested in each asset-classes and derive the procedure  to update the book values of each asset class. The case $MV_t\le 0$ is very unlikely but may theoretically happen, for example if the surrendering proportion $p^e_{t-1}$ is high, the stock value has strongly decreased ($S_t/S_{t-1}<<1$) and interest rates have strongly increased on $(t-1,t)$. In this case we assume that the shareholders of the insurance company directly pay back  $COF_{t}$ to the surrenders and that the portfolio is reallocated according to the target weights with a self-financing strategy, so that its market value remains positive. The book values are modified accordingly and the crediting rate is then $r^G$. Since the case $MV_t\le 0$ never happens in usual conditions and in our simulations, we do not give more detail.

We first consider the equity, where the target is to achieve a proportion $w^s_t$ of the market value $MV_{t}$. This leads to a new position in stock given by:
\begin{align*}
\phi^s_{t_3}=\frac{w^s_tMV_{t}}{S_t}>0
\end{align*}
We note $\Delta\phi^s_{t}=\phi^s_{t_3}-\phi^s_{t-1}$ the variation of the number of equity assets hold by the insurer. If $\Delta\phi^s_{t}\geq 0$ (buy order), the book value in equity is increased by the quantity of stocks that was purchased at the market-value~$S_t$:
\begin{align*}
BV^s_{t_3}=BV^s_{t-1}+\Delta \phi^s_{t} S_t
\end{align*}
If $\Delta\phi^s_{t}< 0$, the insurer sells the quantity $-\Delta\phi^s_{t}$ of equity assets. 
In accounting, a standard inventory valuation method used by practitioners is the First In First Out (FIFO) method where the oldest goods purchased are sold in priority. The realized Capital Gain or Loss (CGL) is then calculated accordingly. However, This procedure requires to record the entire history of all purchases and is computationally demanding. Here, we consider the following approximation  
\begin{align*}
BV^s_{t_3}=BV^s_{t-1}\left(1+\frac{\Delta\phi^s_{t}}{\phi^s_{t-1}}\right)=\frac{\phi^s_{t_3}}{\phi^s_{t-1}}BV^s_{t-1},
\end{align*}
which amounts to say that all the equity asset units hold in the portfolio have the same book value. 
The proportional reduction factor $\frac{\Delta\phi^s_{t}}{\phi^s_{t-1}}\in ]-1,0]$ represents the proportion of sold stock. The capital gain or loss made by the sale is then given by $CGL^s_t=-\Delta\phi^s_{t}(S_t-BV^s_{t-1}/\phi^s_{t-1})$.

Let us recall that for $x\in \R$, $x^+=\max(x,0)$ and $x^-=\max(-x,0)$. We sum up the equity book value and the capital gain or loss (regardless of it is a sale or a purchase):
\begin{align}\label{BV_t3}
  BV^s_{t_3}&=BV^s_{t-1}+\left(\Delta\phi^s_{t}\right)^+S_t-\frac{\left(\Delta\phi^s_{t}\right)^-}{\phi^s_{t-1}}BV^s_{t-1}\\
  CGL^s_t&=(\Delta\phi^s_{t})^-\left(S_t-\frac{BV^s_{t-1}}{\phi^s_{t-1}}\right).
\end{align}

We now focus on the reallocation in bonds and recall that we assume that the insurer only buys bonds at par. Before the reallocation,  the bond portfolio is made with bonds with time to maturity going from $1$ to $n-1$. In order to continue the strategy of matching the cash flows coming from the structural surrenders, the insurer needs to invest in a basket of bonds with longest time to maturity equal to $n$. Thus, the insurer always has to buy the bond with longest time to maturity $n$. Let us introduce the following reference  market value 
\begin{align}
\widehat{MV}^b_t=\phi^b_{t-1}\left(\frac{1}{n}\sum_{i=1}^{n-1} B(t,t+i,c^{i+1}_{t-1})+\frac{1}{n}B(t,t+n,c_{swap}(t,n))\right).
\end{align}
This is the market value of the bond portfolio, if the insurer would buy exactly the same quantity $\phi^b_{t-1}/n$ of bonds with time to maturity $n$. If $w^b_tMV_t=\widehat{MV}^b_t$, the insurer thus only buys $\phi^b_{t-1}/n$ bonds with time to maturity~$n$ to reach the target allocation. If   $w^b_tMV_t>\widehat{MV}^b_t$, he has to buy more bonds and if $w^b_tMV_t<\widehat{MV}^b_t$ he has to sell some bonds. In what follows, we describe how to do it, while keeping an equally weighted basket of bonds with time to maturity  going from~$1$ to $n$.

\subsubsection*{Purchase of bonds ($w^b_tMV_{t}\geq \widehat{MV}^b_t$)}
In this case, the insurer needs to buy more bonds to satisfy the target~$w^b_t$. We note ${\bf c}^{swap}_t=(c_{swap}(t,i))_{\in \{1,\dots, n\}}$ and have $\bar{B}(t,n,{\bf c}^{swap}_t)=1$ from~\eqref{def_Bbar}. We then define
\begin{align}\label{def_deltab}
\delta^b_t=w^b_tMV_{t}-\widehat{MV}^b_t\ge 0,
\end{align}
so that $w^b_tMV_{t}=\widehat{MV}^b_t+\delta^b_t\bar{B}(t,n,{\bf c}^{swap}_t)$. 
The insurer will then 
\begin{itemize}
\item buy $\frac{\delta^b}{n}$ at par bonds for each time to maturity~$i\in \{1,\dots, n-1\}$ with coupon $c^{i}_{swap}(t)$,
\item buy $\frac{\delta^b+\phi^b_{t-1}}{n}$ at par bonds with time to maturity~$n$ and coupon $c^{n}_{swap}(t)$.
\end{itemize}
Let us recall now that holding $\alpha>0$ bonds with coupon $c$ and $\alpha'\ge 0$ bonds with coupon $c'$ and the same payment schedule is equivalent to hold $\alpha+\alpha'$  bonds with coupon $\frac{\alpha c + \alpha' c}{\alpha+\alpha'}$. Therefore, after the bond reallocation, the insurance company holds for each $i\in \{1,\dots,n\}$,  $(\delta^b+\phi^b_{t-1})/n$ bonds with time to maturity $i$ and coupon
\begin{align}\label{coupon_rollup}
c^{i}_t=\mathbf{1}_{i\le n-1}\frac{\phi^b_{t-1}c^{i+1}_{t-1}+\delta^b c^{i}_{swap}(t)}{\phi^b_{t-1}+\delta^b} + \mathbf{1}_{i= n}c^{n}_{swap}(t).
\end{align}
We can therefore write the market value of the bond portfolio as
\begin{equation}\label{bond_realloc_achat}
  MV^b_t=\phi^b_{t_3}\bar{B}(t,n, {\bf c}_t), \text{ with } \phi^b_{t_3}=\delta^b+\phi^b_{t-1},
\end{equation}
and in particular we have $ \phi^b_{t_3}\ge \phi^b_{t-1}$.

We now have to update the book value of bonds. Since there are only purchases, the book value of the bought bonds is their market value. We thus set 
\begin{align*}
  BV^b_{t_3}&=BV^b_{t_1}+\frac{\delta^b}{n}\sum_{i=1}^{n-1}B(t,t+i,c^{i}_{swap}(t))+\frac{\delta^b+\phi^b_{t-1}}{n}B(t,t+n,c^{n}_{swap}(t)) \\
  &=BV^b_{t_1}+\delta^b+\frac{\phi^b_{t-1}}{n}.
\end{align*}
\subsubsection*{Sale of bonds ($w^b_tMV_{t}< \widehat{MV}^b_t$) }
When $w^b_tMV_{t}< \widehat{MV}^b_t$, the insurer still has to buy bonds with time to maturity~$n$, but he has to sell the other bonds to get an equally weighted bond portfolio. Thus, he has to find a position such that
\begin{align*}
w^b_tMV_{t}=\phi^{b}_{t_3}\left(\frac{1}{n}\sum_{i=1}^{n-1}B(t,t+i,c^{i+1}_{t-1})+\frac{1}{n}B(t,t+n,c^{n}_{swap}(t))\right).
\end{align*}
Note that we necessarily have  $\phi^{b}_{t_3}<\phi^{b}_{t-1}$, since the right-hand side corresponds to $\widehat{MV}^b_t$ for $\phi^{b}_{t_3}=\phi^{b}_{t-1}$. This gives 
\begin{align*}
\phi^{b}_{t_3}=\frac{w^b_tMV_{t}}{\frac{1}{n}\sum_{i=1}^{n-1}B(t,t+i,c^{i+1}_{t-1})+\frac{1}{n}B(t,t+n,c^{n}_{swap}(t))}>0,
\end{align*}
and the market value of the bond portfolio can be written as
$$ MV^b_t=\phi^b_{t_3}\bar{B}(t,n, {\bf c}_t), \text{ with } c^{i}_t=\mathbf{1}_{i\le n-1}c^{i+1}_{t-1} +\mathbf{1}_{i= n}c^{n}_{swap}(t) \text{ for } i=1,\dots,n.$$
Let $\Delta\phi^{b}_{t}=\phi^{b}_{t_3}-\phi^{b}_{t-1}<0$.  
The insurer has thus to  buy $\frac{\phi^{b}_{t_3}}{n}$ at par bonds with time to maturity~$n$ and to sell, for each time to maturity $i\in\{1,\dots, n-1\}$,  $\frac{\Delta\phi^{b}_{t}}{n}$ bonds with coupon $c^{i+1}_{t-1}$.

We now have to update the book values. We use the same  approximation method as for the equity to evaluate the book value of the sold bonds. We thus set
\begin{align*}
BV^b_{t_3}=BV^b_{t_1}\left(1+\frac{\Delta\phi^{b}_{t}}{\phi^b_{t-1}}\right)+\frac{\phi^{b}_{t_3}}{n}B(t,t+n,c^{n}_{swap}(t))=BV^b_{t_1}\left(1+\frac{\Delta\phi^{b}_{t}}{\phi^b_{t-1}}\right)+\frac{\phi^{b}_{t_3}}{n},
\end{align*}
and the capital gain or loss on bond products is then given by
\begin{align*}
CGL^b_t=-\Delta\phi^{b}_{t}\left(\frac{1}{n}\sum_{i=1}^{n-1}B(t,t+i,c^{i+1}_{t-1})-\frac{BV^b_{t_1}}{\phi^b_{t-1}}\right)
\end{align*}

In the two following formulas, we sum up the book value update and the capital gain and loss on bonds in both selling and buying cases 
\begin{align}\label{BVb_update_t3}
  BV^b_{t_3}&=BV^b_{t_1}\left(1-\frac{(\Delta\phi^{b}_{t})^-}{\phi^b_{t-1}}\right)+\frac{n-1}{n}(\Delta\phi^{b}_{t})^+ + \frac{\phi^{b}_{t_3}}{n},\\
  CGL^b_t&=(\Delta\phi^{b}_{t})^-\left(\frac{1}{n}\sum_{i=1}^{n-1}B(t,t+i,c^{i+1}_{t-1})-\frac{BV^b_{t_1}}{\phi^b_{t-1}}\right),
\end{align}
since $\Delta\phi^{b}_{t}=\delta_b$ and $\phi^b_{t_3}=\phi^{b}_{t-1}+\delta_b$ in the buying case.

The capital gain or loss on equity $CGL^s_t$ is directly taken into account for the profit sharing mechanism. Instead, the capital gain and loss on bonds $CGL^b_t$ is handled separately in the French legislation and supply the capitalization reserve. Precisely, the capitalization reserve at time~$t$ is defined by
\begin{align}\label{def_RCt}
\RC_t=\left(\RC_{t-1}+ CGL^b_t\right)^+.
\end{align}
If $\RC_{t-1}+ CGL^b_t<0$, this quantity reduces the insurer return of the period. Since the capitalization reserve is managed with a separate accounting, only 
\begin{align}
\Delta \RC_t=CR_t-CR_{t-1}
\end{align}
appears in the balance sheet at step~3, see Table~\ref{table_BV_t3}. 
\begin{table}[htp]
\centering
\begin{tabular}{|l|l|} \hline
Assets & Liabilities\\ \hline\hline
$BV^s_{t_3}$ &$MR_{t_2}$\\ 
$BV^b_{t_3}$ &$PSR_{t-1}$\\
$CGL^s_t$ &$\Delta CR_t$\\ 
$CGL^b_t$ &\\ 
$G_{t}$ &$\widetilde{FI}_{t}+CGL^s_t-\left(CR_{t-1}+CGL^b_t \right)^-$\\ \hline
\end{tabular}
\caption{Book value balance sheet after step~3.}
\label{table_BV_t3}
\end{table}
\subsubsection{Step 4: determination of the crediting rate}\label{paragraphe_Step4}
In order to determine the policyholder's earning rate $r_{ph}(t)$ on the period $(t-1,t)$, we propose a management decision that follows the regulatory constraints and is a reasonable trade-off between policyholders and shareholders interests. Most of the existing ALM model use a crediting rate that has been proposed by  Grosen and J{\o}rgensen~\cite{GrJo} which is the minimal regulatory rate.  Here, we propose a more sophisticated model for the crediting rate that we believe closer to the practice. It involves a competitor rate and a control of the Latent Gain or Loss (LGP) and Profit Sharing Reserve.

The existence of LGL, sometimes also called hidden reserve,  results from the difference between market and book values. Formally they can be realized by selling and buying instantly the same amount of assets, but in practice this is just an account entry.  It is a variable that the insurer can use as a control to determine the crediting rate for policyholders. In what follows we consider only LGL for stock assets since CGL on bond are constrained by the capitalization reserve and cannot be redistributed to policyholders nor shareholders.
Let $MV^s_{t}=w^s_tMV_{t}$ the current market value of equity assets. The range of latent gain or loss can be described by the following interval
$$ [-(MV^s_{t}-BV^s_{t_3})^-,(MV^s_{t}-BV^s_{t_3})^+].$$
There is a latent gain if $MV^s_{t}>BV^s_{t_3}$, and a latent loss if $MV^s_{t}<BV^s_{t_3}$.
We define the latent gain or loss function by
\begin{align}
LGL^s_t(\alpha)=-\left(1-\alpha\right)(MV^s_{t}-BV^s_{t_3})^-+\alpha(MV^s_{t}-BV^s_{t_3})^+\quad\alpha\in[0,1]
\end{align}
This function determines the amount of hidden reserve to distribute. Let us note that $LGL^s_t(\alpha)$ is non-decreasing with respect to $\alpha$. The control $\alpha\in[0,1]$ models the fraction of LGL to register on the balance-sheet. The choice $\alpha=1$ amounts to take all the gain or zero loss. The control $\alpha=0$ takes all the loss or zero gain.
\begin{remark}\label{rk_samesign}
  We have  $CGL^s_t LGL^s_t(\alpha)\ge 0$, i.e. $CGL^s_t$ and $LGL^s_t(\alpha)$ have the same sign. If $\Delta \phi^s_t\ge 0$, this is true since $CGL^s_t=0$. If  $\Delta \phi^s_t< 0$, one has to notice that we have
  $$\frac{BV_{t-1}^s}{\phi^s_{t-1}}=\frac{BV^s_{t_3}}{\phi^s_{t_3}}$$
  from~\eqref{BV_t3}. Therefore $S_t-\frac{BV_{t-1}^s}{\phi^s_{t-1}}$ is equal to $S_t-\frac{BV_{t_3}^s}{\phi^s_{t_3}}=(MV^s_t-BV_{t_3}^s)/\phi^s_{t_3}$, and these quantity have the same sign.
\end{remark}

Another control for the insurer is the proportion~$\rho \in (0,1]$ of profit sharing reserve to distribute. For simplicity, we will assume here that
  $$\rho\in\{\bar{\rho},1\},$$
  where $\bar{\rho} \in (0,1]$ is fixed. The insurer has then two possible choices: to use all the profit sharing reserve ($\rho=1$) or to use only a part of it. Let us note that in our model, taking $\bar{\rho}=1$ amounts to have no profit sharing reserve. 
\begin{remark}
    Since $(1/2)^8 \approx 0.004$, we take in our experiments $\bar{\rho}=1/2$  to be in line with the French legislation that requires to redistribute all the profit within 8 years. This is also the choice made by Berdin and Grundl~\cite{BeGr} (see equation~(22) therein) who work under the German rule.
\end{remark}

Due to the participation rate, the minimal crediting rate depends on $\alpha$ and $\rho$. In case of latent gain ($MV^s_{t}\ge BV^s_{t_3}$), we note
 \begin{align}\label{TD_Gain}
   TD_t(\alpha,\rho)=\widetilde{FI}_{t}-\left(RC_{t-1}+CGL^b_t \right)^-&+\rho \left(PSR_{t-1}+CGL^s_t+LGL^s_t(\alpha))\right)
 \end{align}
 the amount that has to be redistributed to policyholders according to the participation rate~$\pi_{pr}$. The first term corresponds to the coupon payment~\eqref{def_FItilde}. The second term is the loss on the bonds that exceeds the capital reserve. The third term corresponds to the aggregated gains on equity and is nonnegative by Remark~\ref{rk_samesign}.

 In case of latent loss ($MV^s_{t}\ge BV^s_{t_3}$), we define
 \begin{align}\label{TD_Loss}
   TD_t(\alpha,\rho)=\widetilde{FI}_{t}-\left(RC_{t-1}+CGL^b_t \right)^-+\rho PSR_{t-1} + CGL^s_t+LGL^s_t(\alpha) 
 \end{align}
 the amount to be redistributed with the participation rate. From Remark~\ref{rk_samesign}, $CGL^s_t$ and $LGL^s_t(\alpha)$ are nonpositive. Contrary to the gains, the insurer does not smooth the losses with a factor~$\rho$.  

 We now sum up the  amount to distribute~\eqref{TD_Gain} by the following formula  that covers both capital gain or loss cases:
 \begin{align}
   TD_t(\alpha,\rho)&=\widetilde{FI}_{t}-\left(RC_{t-1}+CGL^b_t \right)^-+\rho \left(PSR_{t-1} + CGL^s_t+LGL^s_t(\alpha) \right) \label{TD_gen}\\ &\phantom{=\widetilde{FI}_{t}-\left(RC_{t-1}+CGL^b_t \right)^-} -(1-\rho)\left(CGL^s_t+LGL^s_t(\alpha) \right)^-. \notag
 \end{align}
 We have the following straightforward but important property.
 \begin{lem}\label{lemme_TD} The function $(\alpha,\rho)\in [0,1]^2 \mapsto TD_t(\alpha,\rho)$ is continuous and nondecreasing with respect to $\alpha$ and $\rho$. It is constant with respect to $\alpha$ if $S_t=BV^s_{t-1}/\phi^s_{t-1}$, otherwise it is increasing and affine with respect to $\alpha$. 
 \end{lem}

 We are now able to define the minimal distribution of returns that the insurance has to give to the (remaining) policyholders. The minimum guaranteed amount $R^G_t(\alpha,\rho)$ is defined by
\begin{align}
R^G_t(\alpha,\rho)=\max\left\lbrace R^G_t,\pi_{pr}TD_t(\alpha,\rho)\right\rbrace, \text{ with } R^G_t=r^G(MR_{t_2}+PSR_{t-1})
\end{align}
Note that the part of the profit sharing reserve  has to be credited exactly as the mathematical reserve, since this reserve belongs to policyholders. 
Here, $R^G_t$ is the minimum regulatory amount corresponding to the minimum rate $r^G$. Note that in practice, the  minimum regulatory rate is the maximum of~$r^G$ and of 60\% of a technical rate called ``taux moyen des emprunts d'Etat'' that is an average of French sovereign bond rates, see for example paragraph A.6.2 of~\cite{BeKoPa}. Here, we assume for simplicity that $r^G$ remains above this technical rate.

Beyond the minimum rate, the insurance company wants to credit at least the same rate than the other insurance companies in order to keep its policyholders. In fact, the surrender proportion $p^e_t$ on $(t,t+1)$ usually depends on the difference between the crediting rate and the one of the other insurers. 
Thus, we assume that $r^{comp}_t$ is a competitor rate. Typical choices can be \begin{equation}\label{def_r_comp}r^{comp}_t=r_t \text{ or }r^{comp}_t=\max(r_t,\eta r_{ph}(t-1)) \text{ with } \eta\in (0,1),
\end{equation}
where $r_{ph}(t-1)$ is the crediting rate of the past period and $r_t$ is the short interest rate. We define the target crediting amount by
\begin{align}
R^\sigma_t(\alpha,\rho)=\max\left\lbrace R^G_t(\alpha,\rho); R^{comp}_t\right\rbrace,  \text{ with } R^{comp}_t=r^{comp}_t(MR_{t_2}+PSR_{t-1}). 
\end{align}
This is the amount that the insurance company try to distribute if possible, which we discuss now.\\

We now determine $\alpha_t$, $\rho_t$ and the amount $R^{ph}_t$ to be credited to policyholders. By Lemma~\ref{lemme_TD}, we know that we are in one of the following four distinct cases, going from the more to the less favorable case for the insurer and the policyholder.
\begin{enumerate}
  \item[Case~A:] $\pi_{pr}TD_t(0,\bar{\rho})\geq \max\left\lbrace R^G_t;R^{comp}_t\right\rbrace$.\\
This means that the target amount can be credited to policyholders without dissolving unrealized gains if any or by realizing all the latent losses. The insurer decides then to take 
 $$\alpha_t =0, \ \rho_t=\bar{\rho},$$
 and credit the target amount $R^{ph}_t=R_t^\sigma(\alpha_t,\rho_t)=\pi_{pr}TD_t(\alpha_t,\rho_t)$  to policyholders.
\item[Case~B:]   $\pi_{pr}TD_t(1,\bar{\rho})\geq \max\left\lbrace R^G_t;R^{comp}_t\right\rbrace$  and $\pi_{pr}TD_t(0,\bar{\rho})< \max\left\lbrace R^G_t;R^{comp}_t\right\rbrace$.\\
  This means that the target amount can be credited to policyholders, but the insurer has to realize some latent gain or cannot realize all the latent loss. We assume that the insurer decides to realize as less (resp. much) as possible latent gains (resp. losses). Note that by Lemma~\ref{lemme_TD}, the function $\alpha\mapsto TD_t(\alpha,\bar{\rho})$ cannot be constant in Case~B, and the insurer has to find  the value $\alpha$ such that $\pi_{pr}TD_t(\alpha,\bar{\rho})= \max\left\lbrace R^G_t;R^{comp}_t\right\rbrace$. Lemma~\ref{lemme_TD} gives that the function is affine with respect to $\alpha$, and therefore
  $$\alpha_t= \frac{\frac{1}{\pi_{pr}}\max\left\lbrace R^G;R^{comp}\right\rbrace-TD_t(0,\bar{\rho})}{TD_t(1,\bar{\rho})-TD_t(0,\bar{\rho})}.$$
  The insurer also takes $\rho_t=\bar{\rho}$ and credits then  $R^{ph}_t=R_t^\sigma(\alpha_t,\rho_t)=\pi_{pr}TD_t(\alpha_t,\rho_t)$ to policyholders.
\item[Case~C:]  $ R^G_t \le \pi_{pr}TD_t(1,\bar{\rho})< \max\left\lbrace R^G_t;R^{comp}_t\right\rbrace$ \\
 The target amount cannot be reached with the available latent resources, but the minimal guaranteed rate can be reached. We assume then that the insurer makes its best effort on the latent gains or losses by taking
 $$\alpha_t=1 \text{ and } \rho_t=\bar{\rho}.$$
 The amount $R^{ph}_t=\pi_{pr}TD_t(1,\bar{\rho})$ is thus credited to policyholders.
\item[Case~D:]  $\pi_{pr}TD_t(1,\bar{\rho})<R^G_t$. \\
  In this case, the insurance company uses the whole profit sharing reserve and takes $\rho_t=1$. It also takes $\alpha_t=1$ and credits then $R^{ph}_t=\max(\pi_{pr}TD_t(1,1),R^G_t)$ to the policyholders.
\end{enumerate}
Thus, in all cases, the minimum guaranteed amount $R^G_t(\alpha_t,\rho_t)$ is given to policyholders. We define then the crediting rate and update the mathematical and profit sharing reserves as follows:
\begin{align}\label{tx_servi}
  r_{ph}(t)&=\frac{R^{ph}_t}{MR_{t_2}+PSR_{t-1}}, \\
  MR_t&=MR_{t_2}\left(1+r_{ph}(t)\right), \label{dyn_MR}\\
  PSR_t&= PSR_{t-1}r_{ph}(t)+(1-\rho)\left(PSR_{t-1} +(CGL^s_t+LGL^s_t(\alpha))^+ \right). \label{dyn_PSR} 
\end{align}
The profit sharing reserve at time $t$ is thus obtained as the proportion $1-\rho$ of the realized gains and of the updated profit sharing reserve. 
We also update the book value of stock assets to take into account the realized gain or loss:
$$BV^s_{t_4}=BV^s_{t_3}+LGL(\alpha_t).$$
  The shareholder's margin comprises a percentage $1-\pi_{pr}$ on the amount to be distributed $TD_t(\alpha_t,\rho_t)$ minus its contribution to bail out the company when the minimal amount cannot be met:
 \begin{align}
 AM_t=\left(1-\pi_{pr}\right) TD_t(\alpha_t,\rho_t)-(R^G_t(\alpha_t,\rho_t)-\pi_{pr}TD_t(\alpha_t,\rho_t))^+.
 \end{align}
Note that this contribution is only needed in Case~D when $R^G_t>\pi_{pr}TD_t(1,1)$.
 
 Table 5 details the composition of the Book value Balance sheet after the crediting operation.
 \begin{table}[htp]
\centering
\begin{tabular}{|l|l|} \hline
Assets & Liabilities\\ \hline\hline
 &$MR_{t}$\\ 
 $BV^s_{t_4}$ &$PSR_{t}$\\ 
$BV^b_{t_3}$ &$\Delta CR_t$\\ 
&$AM_t$ \\ \hline
\end{tabular}
\caption{Book value balance sheet after step 4.}
\label{Economic balance sheet}
\end{table}
 As mentioned previously, the capitalization reserve is managed separately from other technical reserves. While the mathematical provision and the profit sharing reserve are linked to the performance of the portfolio, regulatory constraints require to invest the capitalization reserve in sovereign bonds Here, we assume that it is invested in a one period zero-coupon bond. Since the capitalization reserve belongs to the equity capital of the insurance company, the interests coming from the capitalization reserve are given to shareholders. Their  cash flow is then the sum of the accounting margin $AM_t$ and the yield of the capitalization reserve:
 \begin{align}\label{PandL}
 P\&L_t=AM_t+\RC_{t-1}\left(\frac{1}{P(t-1,t)}-1\right).
 \end{align}
 
 \subsubsection{Step 5: externalization of the shareholders' margin and of the capitalization reserve from the accounting}\label{paragraphe_Step5}
 The margin $AM_t$ that determines the accounting return on capital invested by shareholders on $(t-1,t)$ must be removed from the balance-sheet. The same has to be done for the capitalization reserve movement $\Delta\RC_t$, since the capitalization reserve is handled separately. One has then to clear the amount $AM_t+\Delta CR_t$ from the balance-sheet. If this amount were externalized in cash, one would have to calculate the capital gains made on it: the gain on equity assets should then to be distributed to policyholders with the participation rate~$\pi_{pr}$ and the gain on bonds should modify the capitalization reserve. Thus, one would have to repeat the previous steps indefinitely. To avoid this difficulty, we assume if $AM_t+CR_t>0$ that a fraction of the assets corresponding to the accounting value $AM_t+CR_t$ is  removed. This amounts to fund the shareholders' margin and the capitalization reserve with a fraction of the portfolio instead of cash. If $AM_t+CR_t<0$, we simply buy the quantity of assets and bonds with weights $w^s_t$ and $w^b_t$ that corresponds to this book value.

 This procedure leads to the following update of the stock book value 
\begin{align*}
BV^s_{t}=BV^s_{t_4}\left(1-\frac{(AM_t+CR_t)^+}{BV_{t_4}}\right)+w^s_t (AM_t+CR_t)^-,
\end{align*}
where $BV_{t_4}=BV^s_{t_4}+BV^b_{t_3}$ since the bond book value is unchanged at step~4. The corresponding position is
\begin{align*}
\phi^s_{t}=\phi^s_{t_3}\left(1-\frac{(AM_t+CR_t)^+}{BV_{t_4}}\right)+\frac{w^s_t (AM_t+CR_t)^-}{S_t}
\end{align*}
We do the same for the bonds and have
\begin{align*}
BV^b_{t}=BV^b_{t_3}\left(1-\frac{(AM_t+CR_t)^+}{BV_{t_4}}\right)+w^b_t (AM_t+CR_t)^-
\end{align*}
with the corresponding position
\begin{align*}
\phi^b_{t}=\phi^b_{t_4}\left(1-\frac{(AM_t+CR_t)^+}{BV^G_{t_4}}\right)+\frac{w^b_t (AM_t+CR_t)^-}{\bar{B}(t,n,{\bf c}_t)}.
\end{align*}
Note that for simplicity, we assume here that we can buy (when $AM_t+CR_t<0$) bonds with time to maturity~$i$ and coupon $c^i_t$. It would have been possible to buy bonds at par, but this would require then to modify again the coupon rates accordingly, similarly as in step~3.
Table~\ref{BS_time_t} represents the balance-sheet at the end of step 5 and thus at the and of the whole reallocation procedure. 

\subsection{Closing of the strategy at time~$T$}
We now describe how the ALM portfolio is closed at time~$T$. The insurance company starts with the implementation of Step~1 described in paragraph~\ref{paragraphe_Step1}. Things change then since the insurer has to liquidate the portfolio. Since the insurer closes its portfolio, we consider the policyholders that exit on $(T-1,T)$ and the others that exits at time~$T$ in the same way.  All the assets are sold and  all the capital gains or  losses are realized, and the profit sharing reserve is released. The capital gain or loss realized when liquidating the stock portfolio is given by:
\begin{align*}
CGL^s_T=\phi^s_{T-1}\left(S_T-\frac{BV^s_{T-1}}{\phi^s_{T-1}}\right).
\end{align*}
We now focus on the equally weighted basket of bonds. Keeping in mind that the bond with shortest time to maturity has come due, the bond portfolio comprises bonds from maturity $1$ to $n-1$, and the capital gain or loss is therefore given by:
\begin{align*}
CGL^b_T=\phi^b_{T-1}\left(\frac{1}{n}\sum_{i=1}^{n-1} B(T,T+i,c^{i+1}_{T-1}) \right)-\left(BV^b_{T-1}-\frac{\phi^b_{T-1}}{n}\right).
\end{align*}
This quantity impacts the capitalization reserve level as follows
\begin{align*}
CR_T=(CR_{T-1}+CGL^b_T)^+.
\end{align*}
The terminal bonus declaration is rather simple. The insurer must liquidate the profit sharing reserve since it belongs to its policyholders and comply with the minimum guaranteed rate of return $r^G$.
Let us define
\begin{align}
TD_T=FI_{T}-(CR_{T-1}+CGL^b_T)^- +PSR_{T-1}+CGL_T^s,
\end{align} the amount to distribute to policyholders. The credited amount to policyholders is:
\begin{align}
R^G_T=\max\left\lbrace \pi_{pr}TD_T, r^G (MR_{T-1}+PSR_{T-1})\right\rbrace.
\end{align}
Note that we do not consider for the final date a competitor rate since all the contracts terminate. We then define the crediting rate $r_{ph}(T)=\frac{R^G_T}{MR_{T-1}+PSR_{T-1}}$, the mathematical reserve $MR_{T}=MR_{T-1}(1+r_{ph}(T))$ and $PSR_T=r_{ph}(T)PSR_{T-1}$, exactly as in equations~\eqref{tx_servi}, \eqref{dyn_MR} and~\eqref{dyn_PSR}.

We define then the final accounting margin of the shareholders by
\begin{align*}
AM_T=(1-\pi_{pr})TD_T-(R^G_T-\pi_{pr}TD_T)^+. 
\end{align*}
Since the  capitalization reserve $RC_T$ is a part of the equity of the insurance company, it is given to shareholders. The terminal shareholders P\&L is then:
\begin{align}
P\&L_T=AM_T+CR_{T-1}\left(\frac{1}{B(T-1,T)}-1\right)+ CR_T
\end{align}
At the maturity of the contract, the mathematical and profit sharing reserves $MR_T$ must be payed to policyholders.  The terminal cash outflow is thus
\begin{align}
COF_T=MR_T+PSR_{T}.
\end{align}

\subsection{Overall performance of the ALM}
To assess the solvency situation, Solvency II regulation requires to value assets and liabilities on a market consistent basis. It 
prescribes to use the Best Estimates Liabilities and the Basic Own-Funds (also called Net Asset Value) to value the liability of the company. We explicit these key quantities in our framework.

The Basic Own-Funds (BOF) corresponds to the market-consistent valuation of the equity  capital of the firm. It is determined as the present-value of future shareholders P\&L cash flows under the pricing measure $\mathbb{Q}$. If we consider a short interest rate model $(r_t,t\ge 0)$, the BOF is given as follows:
\begin{align}\label{def_BOF}
BOF_0=\E^{\Q}\left[\sum_{t=1}^{T} e^{-\int_0^{t}r_s ds}(P\&L_t)\right]. 
\end{align}
More generally, if $(\cF_t,t\ge 0)$ is the filtration representing the market information, we can define the Basic Own-Funds at time~$t\in\{0,\dots,T-1\}$ as
$BOF_t=\E^{\Q}\left[\sum_{u=t+1}^{T} e^{-\int_{t}^{u}r_s ds}(P\&L_u)\bigg|\cF_t\right]$.

The Best Estimates Liability (BEL) represents the total debt of the insurer. It corresponds to the discounted sum (present-value) of future surrender cash outflows and terminal liability payment 
\begin{align}\label{def_BEL}
BEL_0=\E^{\Q}\left[\sum_{t=1}^{T} e^{-\int_0^{t}r_s ds} COF_t\right].
\end{align}
We more generally define $BEL_t=\E^{\Q}\left[\sum_{u=t+1}^{T} e^{-\int_t^{u}r_s ds} COF_u \bigg| \cF_t\right]$ for $t\in\{0,\dots,T-1\}$.

Since we are considering a pool of policyholders running off, all these cash flows are generated from the initial Mathematical Reserve. We therefore get the so-called "no-leakage" condition:
$$MR_0=BOF_0+BEL_0.$$
More generally, we have
$$\forall t \in\{0,\dots, T-1\}, \ MR_t+PSR_t+CR_t=BOF_t+BEL_t.$$ 
\subsection{Solvency Capital Requirement of the ALM with the standard formula}
To determine the Solvency Capital Requirement (SCR), the supervision authority  provides a standard formula that consists in various stress tests for different type of risks. The risks are divided between modules and sub-modules and  combined into a global SCR for market risk according to an aggregation formula. The detailed description of the SCR calculation can be found in the note written by the European  Insurance and Occupational Pensions Authority (EIOPA), Section~2 of~\cite{EIOPA}. Short descriptions of the standard formula can be found in the appendix of~\cite{BrScSc} or in Subsection~3.2 of~\cite{Boonen}. In our model, we only have to consider the equity and interest rate modules and we briefly explain the standard formula in our framework.

\subsubsection{Equity module}
The SCR for equity risk $SCR_{eq}$ is determined by the variation of the Basic Own-Funds $BOF_0$ after a negative shock on the equity asset class that occurs immediately after time~$0$, i.e. after the first asset allocation. The negative shock $s^{eq}\in(-1,0)$ assumes that the value of stock assets decreases with a certain percentage. The shock prescribed by the EIOPA may differ according to the type of equity, see~\cite{EIOPA} p.~140 and Section~3 of~\cite{GaMa}. Here, we recall that $S$ should be seen as a weighted average of stocks (like indices) in which the insurance company invests. Thus, we assume that 
\begin{align*}
S_{0+}^{shock}=S_0\left(1+s^{eq}\right),
\end{align*}
where $s^{eq}$ is the corresponding average of the shocks prescribed by the EIOPA. 
We note $BOF^{eq\_shock}_0$ the Basic Own-Funds calculated with this shock. The SCR for equity risk is then defined  by:
\begin{align}\label{def_SCR_eq}
SCR_{eq}=(BOF^{eq\_shock}_0-BOF_0)^-=(BOF_0-BOF^{eq\_shock}_0)^+
\end{align}
\subsubsection{Interest rate module}\label{IR_module}
To estimate the solvency capital for the risk on interest rates, the EIOPA provides upward and downward shocks to the initial term-structure. As for the equity, the shocks are assumed to occur  immediately after the first allocation at time~$0$. Let us suppose that we observe at time~$0$ market prices of zero-coupon bonds $t \mapsto P^{mkt}(0,t)$ and we note $R^{mkt}(0,t)=-\frac 1 t \log( P^{mkt}(0,t))$ the corresponding yield curve. The shifted yield curves are then given by:
\begin{align}\label{stress_multiplicatif}
R^{up/down}(0,t)=(1+s_t^{up/down})R^{mkt}(0,t)
\end{align}
where $s^{up}_t$  (resp. $s^{down}_t$) is the upward (resp. downward) shock to the yield with maturity $t$. These coefficients have been recommended in p.~137 of~\cite{EIOPA} and implemented by the European Commission in the Articles 166 and 167 of the Delegated Regulation~\cite{DelReg}\footnote{Besides, a minimal increase (resp. decrease) of $1\%$ is assumed for $R^{up}(0,t)$ (resp. $R^{down}(0,t)$), see also Boonen~\cite{Boonen}, p.~411.}. They are summarized in the Table~\ref{table_EIOPA} below.
\begin{table}[H]
\vspace*{-0cm}
\begin{center}
\begin{tabular}{ |c|c|c|c|c|c|c|c|c|c|c|c| } 
\hline
$t$ & 1 & 2& 3&4&5&6&7&8&9&10 \\
\hline
 $s^{up}_t$& $70\%$ & $70\%$&64\%&59\%& 55\%& 52\%&49\%&47\%&44\%& 42\%\\ 
$s^{down}_t$ & $-75\%$ & $-65\%$& -56\%&-50\%&-46\%&-42\%&39\%&-36\%&-33\%&-31\% \\

\hline

\end{tabular}
\begin{tabular}{ |c|c|c|c|c|c|c|c|c|c|c|c| } 
\hline
$t$ & 11 & 12& 13&14&15&16&17&18&19&20 \\
\hline
 $s^{up}_t$& $39\%$ & $37\%$&35\%&34\%& 33\%& 31\%&30\%&29\%&27\%& 26\%\\ 
$s^{down}_t$ & $-30\%$ & $-29\%$& -28\%&-27\%&-28\%&-28\%&-28\%&-28\%&-29\%&-29\% \\

\hline

\end{tabular}
\end{center}
\caption{Stress-Factors of the standard formula given by the EIOPA~\cite{EIOPA} in December 2012 } \label{table_EIOPA}
\end{table}
For years $t\geq 90$, the regulator prescribes the shocks $s^{up}_t=0.2$ and $s^{down}_t=-0.2$.
Between $t_a=20$ and $t_b=90$ years, a the linear interpolation method has to be used to get the shocks:
\begin{equation}\label{interp_shocks}\forall t\in[t_a,t_b], \ s_t^{up/down}=s^{up/down}_{t_{a}}+(s^{up/down}_{t_{b}}-s^{up/down}_{t_{a}})\frac{t-t_a}{t_b-t_a}.
\end{equation}

The SCR for up and down shock are determined by the variation of the basic own-funds if the stressed yield-curve is used instead of the initial term-structure. Namely, we set $SCR_{up}=\left(BOF_0-BOF^{up}_0\right)^+$ and $SCR_{down}=(BOF_0-BOF^{down}_0)^+$. 
The SCR for the risk on  interest rates is defined as the worst one of the two shocks: 
\begin{align}\label{def_SCR_int}
SCR_{int}=\max ( SCR_{up},SCR_{down}) .
\end{align}

\begin{remark}\label{rk_fwd_rate} Let us note $f^{mkt}(0,t)=-\log\left(\frac{P^{mkt}(0,t+1)}{P^{mkt}(0,t)} \right)=(t+1)R^{mkt}(0,t+1)-t R^{mkt}(0,t)$, that can be seen as a forward rate on $(t,t+1)$. Let $(s_t)_{t\in \N^*}$ be prescribed (deterministic) shocks and $R^{shock}(0,t)=R^{mkt}(0,t)(1+s_t)$. Let $f^{shock}(0,t)=(t+1)R^{shock}(0,t+1)-t R^{shock}(0,t)$ the stressed forward rate. We then have
  \begin{align*}
    f^{shock}(0,t)-f^{mkt}(0,t) =(t+1)s_{t+1}R^{mkt}(0,t+1)-ts_t R^{mkt}(0,t).
  \end{align*}
  For large maturity~$t$, it is likely to have $R^{mkt}(0,t+1) \approx R^{mkt}(0,t)$, which gives $$ f^{shock}(0,t)-f^{mkt}(0,t)\approx s_{t+1}R^{mkt}(0,t+1)+t(s_{t+1}-s_t) R^{mkt}(0,t).$$
  Due to the multiplication by~$t$, even small variations of the stress factor (and also of $s_{t+1}-s_t$) may lead to important variations of the shocked forward rate. For example, if we assume for simplicity $R^{mkt}(0,t)=r$ for all~$t$, the downward shock of Table~\ref{table_EIOPA} gives
  $f^{down}(0,13)=r -\frac{27}{100} r + \frac{13}{100} r=\frac{86}{100} r$ and $f^{down}(0,14)=\frac{58}{100} r$, leading to an important variation of the downward shocked forward rate between maturities $13$ and $14$. The same is observed with    $f^{down}(0,18)=r -\frac{28}{100} r + \frac{18}{100} r=\frac{90}{100} r$  and $f^{down}(0,19)=\frac{71}{100} r$. 
\end{remark}

This methodology has been set up when the interest rates were around $2\%$ or $3\%$, but is no longer relevant for very low or even negative interest rates. If we suppose for simplicity that $R^{mkt}(0,t)=0$, then the multiplicative stress rule~\eqref{stress_multiplicatif} leaves rates unchanged and therefore leads to a null SCR. Also, for negative rates, the formula~\eqref{stress_multiplicatif} inverts the sign of the stress: the upward stress factor leads to a decrease of the interest rate and conversely. To bypass this issue, the EIOPA has recently recommended in 2018~\cite{EIOPA2} to add an additive factor, i.e. to replace~\eqref{stress_multiplicatif} by
\begin{align}\label{stress_2018}
R^{up/down}(0,t)=(1+s_t^{up/down})R^{mkt}(0,t)+b_t^{up/down}.
\end{align}
Between $t_a=20$ and $t_b=90$ years, the interpolation formula~\eqref{interp_shocks} is kept for $s_t$. An analogous formula is used for $b_t^{up/down}$:
$$\forall t\in[t_a,\widetilde{t_b}], \ b_t^{up/down}=b^{up/down}_{t_{a}}+(b^{up/down}_{\widetilde{t_b}}-b^{up/down}_{t_{a}})\frac{t-t_a}{\widetilde{t_b}-t_a},$$
with $t_a=20$, $\widetilde{t_b}=60$ years and $b_{t}^{up/down}=0$ for $t\ge\widetilde{t_b}$.

\begin{table}[H]
\vspace*{-0cm}
\begin{center}
\begin{tabular}{ |c|c|c|c|c|c|c|c|c|c|c|c| } 
\hline
$t$ & 1 & 2& 3&4&5&6&7&8&9&10 \\
\hline
 $s^{up}_t$& 61\% & 53\%& 49\%&46\%& 45\%& 41\%&37\%&34\%&32\%& 30\%\\ 
$s^{down}_t$ & -58\% & -51\% & -44\%&-40\%&-40\%&-38\%&-37\%&-38\%&-39\%&-40\% \\
\hline
$b^{up}_t$& 2.14\% & 1.86\%&1.72\%&1.61\%& 1.58\%& 1.44\%&1.30\%&1.19\%&1.12\%& 1.05\%\\ 
$b^{down}_t$ & -1.16\% & -0.99\%& -0.83\%&-0.74\%&-0.71\%&-0.67\%&-0.63\%&-0.62\%&-0.61\%&-0.61\% \\
\hline

\end{tabular}
\begin{tabular}{ |c|c|c|c|c|c|c|c|c|c|c|c| } 
\hline
$t$ & 11 & 12& 13&14&15&16&17&18&19&20 \\
\hline
 $s^{up}_t$& 30\% & 30\% &30\%&29\%& 28\%& 28\%&27\%&26\%&26\%& 25\%\\ 
$s^{down}_t$ & -41\% & -42\% & -43\%&-44\%&-45\%&-47\%&-48\%&-49\%&-49\%&-50\% \\
\hline
$b^{up}_t$& 1.05\% & 1.05\%&1.05\%&1.02\%& 0.98\%& 0.98\%&0.95\%&0.91\%&0.91\%& 0.88\%\\ 
$b^{down}_t$ & -0.60\% & -0.60\%& -0.59\%&-0.58\%&-0.57\%&-0.56\%&-0.55\%&-0.54\%&-0.52\%&-0.50\% \\
\hline

\end{tabular}
\end{center}
\caption{Stress-Factors of the standard formula given by the EIOPA~\cite{EIOPA2} in February 2018. } \label{table_EIOPA2}
\end{table}

\subsubsection{Aggregation Formula}
The SCR for market-risk is a combination between the equity and interest-rate risk in our framework. It is defined as follows (see Articles 164 and 165 of~\cite{DelReg}):
\begin{align}\label{def_SCR_mkt}
SCR_{mkt}=\sqrt{SCR^2_{eq}+SCR^2_{int}+2\varepsilon SCR_{eq}SCR_{int}}
\end{align}
where the ``correlation factor'' $\varepsilon=0$ if the interest-rate exposure is due to the upward-shock and $\varepsilon=\frac{1}{2}$ if it is due to the downward shock.

\section{Asset Model}\label{sec_asset_model}

The insurance company invests policyholders deposits between two asset classes: riskless bonds and stocks. Therefore, we have to model the equity asset~$S_t$ and the interest rates, for which we choose a short rate model~$(r_t,t\ge 0)$. We denote $(\Omega,\cF,\Q)$ a risk neutral probability space and $(\cF_t,t\ge 0)$ the filtration that represents market information. Let $(W_t,Z_t)_{t\ge 0}$ be a standard two-dimensional Brownian motion under~$\Q$ and we set $Z_t^\gamma=\gamma W_t+\sqrt{1-\gamma^2}Z_t$ for $\gamma\in [-1,1]$. Since we will mainly focus in this paper on the SCR valuation with the standard formula that has to be made under a risk-neutral measure, we will not model the asset dynamics under the real world probability. However, modeling for both risk-neutral and real world probabilities is relevant for ALM to determine, for example, an optimal asset allocation under SCR constraints. This is however beyond the scope of the paper.

Before specifying the equity and interest rate dynamics, we first describe the surrender rate model for policyholders. We consider that the surrender rate is the sum of a component $\underline{p}\in(0,1)$ quantifying structural surrenders and a market contingent surrender rate $\RD(\Delta_t)$, where  $\Delta_t=r_{ph}(t)-\CompRate$ is the spread between  the crediting rate to policyholders $r_{ph}(t)$ defined in~\eqref{tx_servi} and the competitor rate $\CompRate$ defined by~\eqref{def_r_comp}. The function $\RD$ is defined as follows
\begin{numcases}{\RD(\Delta)=}
    \RD_{max} & for $\Delta<\alpha$,\notag\\
    \RD_{max}\frac{\beta-\Delta}{\beta-\alpha} & for $ \alpha\leq \Delta\leq\beta$,\label{def_DSR}\\
     0 & for $\Delta>\beta$,\notag
\end{numcases}
where $\RD_{max}\in (0,1-\underline{p})$ is the maximum surrender rate, $\alpha$ the massive surrender threshold and $\beta$ the triggering surrender threshold.
Therefore, surrenders occur with a proportion $p^e_t$ also called exit rate:
\begin{equation}\label{model_exit_rate}
  p^e_t=\underline{p}+\RD(\Delta_t) .
\end{equation}
We assume that the equity asset follows a Black-Scholes model:
$$S_t=S_0 \exp\left(\int_0^t r_s ds +\sigma_S W_t -\frac{\sigma_S^2}{2}t \right),$$
where $\sigma_S>0$ is the volatility. Concerning interest rates, we consider a priori two different models: the shifted Vasicek model (called Vasicek++ later on, see Brigo and Mercurio~\cite{BrMe} paragraph 3.2.1 and Section 3.8) and the Hull and White model (see e.g.~\cite{BrMe} Section 3.3). The Vasicek++ model assumes that
\begin{align}
  r_t&=x_t+\varphi(t), \label{Vasicek++}\\
  x_t&=x_0+\int_0^t k(\theta-x_s)ds+\sigma_r Z^\gamma_t, \notag
\end{align}
for some parameters $x_0,\theta \in \R$, $k,\sigma_r>0$, and $\varphi:\R_+ \rightarrow \R$. The Hull and White model assumes that 
\begin{equation}\label{HW}
  r_t=r_0+\int_0^tk(\vartheta(s)-r_s)ds+\sigma_r Z^\gamma_t,
\end{equation}
with parameters $r_0 \in \R$, $k,\sigma_r>0$, and $\vartheta:\R_+ \rightarrow \R$.
Both models have very similar properties: these are Gaussian models with explicit zero-coupon bond prices and closed formula for caplets, and we refer to~\cite{BrMe} for further details. In fact, as noticed by Brigo and Mercurio (\cite{BrMe}, p.~101) the two models are identical if we take
\begin{equation}\label{equiv_HW_V++}\vartheta(t)=\theta + \varphi(t)+\frac 1 k \varphi'(t).
\end{equation}
Let us note however that the Vasicek++ parametrization offers a slightly larger class of dynamics: for any piecewise continuous function $\vartheta$, we can find a piecewise $C^1$ function $\varphi$ such that~\eqref{equiv_HW_V++} holds. Instead, there is no $\vartheta$ satisfying~\eqref{equiv_HW_V++} when $\varphi$ is piecewise continuous and not differentiable. 
We just recall the zero-coupon bond prices at time~$t$ with maturity~$T\ge t$ in the Vasicek++ model:
\begin{align*}
  P^{V++}(r_t,t,T)&=A(T-t)\exp\left(-\int_t^T\varphi(s)ds -(r_t-\varphi(t))g_k(T-t)-\theta(T-t-g_k(T-t)) \right),\\
  g_k(t)&=\frac{1-e^{-kt}}k, \quad A(t)=\exp\left( \frac{\sigma_r^2}{2k^2} g_k(t)- \frac{\sigma_r^2}{4k}g^2_k(t)\right)
\end{align*}
and in the Hull and White model:
\begin{align*}
  P^{HW}(r_t,t,T)&=A(T-t)\exp\left(-r_t g_k(T-t)-\int_t^T(1-e^{-k(T-s)})\vartheta(s)ds \right).
\end{align*}

The methodology to calibrate these models to market data (as required by the regulation, see e.g.~\cite{VeEKLoPr} p.8) is the same. For each parameters $(x_0,\theta,k,\sigma_r)$ (resp. $(r_0,k,\sigma_r)$), there exists a unique deterministic function $\varphi$ (resp. $\vartheta$) that perfectly fits the zero-coupon bond prices $P^{mkt}(0,t)$ observed on the market (or deduced from market data). Therefore, one tries to find the parameters that better fit the market data on options such as caplet or swaption prices, and then pick the corresponding function  $\varphi$ or $\vartheta$. These models comply with the Solvency II regulation that imposes to fit the initial term-structure of interest-rate and to approximate well options market prices. To perform the perfect fit of the the zero-coupon bond prices, one typically assumes some parametrization of the functions $\varphi$ and $\vartheta$. A typical choice is to assume these functions to be piecewise constant or piecewise linear. Once parameterizations are chosen for $\varphi$ and $\vartheta$, the Vasicek++ and Hull and White model may no longer be the same: the Vasicek++ model with piecewise constant~$\varphi$ is not a Hull and White model with piecewise constant $\vartheta$, and conversely. In what follows, we argue that the parametrization of the Vasicek++ model is much more convenient to deal with the standard formula.

\begin{figure}[h]  
\centering
\includegraphics[keepaspectratio=true,scale=0.5]{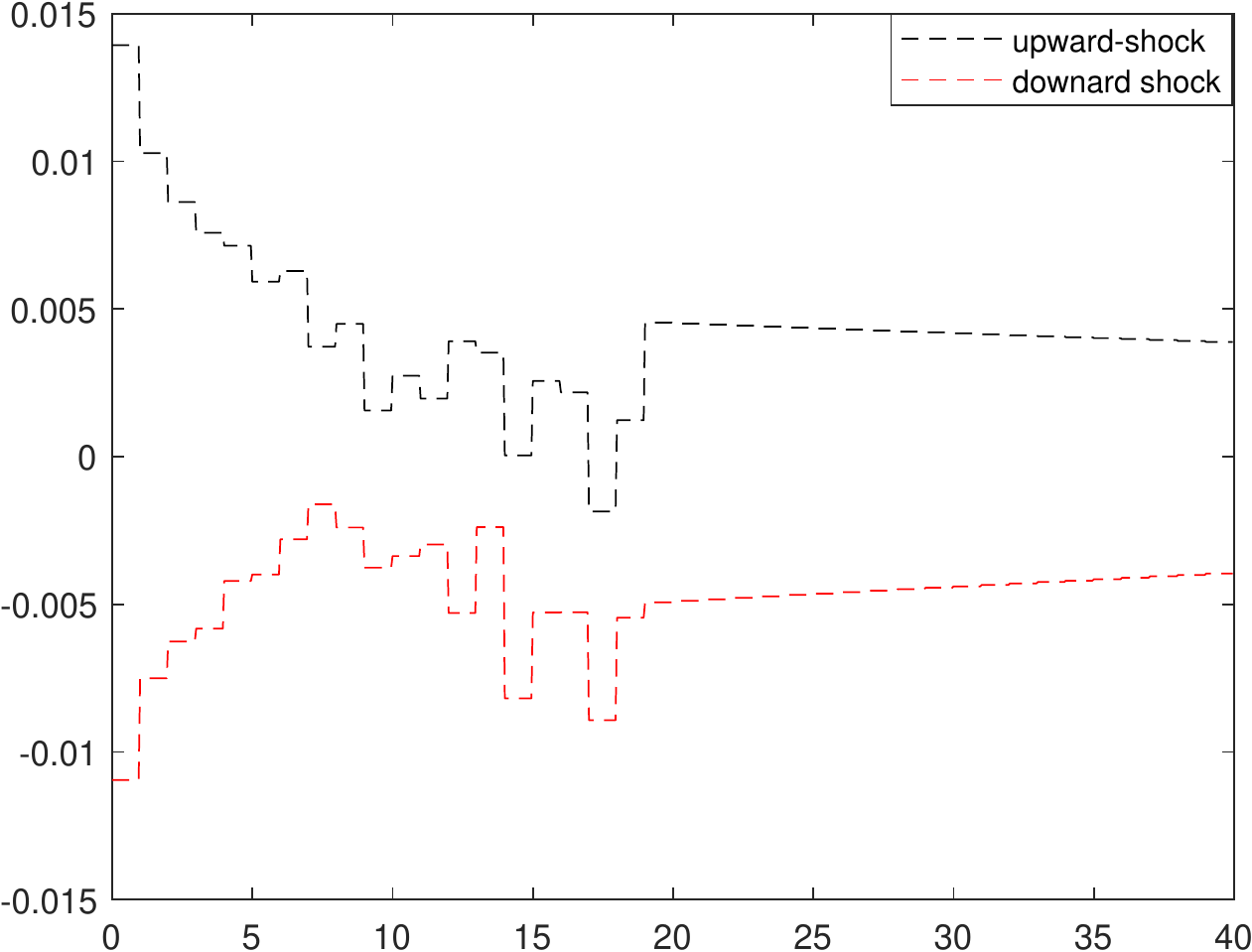}\hspace{2cm}\includegraphics[keepaspectratio=true,scale=0.5]{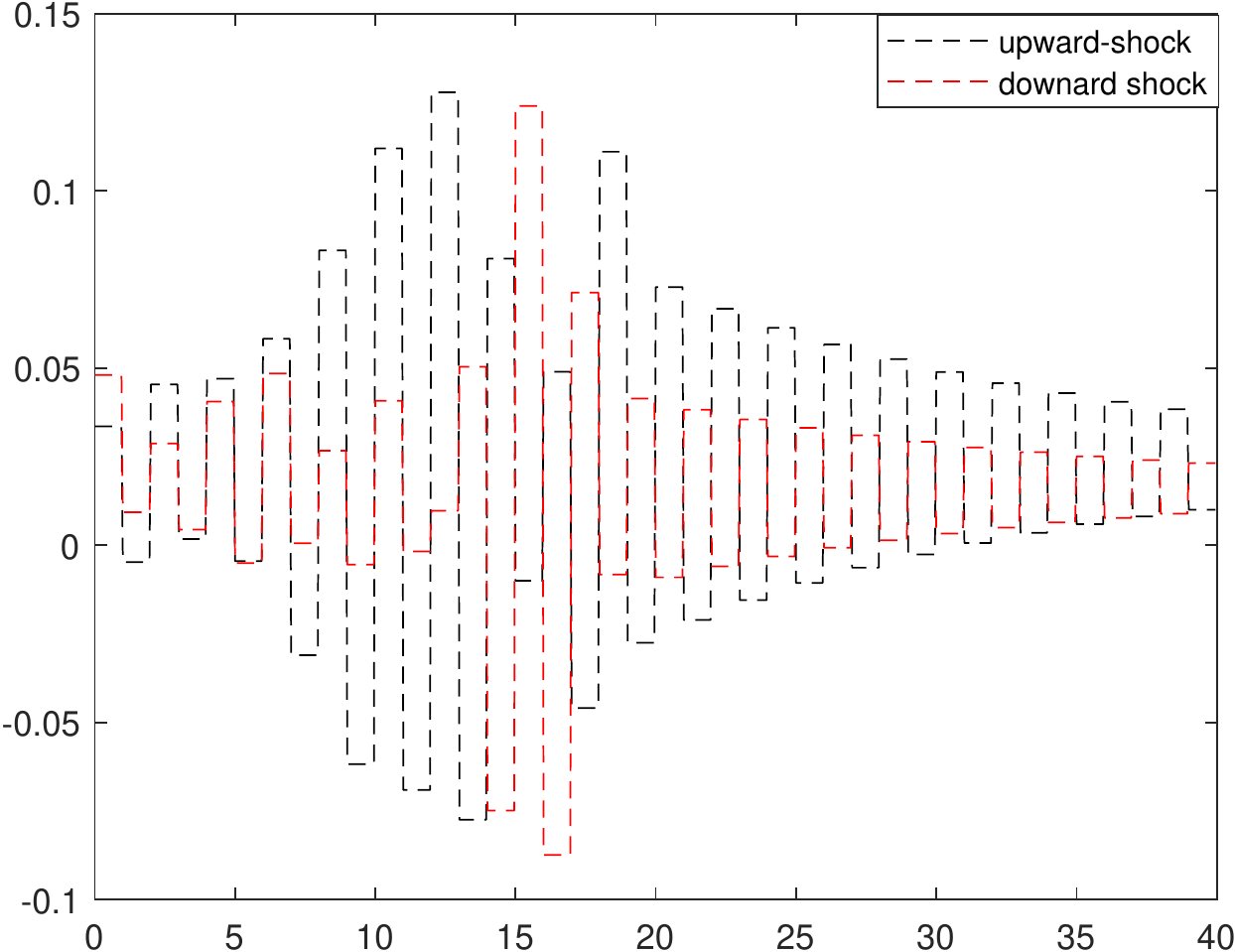}
\caption{Calibrated  piecewise constant functions $t\mapsto \varphi^{shock}(t)$ (left) and $t\mapsto \vartheta^{shock}(t)$ (right) after the upward and downward shocks specified in Table~\ref{table_EIOPA} with no additive shock (i.e. $b_t=0$), $r_0=0.02$ and $k=0.2$.}\label{fitted_thetaphi}
 \end{figure}

To implement the standard formula described in paragraph~\ref{IR_module}, one has to re-calibrate the models to the stressed zero-coupon curve. Since the stressed factors given in Tables~\ref{table_EIOPA} and~\ref{table_EIOPA2} are given on an annual basis, it is rather natural to consider piecewise constant shapes for $\varphi(t)$ and $\vartheta(t)$:
$$\varphi(t)=\sum_{i=0}^{\infty}\varphi_i\mathds{1}_{t\in[i,i+1[}, \ \vartheta(t)=\sum_{i=0}^{\infty}\vartheta_i\mathds{1}_{t\in[i,i+1[}. $$
We note $\varphi^{mkt}$ (resp. $\vartheta^{mkt}$) the function calibrated in the Vasicek++ (resp. Hull and White) model to $P^{mkt}(0,t)$ and $\varphi^{shock}$ (resp. $\vartheta^{shock}$) the function calibrated  to $\exp\left(-t[(1+s_t)R^{mkt}(0,t)+b_t]\right)$.
We keep the parameters  $(x_0,\theta,k,\sigma_r)$ constant (i.e. as before the shock) for the Vasicek++ model. For the Hull and White model, we take $(1+s_1)r_0+b_1$ as the initial short rate value after the shock and keep the parameters $(k,\sigma_r)$ constant: this is to reduce the fluctuations that we already observe on $\vartheta^{shock}$ in Figure~\ref{fitted_thetaphi}. From the zero-coupon bond price formulas, we easily get:
\begin{align*}
\exp\left(-t[s_tR^{mkt}(0,t)+b_t]\right)  &=\exp\left(\int_0^t\varphi^{mkt}(s)-\varphi^{shock}(s)ds\right)\\
  &=\exp\left(-(s_1r_0+b_1) g_k(t)+ \int_0^t(1-e^{-k(t-s)})(\vartheta^{mkt}(s)-\vartheta^{shock}(s))ds  \right).
\end{align*}

\begin{figure}[h]
\centering
\includegraphics[keepaspectratio=true,scale=0.44]{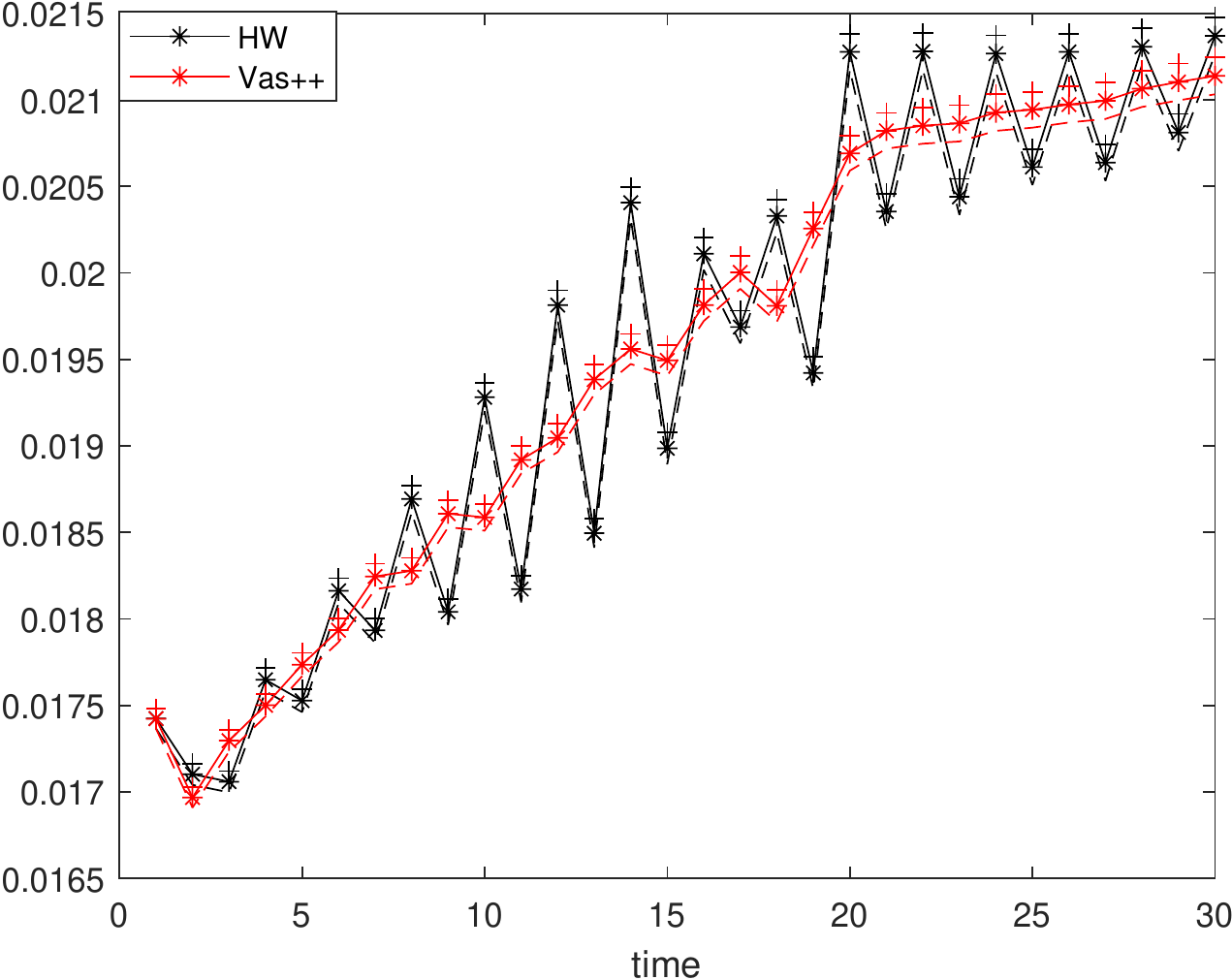}\includegraphics[keepaspectratio=true,scale=0.44]{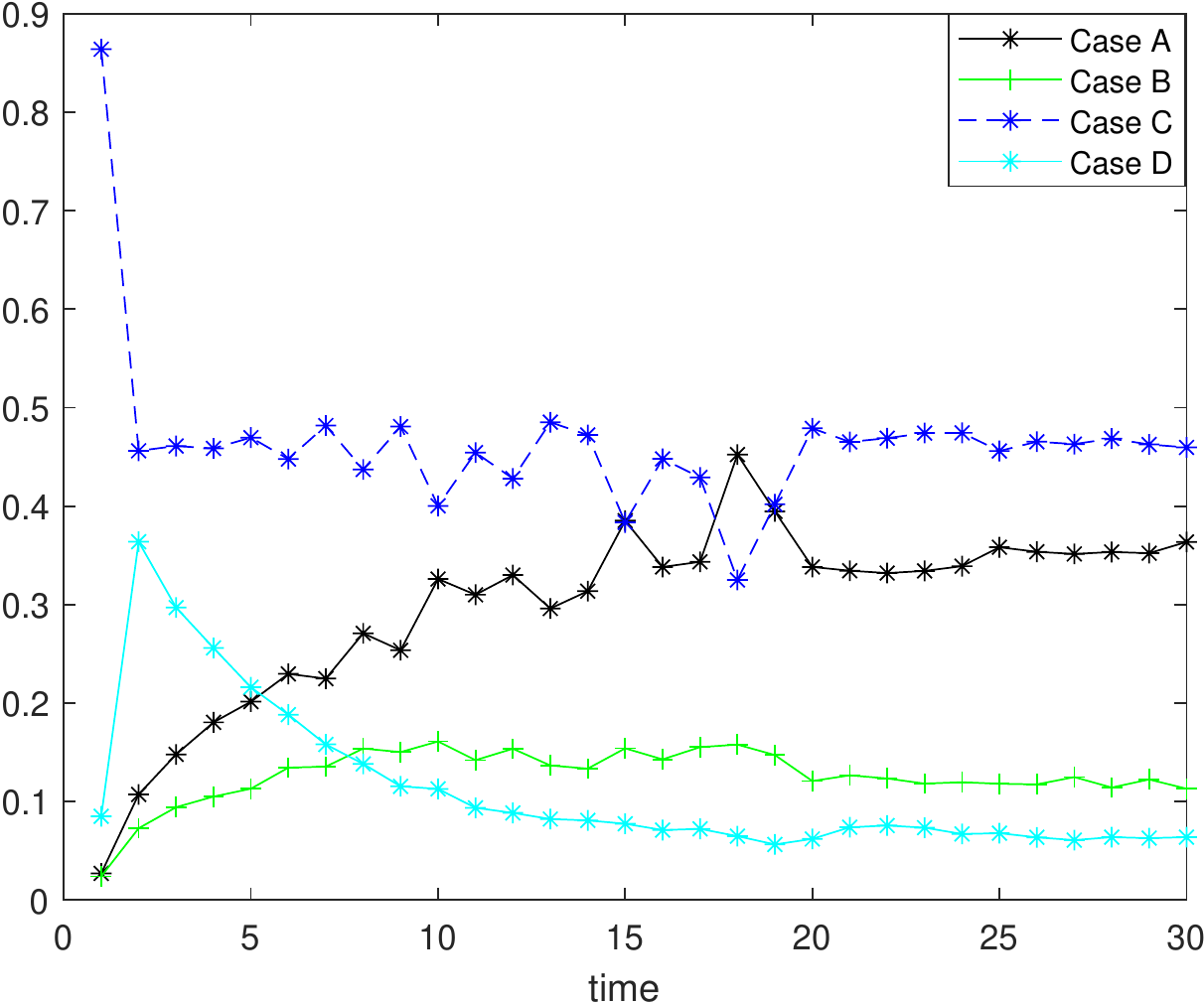}\includegraphics[keepaspectratio=true,scale=0.44]{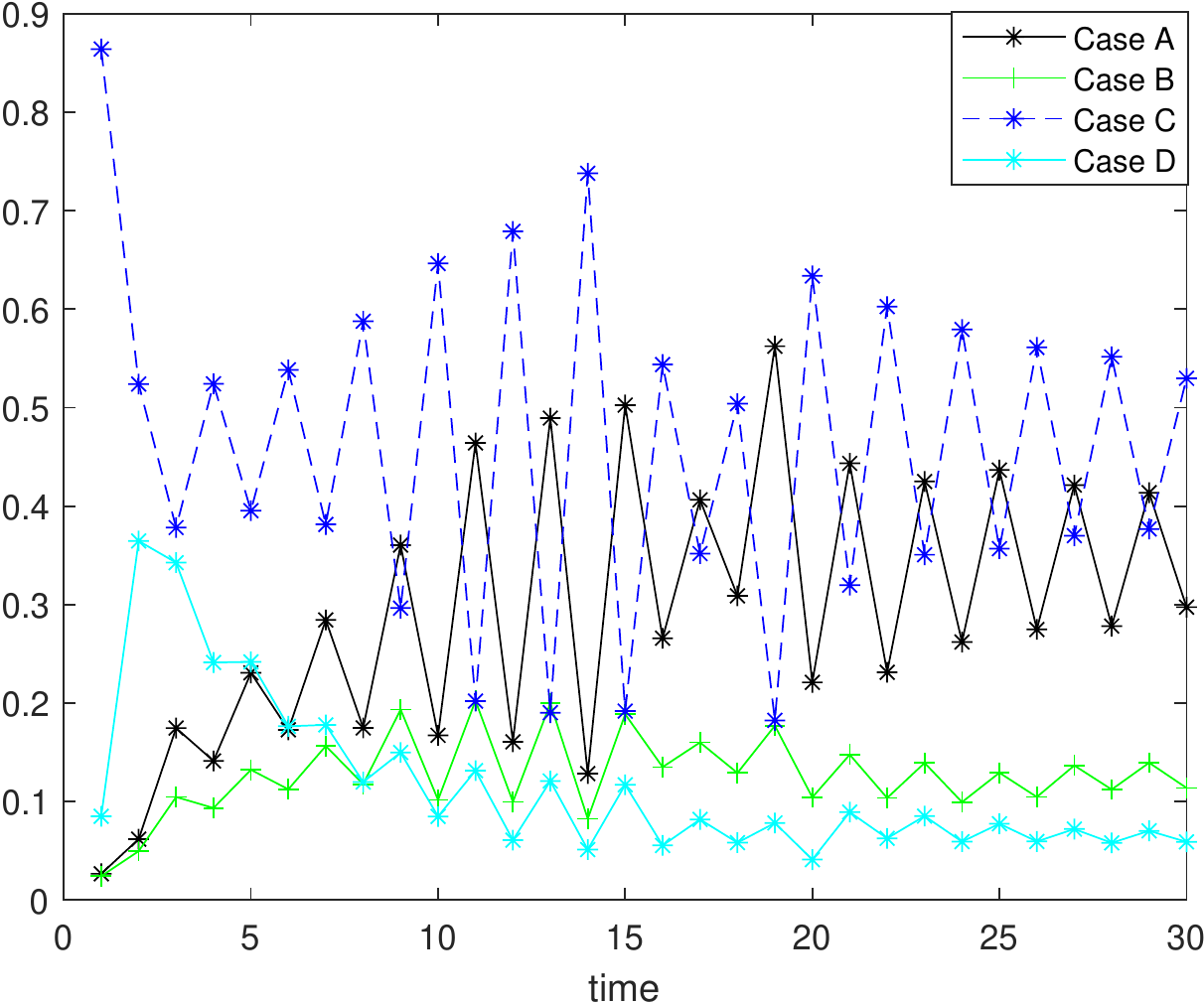}
\caption{Simulations after the upward shock on interest rates described in Table~\ref{table_EIOPA}. Left: mean of the crediting rate $r_{ph}(t)$ defined in~\eqref{tx_servi} with the 95\% confidence interval. Middle (resp. Right): proportions of the cases A, B, C and D in the Vasicek++ (resp. Hull and White) model.}\label{oscillations}
\end{figure}

Figure~\ref{fitted_thetaphi} illustrates the calibrated functions. We have considered the case where the zero-coupon bond prices $P^{mkt}(0,t)$ are given by a Vasicek model with $r_0=\theta=0.02$, $k=0.2$ and $\sigma=0.01$. We assume $r^{comp}_t=r_t$ and constant allocation targets $w^b_t=0.95$ and $w^s_t=0.05$. The first striking point is the oscillations of $ \vartheta^{shock}$, making the Hull and White model poorly realistic after the shock. Instead, the variations of  $ \varphi^{shock}$ are much more reasonable. We still observe some unlikely moves between years 10 and years 20: as explained in Remark~\ref{rk_fwd_rate}, this is due to small variations of the stress factors that are amplified by the maturity. Thus, the Vasicek++ model has much more meaning after the shock than the Hull and White model.  We have plotted in Figure~\ref{oscillations} the crediting rate to policyholders, as well as the empirical distribution of the cases A, B, C and D described in Step~4 that determines the crediting rate to policyholders. We observe significant oscillations of the mean crediting rate for the Hull and White model, that are explained by important oscillations between the proportions of case A and case C. In contrast, the mean crediting rate and the distribution of the cases A, B, C and D is much more regular in the Vasicek++ model. We have also done the same analysis for the downward shocks: oscillations again appears in the Hull and White model, but they are less marked because of the minimum guaranteed rate. 

\begin{figure}[h]
\centering
\includegraphics[keepaspectratio=true,scale=0.44]{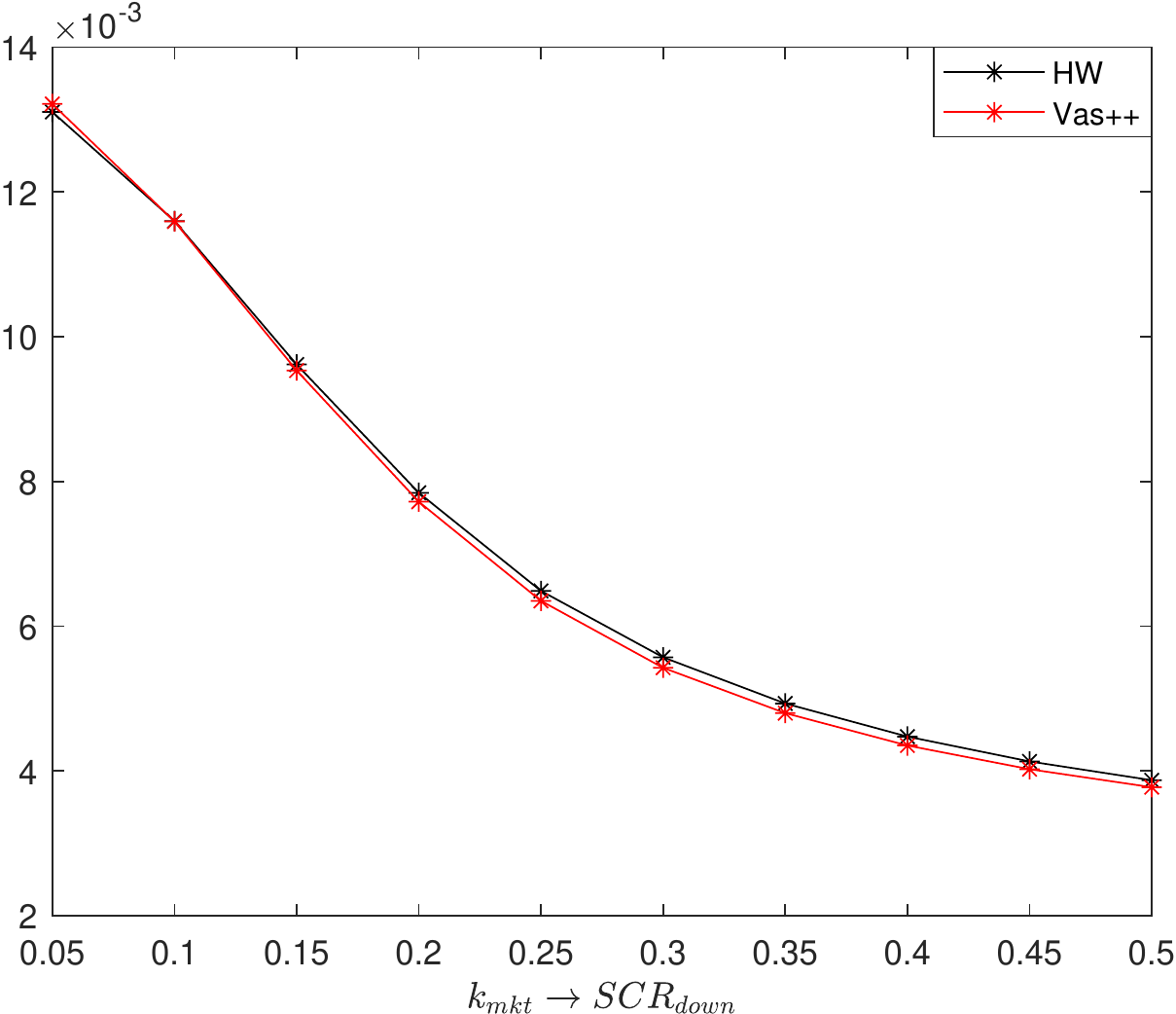}\includegraphics[keepaspectratio=true,scale=0.44]{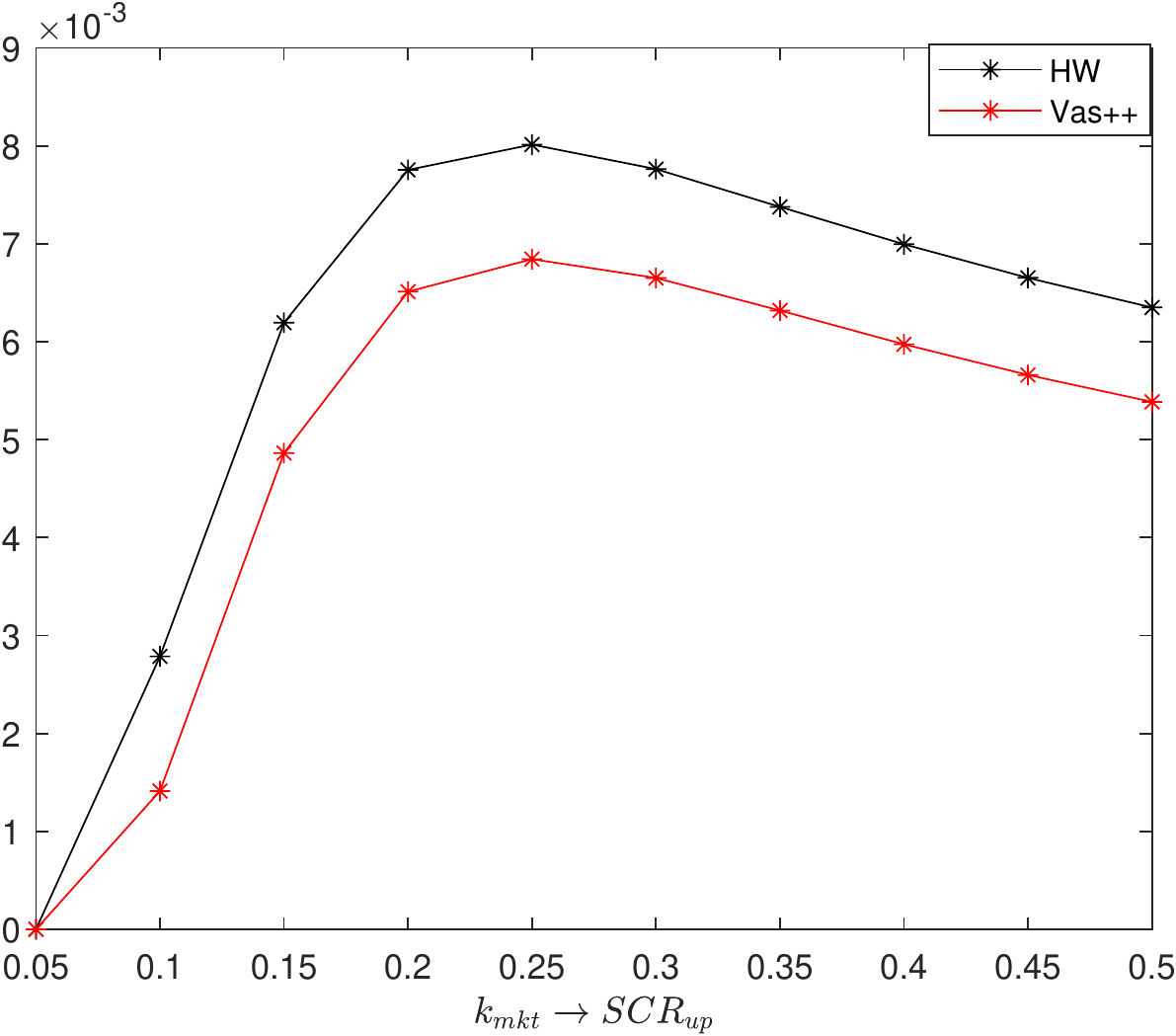}\includegraphics[keepaspectratio=true,scale=0.44]{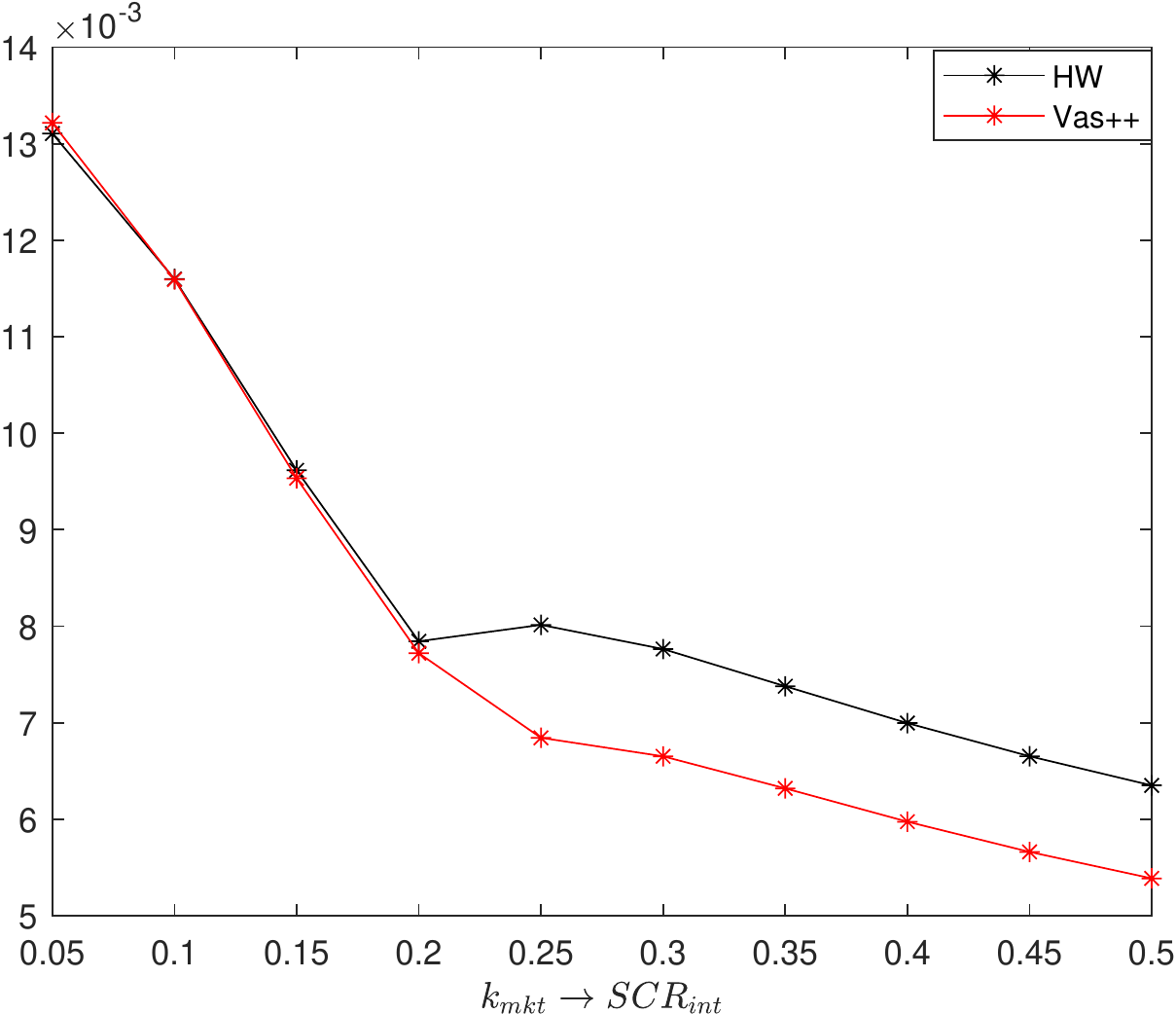}
\caption{SCR values with Vasicek++ and Hull and White models in function of~$k$ for $SCR_{down}$ (left), $SCR_{up}$ (middle) and $SCR_{int}$ (right).}\label{SCR_V++_HW}
\end{figure}

We now focus on $SCR_{int}$ defined by~\eqref{def_SCR_int}.  Figure~\ref{SCR_V++_HW} shows the value of the SCR in function of~$k$ when the central model is a Vasicek model with $r_0=\theta=0.02$, $\sigma=0.01$ and $k$. We observe almost the same value for $SCR_{down}$ between Vasicek++ model and the Hull and White model, but there is a significant difference for $SCR_{up}$ in favor of the Vasicek++ model, which impacts then $SCR_{int}$ when the upward shock has a greater contribution than the downward shock.

\begin{figure}[h]  
\centering
\includegraphics[keepaspectratio=true,scale=0.44]{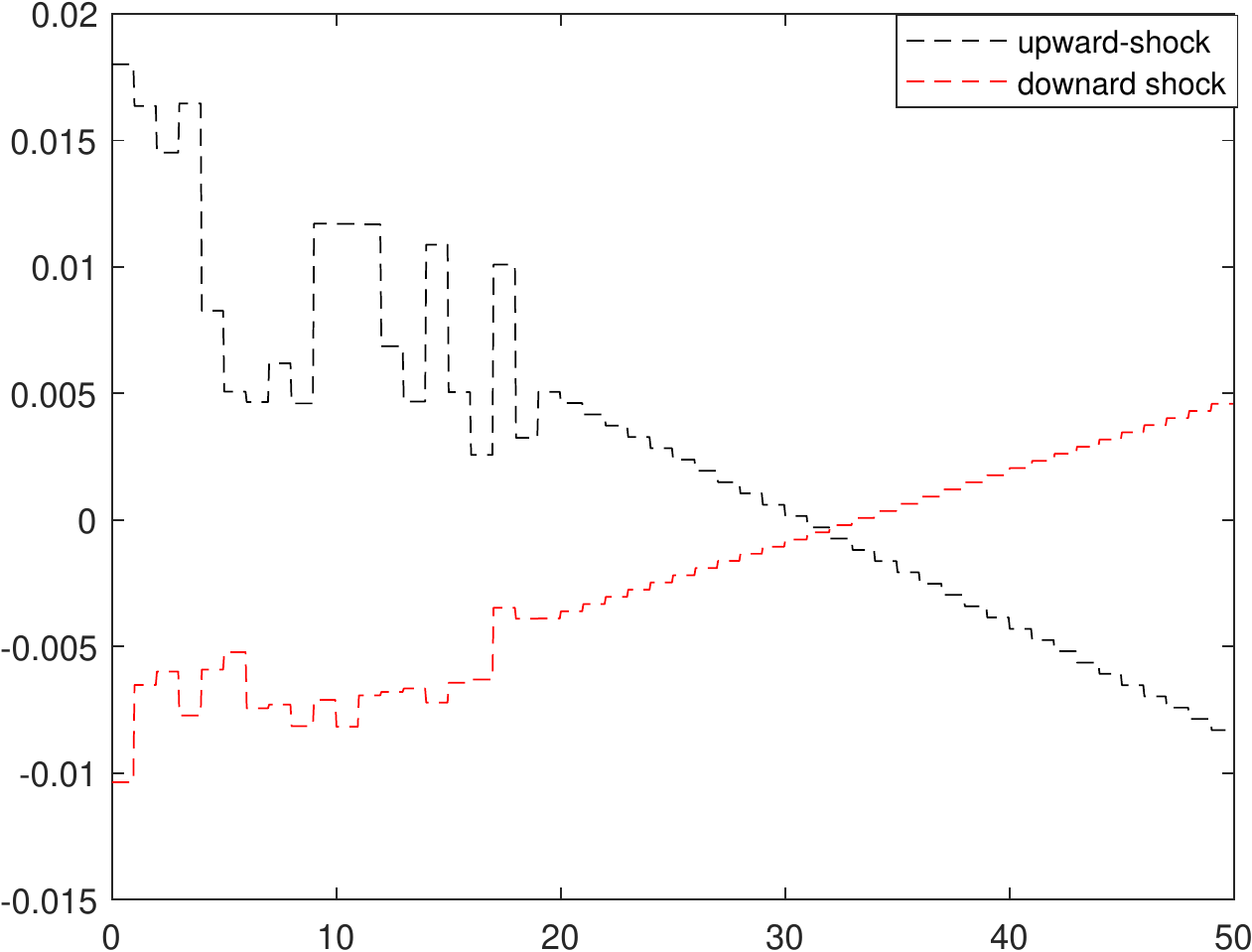}\includegraphics[keepaspectratio=true,scale=0.44]{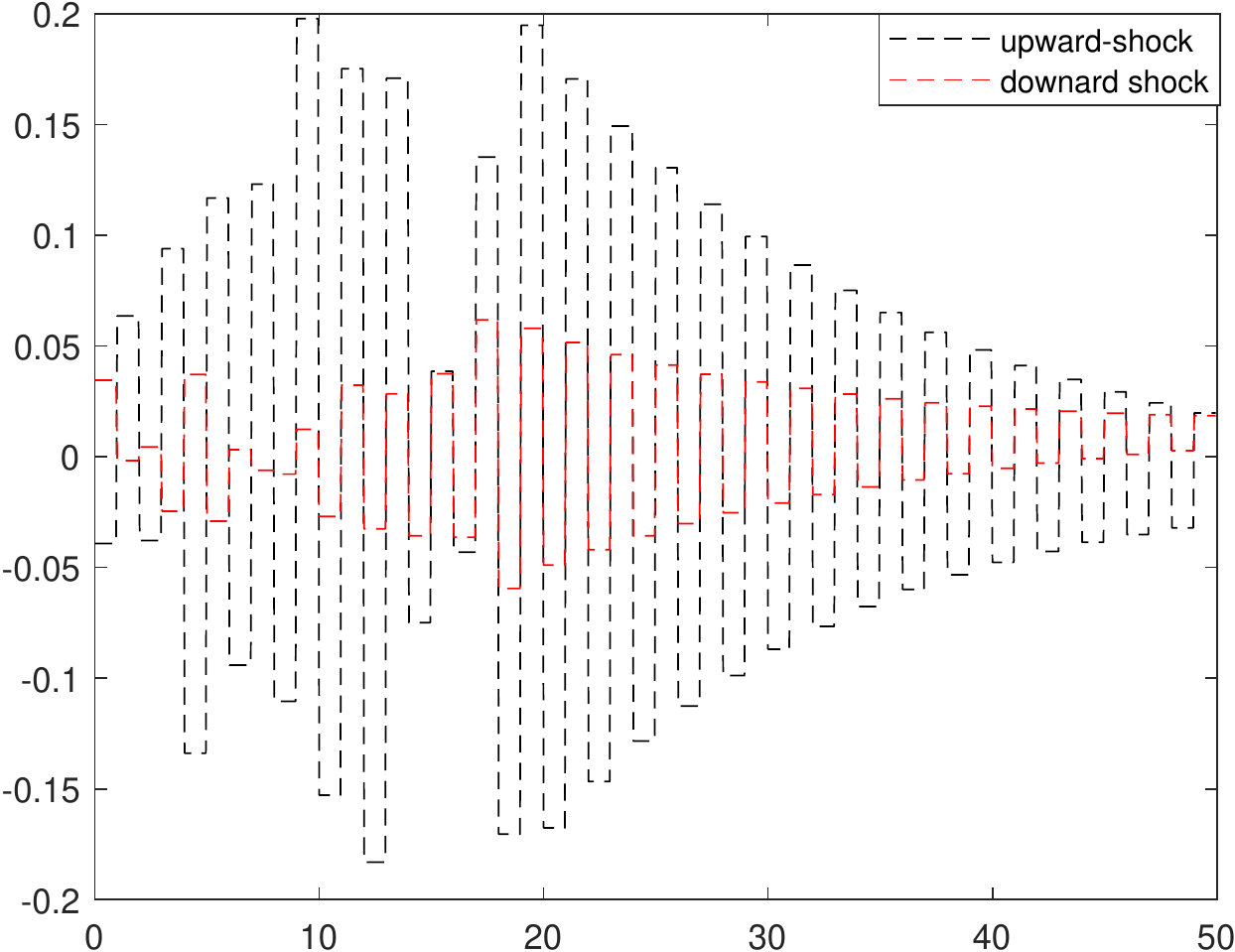}\includegraphics[keepaspectratio=true,scale=0.44]{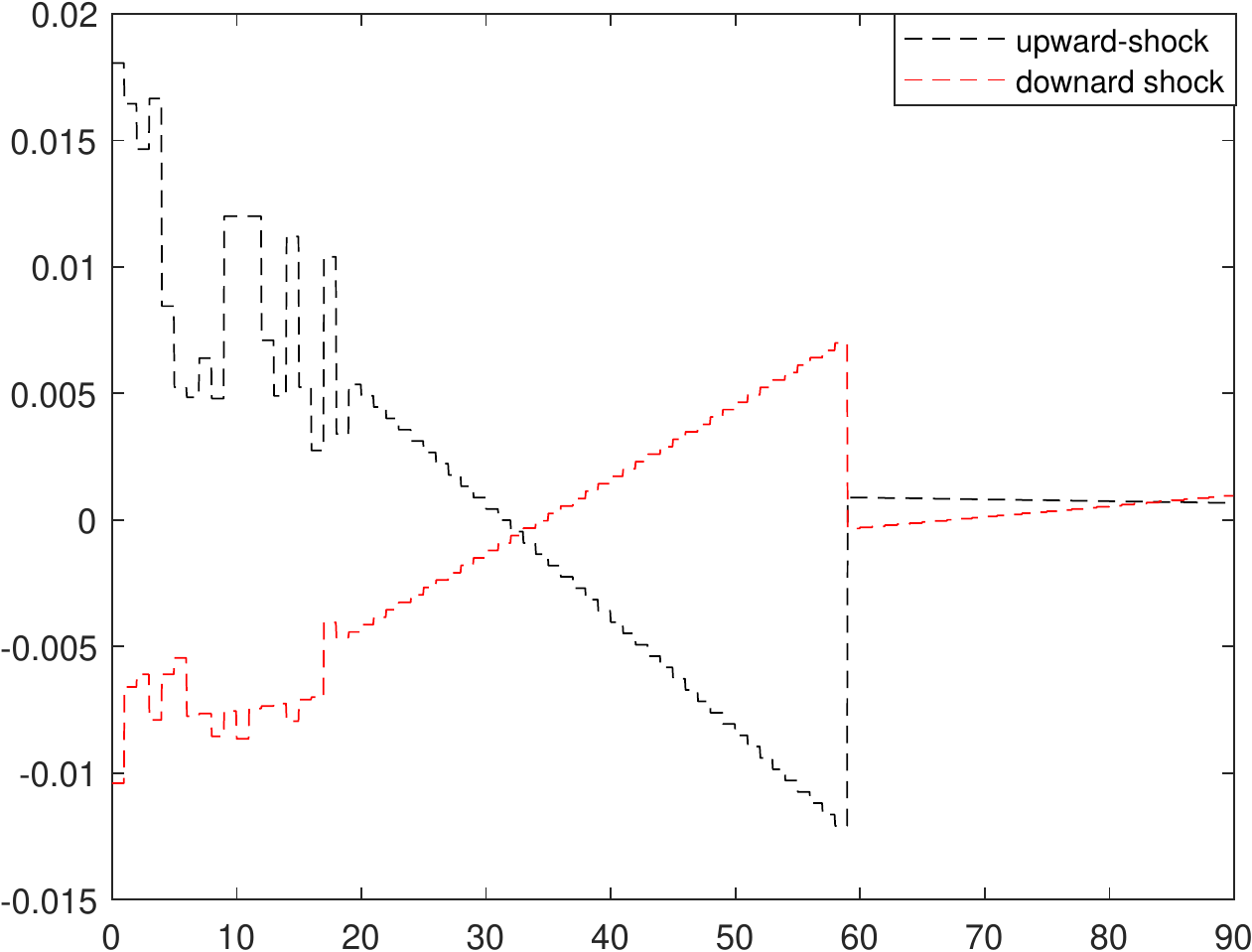}
\caption{Calibrated  piecewise constant functions $t\mapsto \varphi^{shock}(t)$ (left) and $t\mapsto \vartheta^{shock}(t) $ (middle) after the upward and downward shocks specified in Table~\ref{table_EIOPA2}. Right: calibrated  piecewise constant functions $t\mapsto \varphi^{shock}(t)$ for a constant rate model (Vasicek model with  $r_0=\theta=0.005$ and $\sigma=0$). }\label{fitted_thetaphi_rbas}
\end{figure}

Last, we have plotted in Figure~\ref{fitted_thetaphi_rbas} the functions $ \varphi^{shock}$ and $\vartheta^{shock}$ after the upward and downward shocks with the new recommendations of the EIOPA given in Table~\ref{table_EIOPA2}. Here, the zero-coupon bond prices $P^{mkt}(0,t)$ are given by a Vasicek model with $r_0=\theta=0.005$, $k=0.2$ and $\sigma=0.01$. We observe even more oscillations with the Hull and White models, which makes this model irrelevant after the shocks. Surprisingly, for the Vasicek++ model, we notice that the shifted functions cross after 30 years: $\varphi^{up}$ (resp. $\varphi^{down}$) becomes negative (resp. positive). Thus, the upward (resp. downward) shock on the risk-free interest rates leads to a downward (resp. upward) shock on the spot rate after approximately year 35, which is a puzzling. This behavior is mostly due to the phasing out of the additive term that is less innocuous as one may think. In the simplest constant rate model, we observe that stopping the phasing out of~$b$ at time~$60$ has a significant effect. 

This study tends to show that shifted models such as Vasicek++ or CIR++ have much more meaning after the shocks prescribed by the EIOPA than mean-reverting curve models such as Hull and White or Black and Karasinski models. In shifted models, the shock is translated in the shifted functions $\varphi$ that directly impacts $r_t$. Instead, for mean-reverting models, the variations of $\vartheta$ only impacts $r_t$ (and thus $R(0,t)$) after some time (typically $1/k$), it is necessary to have strong variations of $\vartheta$ to follow the variations of $R(0,t)$ at each year~$t$. This explains heuristically why oscillations are observed to fit the shocked curve. Thus, in our numerical experiments, we work then with the Vasicek++ model.

\section{Numerical results}\label{sec_num}
In this section, we provide numerical results for different model parameters. We compute the solvency capital requirement of the insurance company for an ALM portfolio over $T=30$ years by using the standard formula. We study and discuss the impact of the  shocks prescribed by the standard formula, and the corresponding values of the SCR modules. In our simulations, we sample exactly $N$ paths $(S^i_t,x^i_t)_{t\in \{1,\dots, T\}}$, for $1\le i\le N$, and use the same simulations for the central and shocked frameworks: this gives a fair comparison between different settings and models. In the first subsection, we present general results with the shocks given by the present regulation in Europe~\cite{DelReg} and with interest rates around 2\%. Then, we analyze in the second subsection the importance of the cash flow matching in the ALM, and discuss its impact on the SCR. In the third subsection, we do a similar analysis with low interest rates around 0.5\% using the last table of shocks recommended by the EIOPA (Table~\ref{table_EIOPA2}).

\subsection{Analysis of the SCR with the standard formula}

\begin{figure}[h]
\centering
\includegraphics[keepaspectratio=true,scale=0.5]{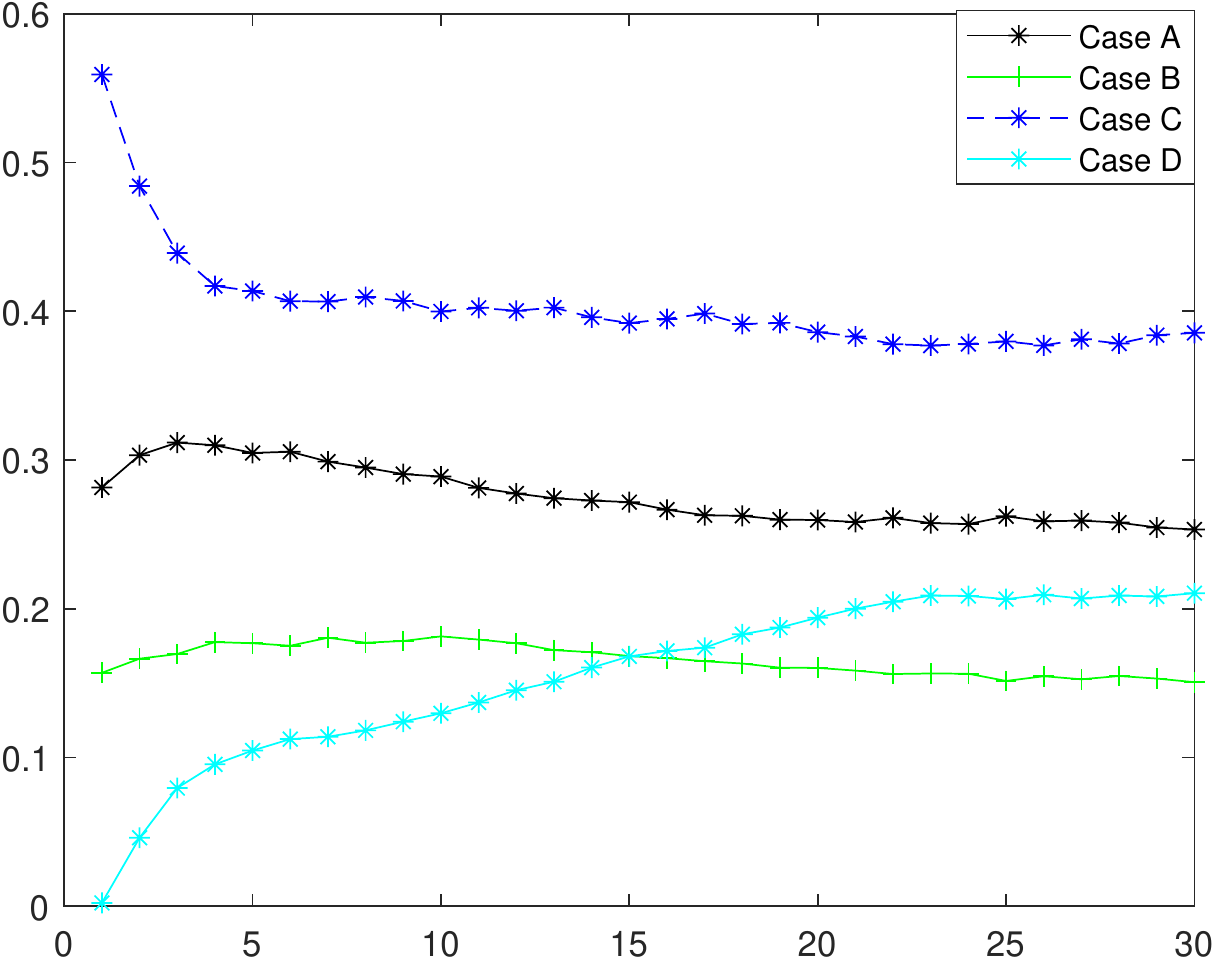}\hspace{2cm} \includegraphics[keepaspectratio=true,scale=0.5]{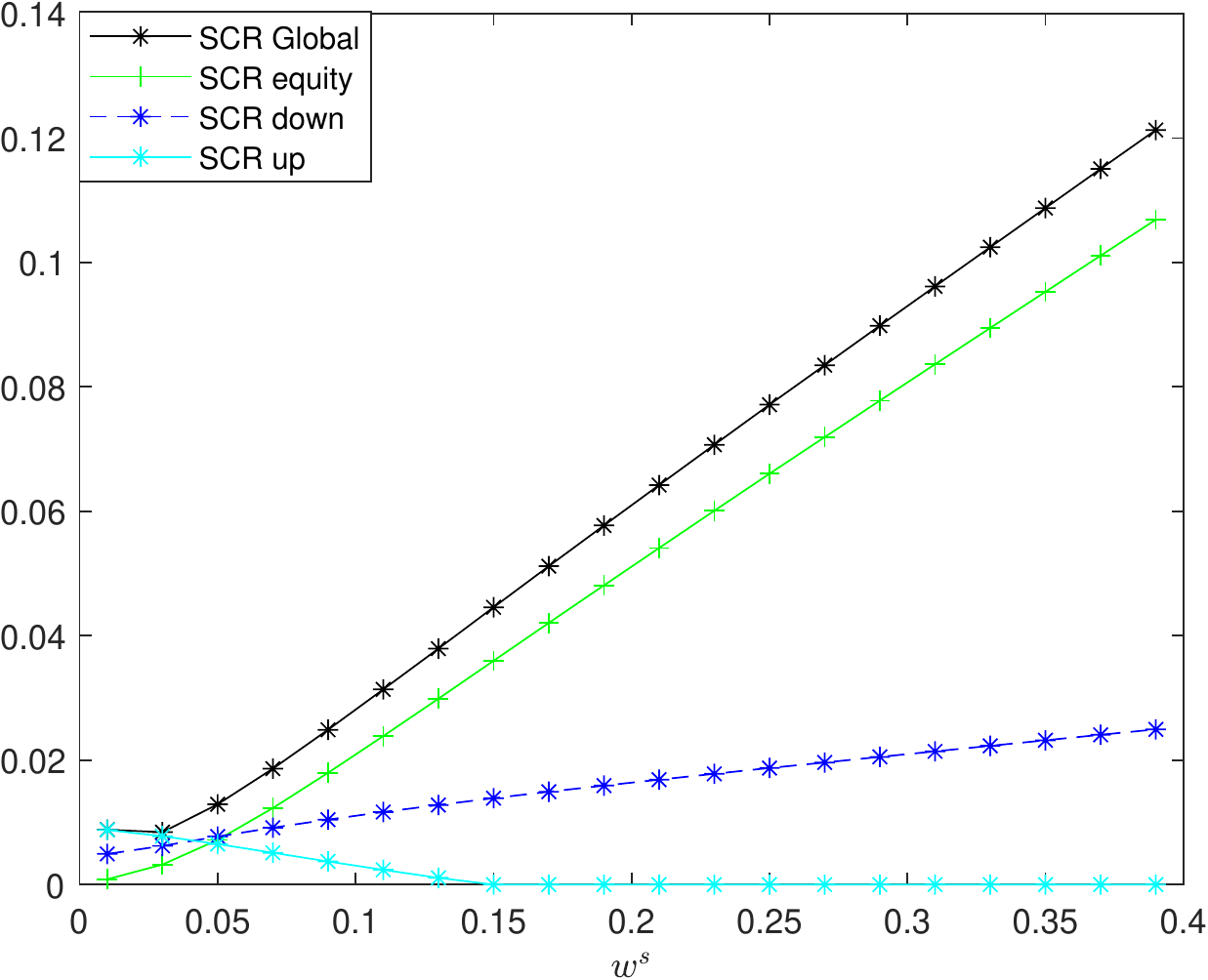}
\caption{Left: Empirical distribution of the cases A, B, C, and D determining the crediting rate, in function of the time~$t$. Right: the SCR modules in function of the constant allocation weight in equity $w^s$.}\label{Central_0.02}
\end{figure}

We work with the parameters describes in Tables~\ref{Model_param} and~\ref{LM_param}: the interest rate model follows a Vasicek model mean reverting around 2\% while the minimum guaranteed rate is set at $r^G=1.5\%$. Thus, this is a setting somehow well balanced, in the sense that the all the cases  A, B, C and D that determine the crediting rate  occur with a significant proportion. This is confirmed by the empirical distribution  plotted at the left of Figure~\ref{Central_0.02}. If $r^G$ were higher (resp. lower) we would observe mostly cases C and D (resp. A and B).

To determine the constant allocation in stock and bond that we consider in our simulations, we have drawn in the right of Figure~\ref{Central_0.02} the different SCR components in function of $w^s$, as well as the global SCR given by formula~\eqref{def_SCR_mkt}. We use the shocks given by Table~\ref{table_EIOPA}. In our simulations, we are looking for an allocation that makes the SCR on equity and the SCR on interest rates of the same order, since the aggregation formula~\eqref{def_SCR_mkt} somehow encourages to diversify the risk components.  This is achieved by $w^s=0.05$. Note that in this case, this is also the allocation that minimizes the SCR. As one may expect, the SCR on equity is increasing with respect to $w^s$. The risk neutral valuation dissuades from taking risk and is questionable in the life insurance context, as pointed by Vedani et al.~\cite{VeEKLoPr}.
The monotonicity is also observed for the downward (resp. upward) shock on the equity: the higher is $w^s$, the less the insurance company has capital gains (resp. loss) from the downward shocks. Here, the curves of $SCR_{up}$ and $SCR_{down}$ cross also around $w^s=0.05$. 
\begin{table}[H]
\centering  
\begin{tabular}{|l|r|}
  \hline
  Stock model & Short-rate model  \\
  \hline
  $S_0=1$ & $r_0=\theta=0.02$  \\
  $\sigma_S=0.1$ & $\sigma_r=0.01$  \\
   $\gamma=0$ & $k=0.2$  \\
  \hline
\end{tabular} 
\caption{Market-model parameters}\label{Model_param}
\end{table}

\begin{table}[H]
\centering  
\begin{tabular}{|l|l|}
  \hline
  Management Parameters & Liability Parameters \\
  \hline
  Allocation in stock $w^s=0.05$ &Lapse triggering threshold $\beta=-0.01$ \\
  Allocation in bond $w^b=0.95$ & Massive lapse triggering threshold $\alpha=-0.05$ \\
  Participation rate $\pi_{pr}=0.9$ & Maximum lapse dynamic lapse rate $DSR_{max}=0.3$  \\
  Minimum guaranteed rate $r^G=0.015$ &Static lapse rate $\underline{p}=0.05$ \\
  Competitor rate $r^{comp}_t=r_t$& \\
  Smoothing coefficient of the PSR: $\bar{\rho}=0.5$ &  \\
  Bond portfolio maximal maturity $n=20$ &  \\ 
  \hline
\end{tabular} 
\caption{Liability and management parameters}\label{LM_param}
\end{table}
\begin{table}[H]
\centering  
\end{table}

\begin{figure}[h]
\centering
\includegraphics[keepaspectratio=true,scale=0.5]{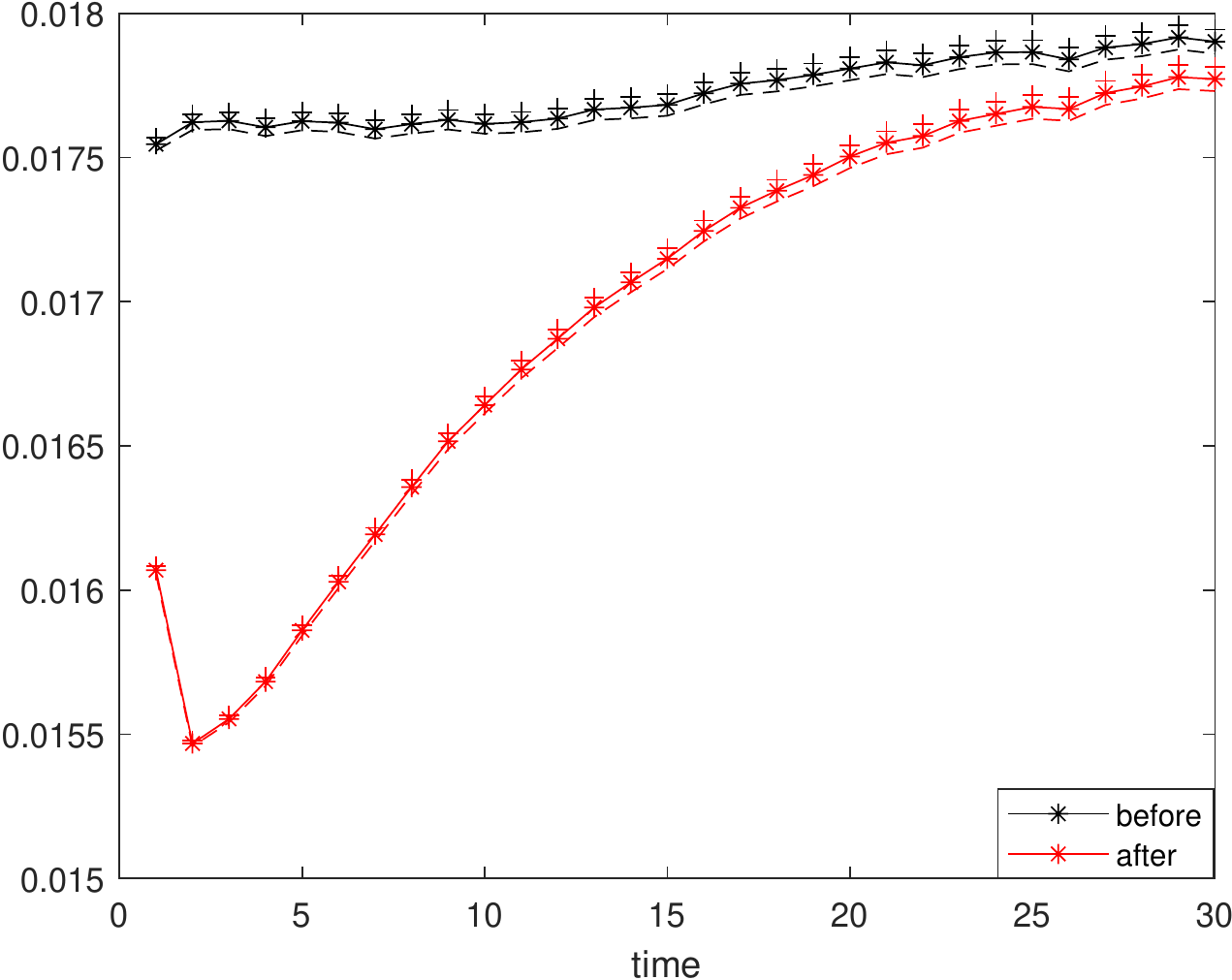}\hspace{2cm}\includegraphics[keepaspectratio=true,scale=0.5]{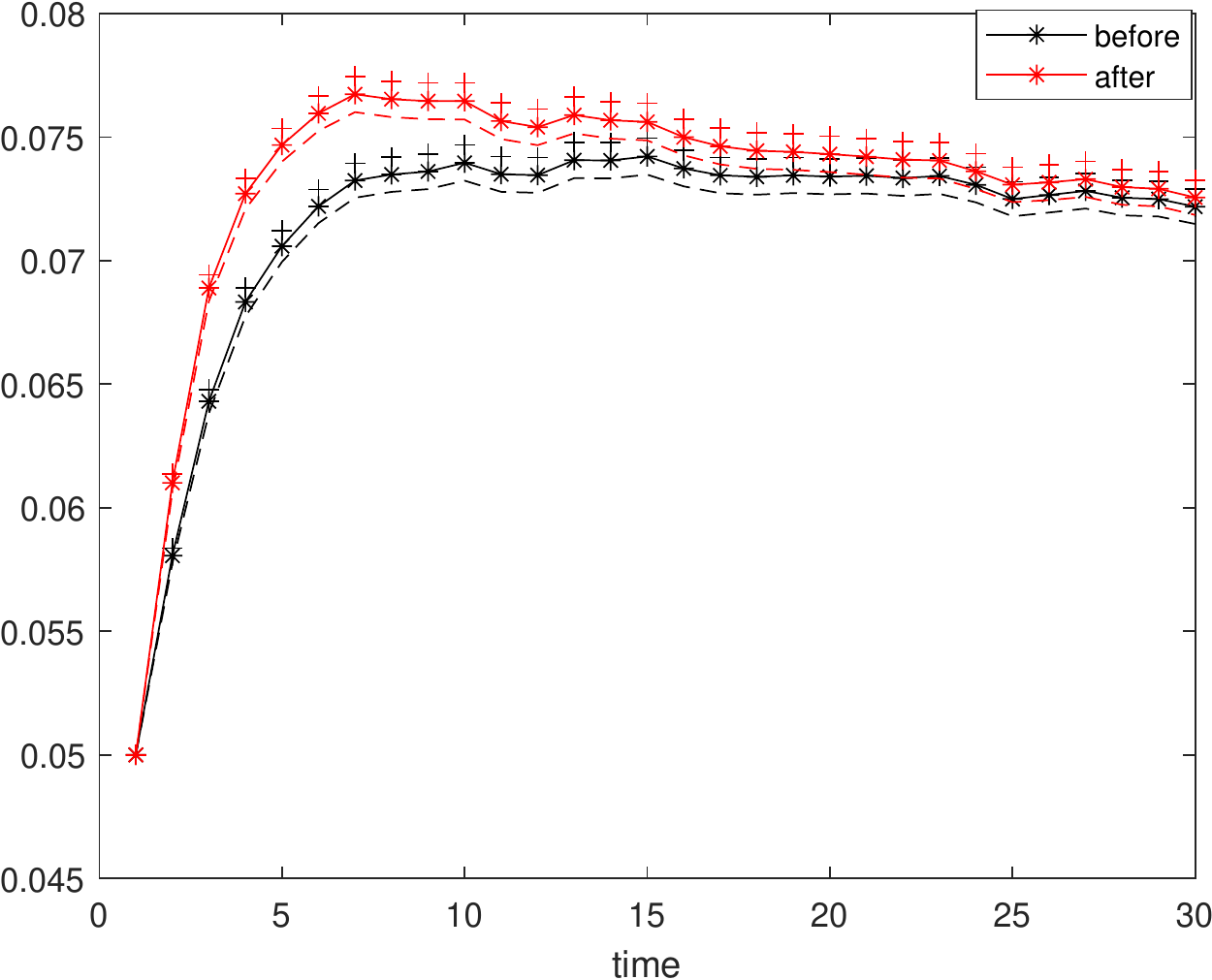}
\caption{Before and after the equity shock of 39\%. Evolution of the mean crediting rate $\E[r_{ph}(t)]$ (left) and of the mean exit rate $\E[p^e_t]$ (right) in function of the time~$t$.}\label{Eq_sh_0.02}
\end{figure}

\begin{figure}[h]
\centering
\includegraphics[keepaspectratio=true,scale=0.44]{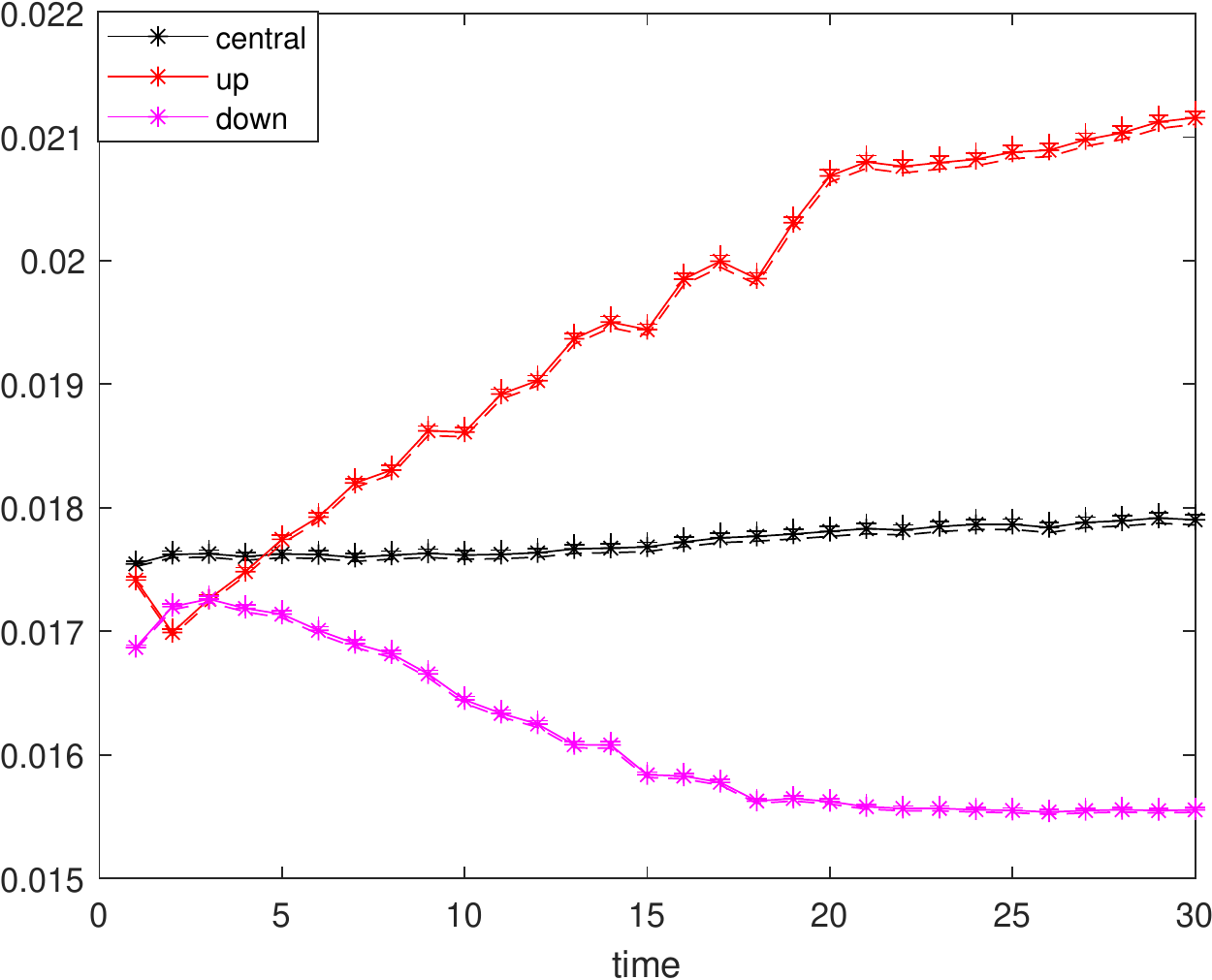}\includegraphics[keepaspectratio=true,scale=0.44]{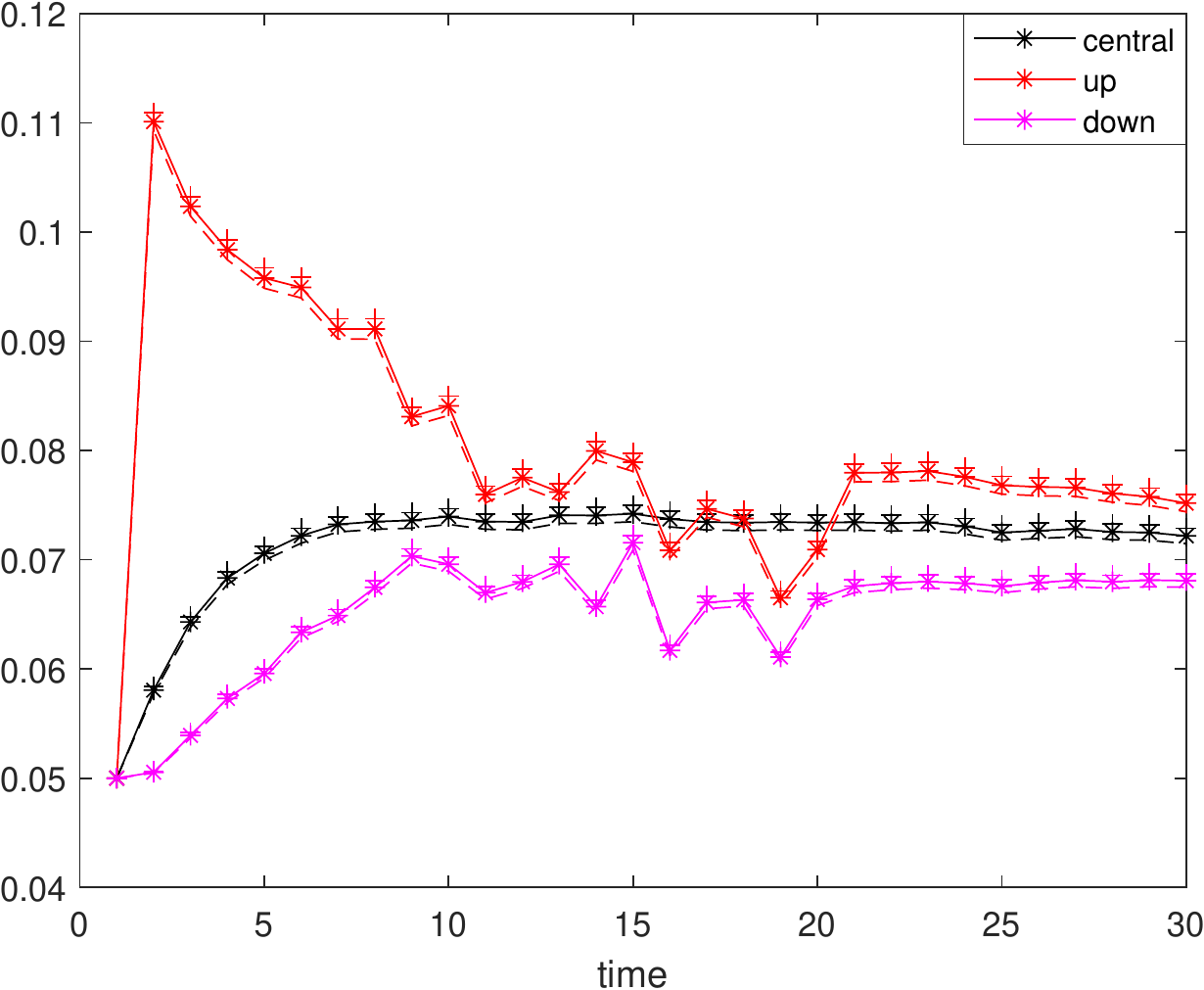}\includegraphics[keepaspectratio=true,scale=0.44]{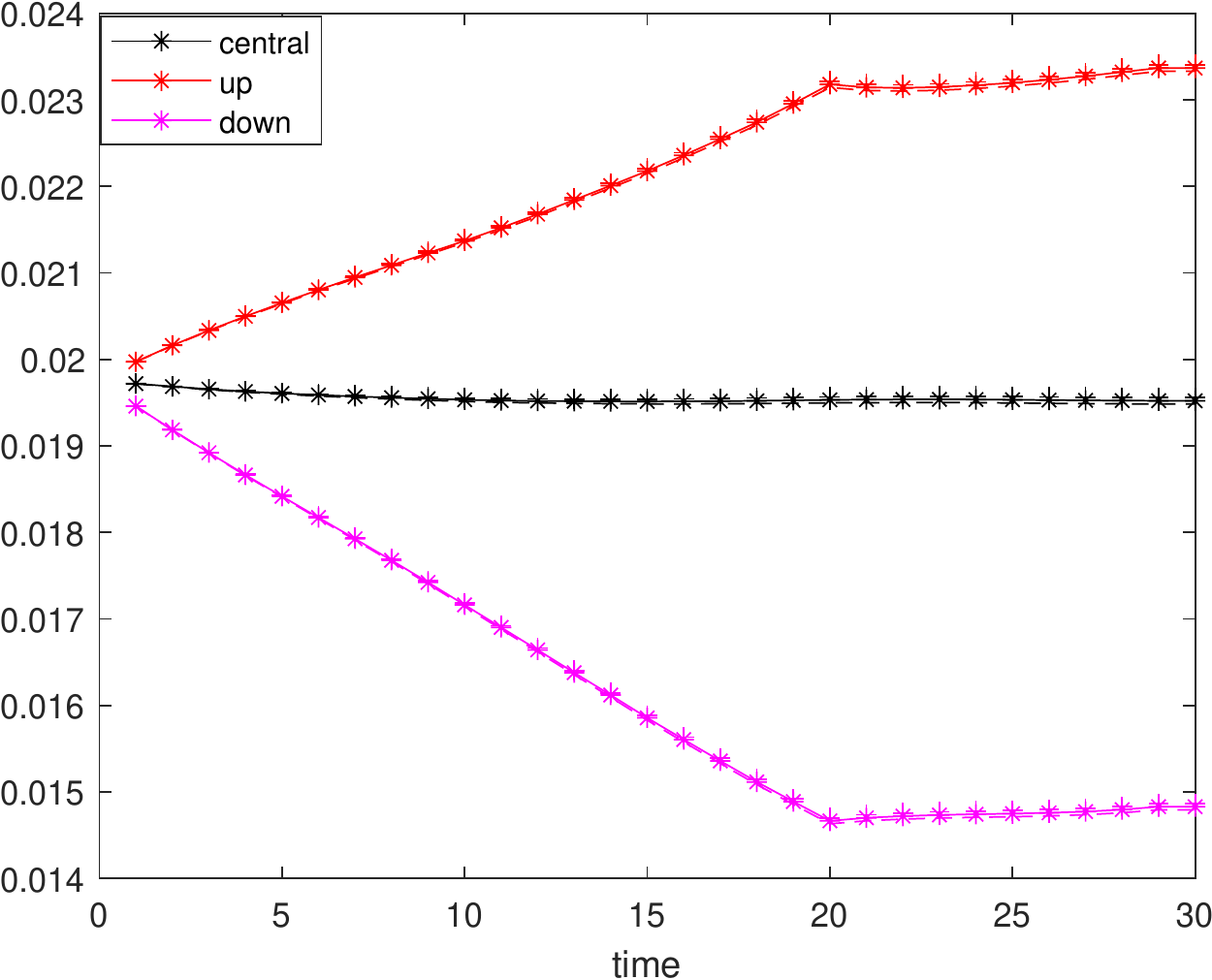}
\caption{Before and after the downward and upward shocks on interest rates. Evolution of the mean crediting rate $\E[r_{ph}(t)]$ (left), of the mean exit rate   $\E[p^e_t]$ (middle) and of the average coupon in the Bond portfolio (right)in function of the time~$t$.}\label{Int_sh_0.02}
\end{figure}

We now analyse the shocks. In Figure~\ref{Eq_sh_0.02}, we have plotted the empirical means of the crediting rate and of the exit rate after the equity shock $s^{eq}=-0.39$. We have also indicated with a plus sign (resp. a dotted line) the upper (resp. lower) bound of the 95\% confidence interval. As one may expect, the equity shock give an important loss resulting in a lower crediting rate and thus a higher exit rate. Nonetheless, the effects on these rates are moderate due to the guaranteed rate: in average, the maximal difference between the competitor rate $r_t$ is about $0.5\%$, and therefore only few scenarios are at some time above the surrender triggering threshold~$\beta$. The shocks on the interest rate, illustrated in Figure~\ref{Int_sh_0.02}  mix different effects. The downward shock gives an important gain at the beginning, but on the long run it makes harder for the insurance company to credit the minimal guaranteed rate. This is known as the reinvestment risk in the literature. This fact is confirmed by the plot of the mean value of the average coupon rate $\frac 1n \sum_{i=1}^n c^i_t$, that is even slightly below~$r^G=1.5\%$ after 20 years. This plot of the average coupon rate also illustrates the rolling mechanism described in equation~\eqref{coupon_rollup}.  Conversely, the upward shock gives an important initial loss, but on the long run it makes much easier for the insurance company to credit the minimal guaranteed rate. Also, because of the initial loss, the insurer tends to credit at the beginning rather low rates to policyholders while the competitor rate~$r_t$ is high: this has an important effect on the surrender rate.

Let us mention here that we have also run the same ALM strategy when the competitor rate is $r^{comp}_t=\max(r_t, 0.9 r_{ph}(t-1))$. We have observed rather minor differences with the case
$r^{comp}_t=r_t$. Therefore for the simplicity of the exposition, we have preferred to keep $r^{comp}_t=r_t$ in this numerical section.
\begin{figure}[h]
\centering
\includegraphics[keepaspectratio=true,scale=0.55]{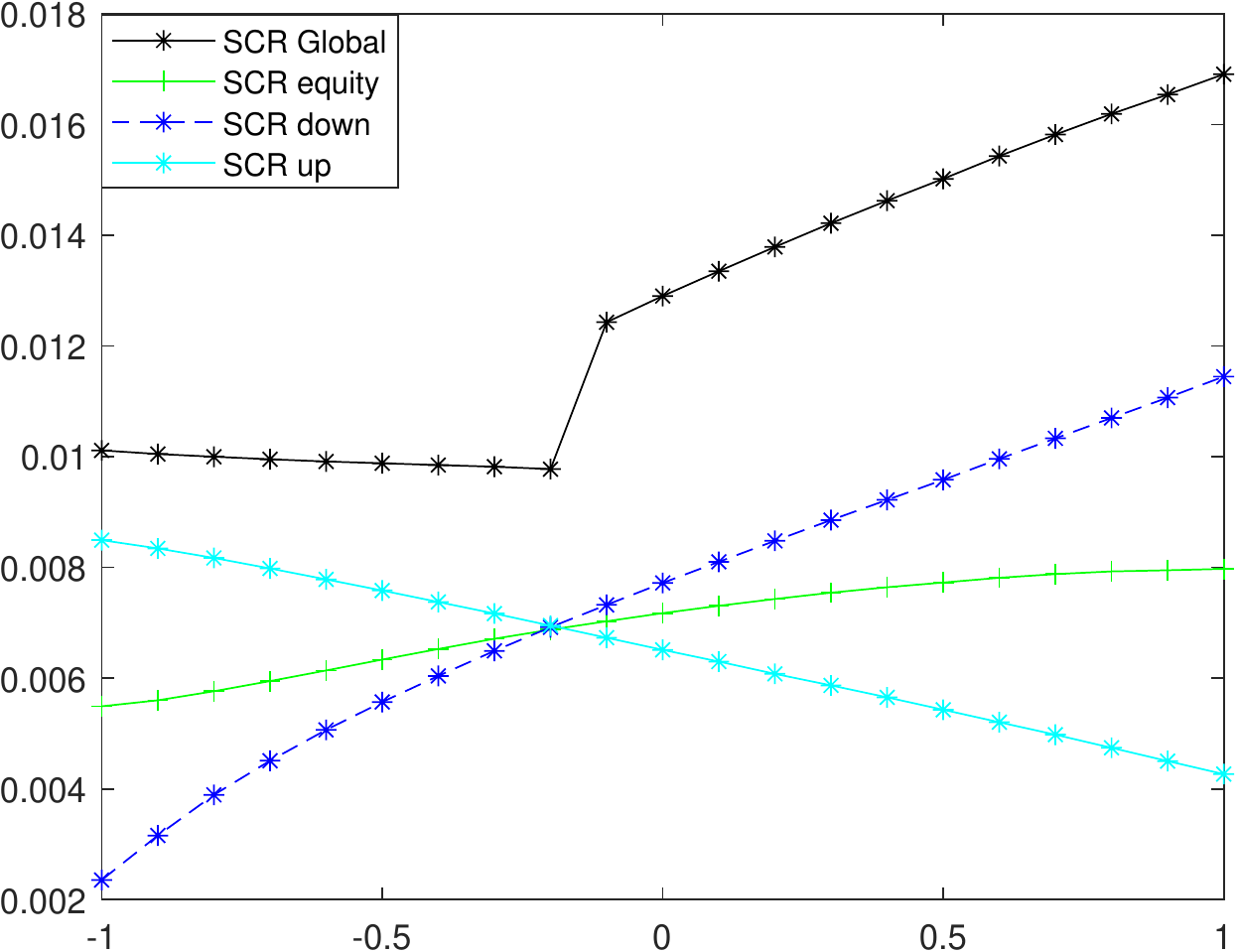}
\caption{Values of the different SCR modules in function of~$\gamma$, the correlation between bonds and stocks.}\label{Correl_gamma}
\end{figure}
Before going further with the analysis of the importance of the cash-flow matching, we have drawn the dependence of the different SCR modules in function of~$\gamma$ that tunes the correlation between the equity and the interest rate. This is the kind of quantitative study an ALM model may help for.  We observe that $SCR_{eq}$  and $SCR_{down}$ are decreasing and $SCR_{up}$ is increasing with respect to~$\gamma$. The aggregated SCR~\eqref{def_SCR_mkt} is first slightly decreasing when $SCR_{down}<SCR_{up}$ and then increasing. We notice an important discontinuity at $SCR_{down}=SCR_{up}$ which is due to the $\varepsilon$ coefficient in formula~\eqref{def_SCR_mkt} that goes from~$0$ to $1/2$ when $SCR_{down}$ goes above $SCR_{up}$.
In the range $[-0.5,0.5]$ usually observed for the correlation between stocks and sovereign bonds (see e.g. Pericoli~\cite{Pericoli} or Rankin and Shah Idil~\cite{RaSI}), we observe an important variation of 50\% of $SCR_{mkt}$, half of which is contained by the discontinuity. Such a discontinuity in the SCR formula is unfair and may incite insurer to be at the edge of this discontinuity: a continuous formula for $SCR_{mkt}$ such as $\max(\sqrt{SCR_{eq}^2+SCR_{up}^2},\sqrt{SCR_{eq}^2+SCR_{down}^2+ SCR_{eq} SCR_{down}})$ would avoid this.

\subsection{Study of the cash-flow matching}

\begin{figure}[h]
\centering
\includegraphics[keepaspectratio=true,scale=0.5]{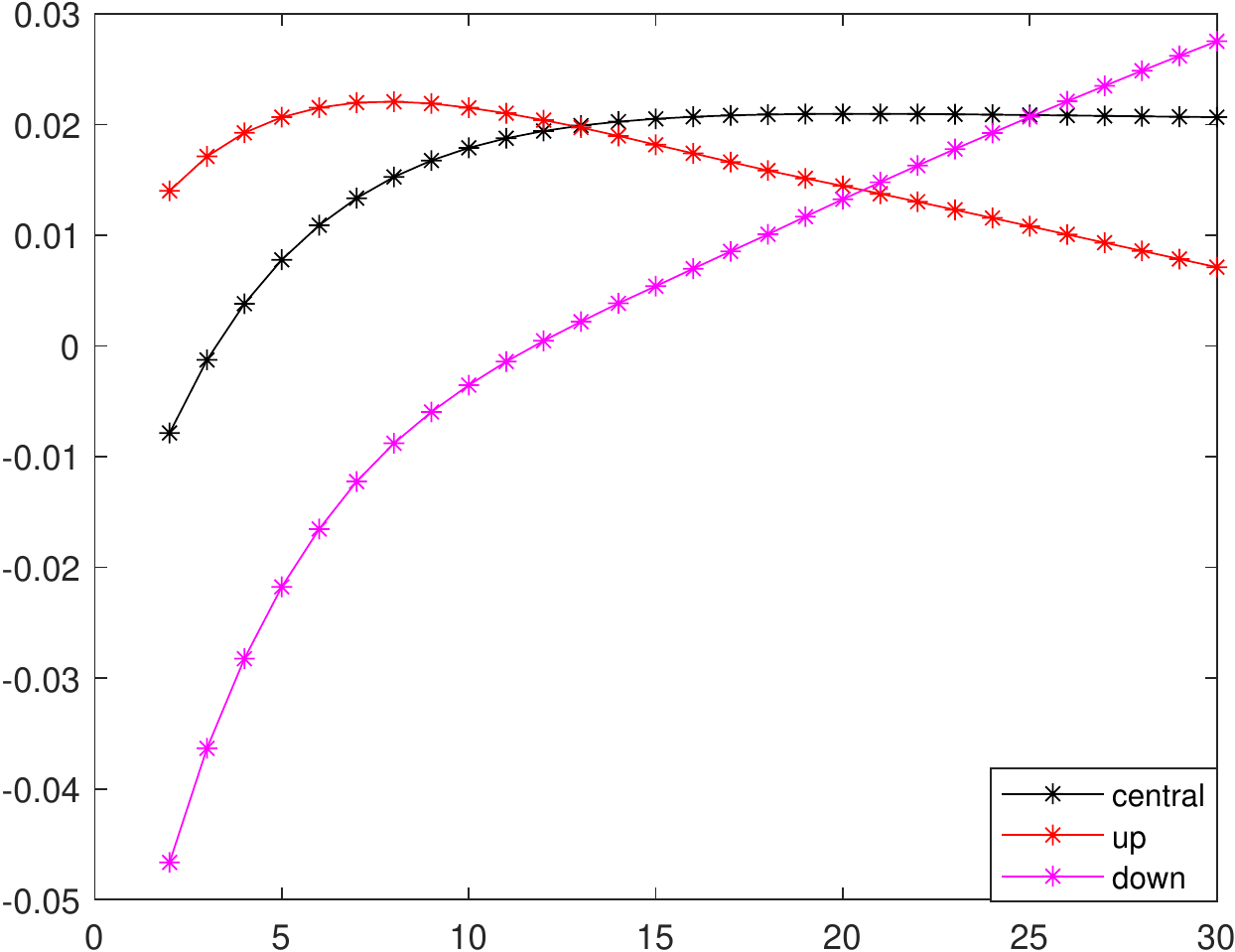}\hspace{2cm}\includegraphics[keepaspectratio=true,scale=0.5]{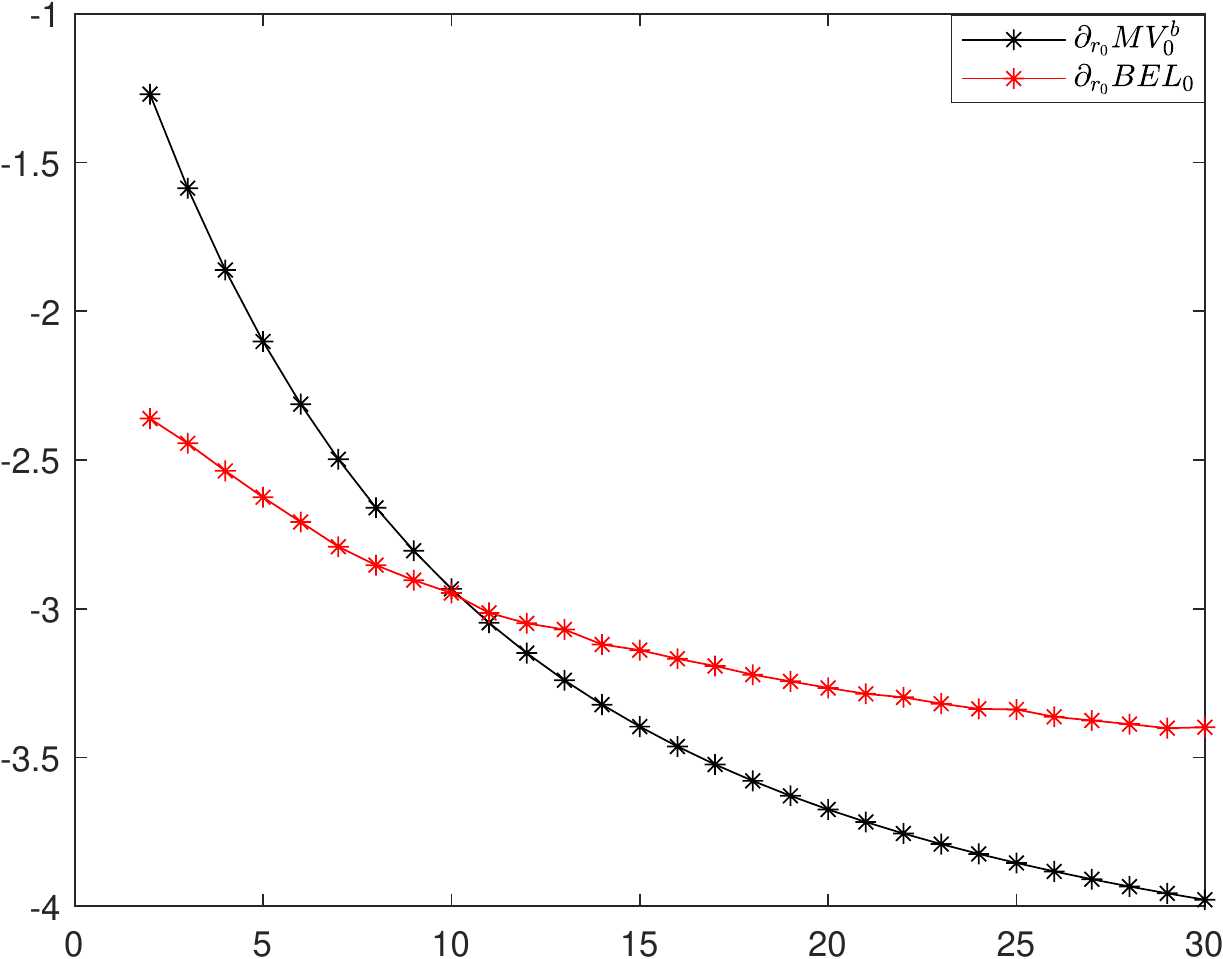}
\caption{Left: mean value of the Basic Own-Funds in function of~$n$ (defining the bond combination~\eqref{def_Bbar} in which bonds are invested) in the central framework  and with the upward and downward shocks on interest rates. Right: Macaulay durations of the assets ($\partial_{r_0}MV^b_{0}$) and of the liabilities ($\partial_{r_0}BEL_0$) in function of~$n$.}\label{BOF_n_0.02}
\end{figure}

In this paragraph, we want to assess the relevance of an original feature of our ALM model: the cash-flow matching between the bond assets and the liabilities. This feature reproduces a common practice of insurance companies. In order to have an idea of a good choice of~$n$ (the maximal maturity of the bond combination~\eqref{def_Bbar}), we have plotted on the left of Figure~\ref{BOF_n_0.02}  the Basic Own-Funds defined in~\eqref{def_BOF} in function of~$n$ for the central and shocked settings. Thus, the difference between the curve of the central case and the shocked cases gives the values of $SCR_{up}$ and $SCR_{down}$. We use here the same parameters as in Tables~\ref{Model_param} and~\ref{LM_param}. We see that in the central case, the BOF is maximized around $n=20$, but is anyway rather flat between $n=15$ and $n=30$. As one may expect, the BOF is increasing with respect to $n$ in the downward shocked scenario: the more the insurer invests in long maturity bonds, the more he benefits from the decrease of the interest rates. For the upward shocks, two effects are mixed. On the one hand, the longer is the bond maturity, the greater is the loss due to the shock. On the other hand, the insurer has interest to match well the  bond assets and the liabilities in order to keep as much as possible policyholders, since the high interest rates will be profitable on the long run. Thus, the higher BOF are obtained for $n=7$ and $n=8$.  
If the insurance company wants to have the minimal $SCR_{int}=\max(SCR_{up},SCR_{down})$ with the standard formula, it has to choose $n$ around where the curves crosses. Thus, $n=20$ appears to be the choice that minimizes $SCR_{int}$. Note that due to the discontinuity of  formula~\eqref{def_SCR_mkt}, this is also a very good choice since we have also $SCR_{up}>SCR_{down}$ : with $n=21$, we would have $SCR_{up}<SCR_{down}$ and thus $\varepsilon=1/2$, leading to a greater $SCR_{mkt}$. Thus, we can tune the value of $n$ to satisfy $SCR_{up}>SCR_{down}$ and benefit from a better diversification coefficient of the standard formula. In comparison, we have also considered the Macaulay duration of the assets and of the liabilities after the initial allocation, i.e. $\partial_{r_0}MV^b_{0}$ and $\partial_{r_0}BEL_0$. They are plotted in the right of Figure~\ref{BOF_n_0.02}, in function of $n$. The curves crosses around $n=10$, which means that to be hedged against small variations of the interest rate at time~$0$, the insurance company should take $n=10$. Note that from the graph on the left, this choice leads to a lower BOF in the central framework and to a much higher SCR with the standard formula. This demonstrates if it were necessary  that hedging small variations is not the same as hedging shocks.

\noindent {\bf A proxy model.} The discussion above already shows the importance of the choice of~$n$. To go further, we would like now to compare with a simpler model where there is no cash-flow matching. This proxy model works as follows. At time~$0$,  the insurer invests in a single at-par coupon bearing bond with yield to maturity $n_p$ and unitary market-value given by:
\begin{align*}
\bar{MV}^b_0=B(n_p,c^{n_p}_{swap}(0))=1
\end{align*}
 At each time $t\in\{1,\dots, T\}$, the insurance company re-balances its according to target weights. The available capital now is reinvested in a single bond with duration $n_p$. However, to approximate the full model, we do not consider that the company sells all of its current bond of maturity $n_p-1$ to buy new bonds with longer term $n_p$: this would imply a tremendous realization of capital gains or losses, leading to important change in book values. To deal with this issue and make a fair comparison with the original model, we propose the following approximation. Before reallocating its portfolio, the insurer adjusts its holding in bonds and compute $\phi^b_{t_2}$ such that:
\begin{align*}
\phi^b_{t-1}B(n_{p}-1,c^b_{t-1})=\phi^b_{t_2}B(n_{p},\tilde{c}^n_{t})
\end{align*} 
where:
\begin{align*}
\tilde{c}^n_{t}=\frac 1 {n}  c^{n}_{swap}(t)+(1-\frac 1 {n} )c^b_{t-1}
\end{align*}
we assume that this procedure does not lead to a realization of CGL. The purpose of this approximation is to adjust the holding in bonds in order to remain unchanged the current market-value $\phi^b_{t-1}B(n_{p}-1,c^b_{t-1})$ of the bond portfolio while taking into account the reinvestment risk of the original model. In particular, the original model  reinvests the  nominal value at the swap rate, which justifies the term $\frac 1 {n} c^{n}_{swap}(t)$, and keeps a fraction $1-\frac 1 {n}$ of bonds with unchanged coupons.

To determine the maturity~$n_p$ of the bond used in the proxy model, we choose the maturity $n_p$ in order to keep approximately the same gain or loss between both models after the downward or upward shock:
\begin{align}
\Delta MV_0\approx \Delta MV_0^{proxy}(n_p)
\end{align}
After numerical investigation, the choice $n_p=\frac{n}{2}$ is satisfactory in practice. Then, to obtain exactly the same size of shocks in terms of loss or gain,  we adjust then the position in bonds in the proxy model:
\begin{align}
\phi^{b,proxy}_{0^+} MV^{b,proxy}_{0+}=\phi^b_{0+} MV^{b}_{0+}.
\end{align}
This adjustment is important to compare fairly the two models: thus, the shocks induces the same initial loss or gain.  
\begin{figure}[h]
\centering
\includegraphics[keepaspectratio=true,scale=0.44]{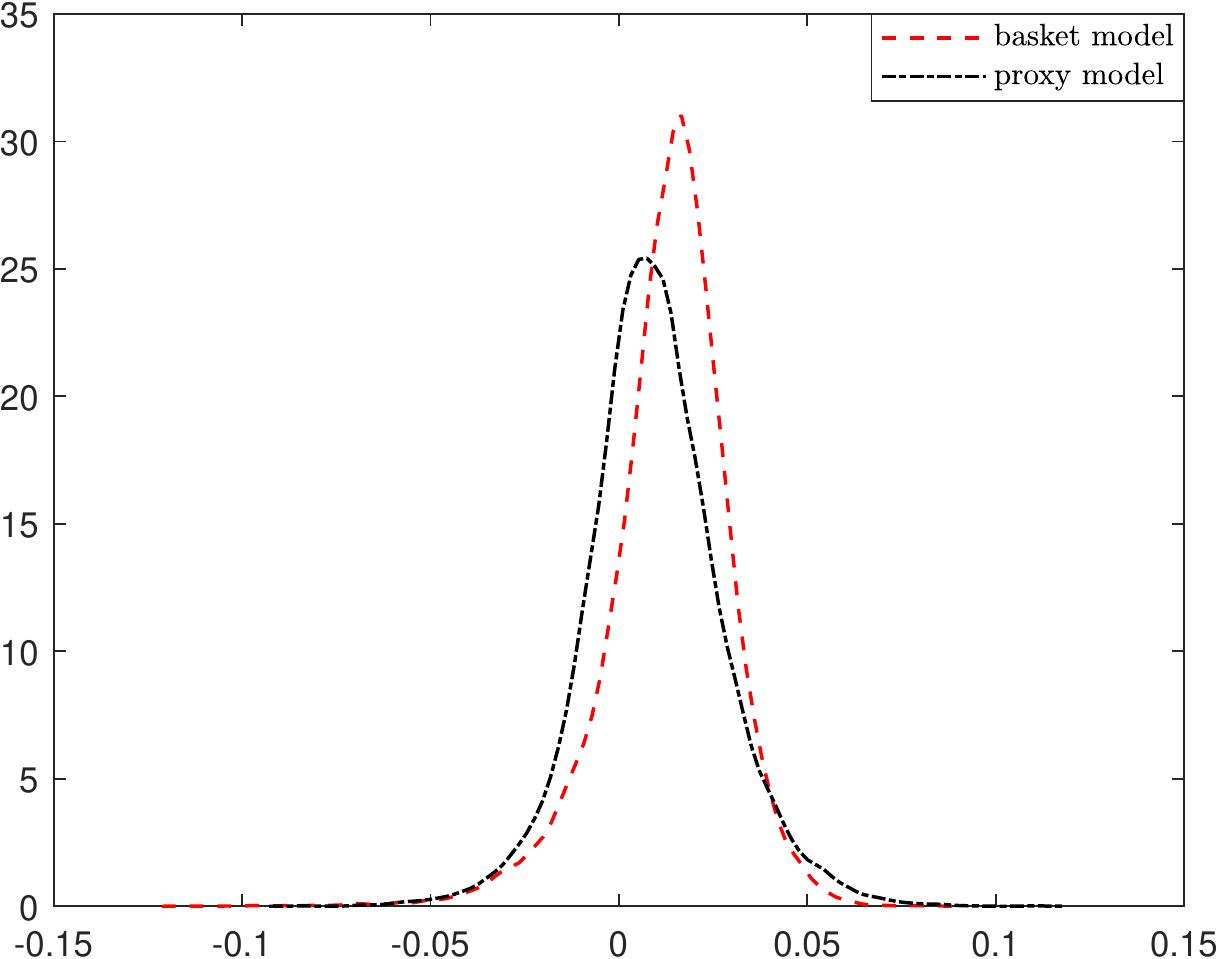}\includegraphics[keepaspectratio=true,scale=0.44]{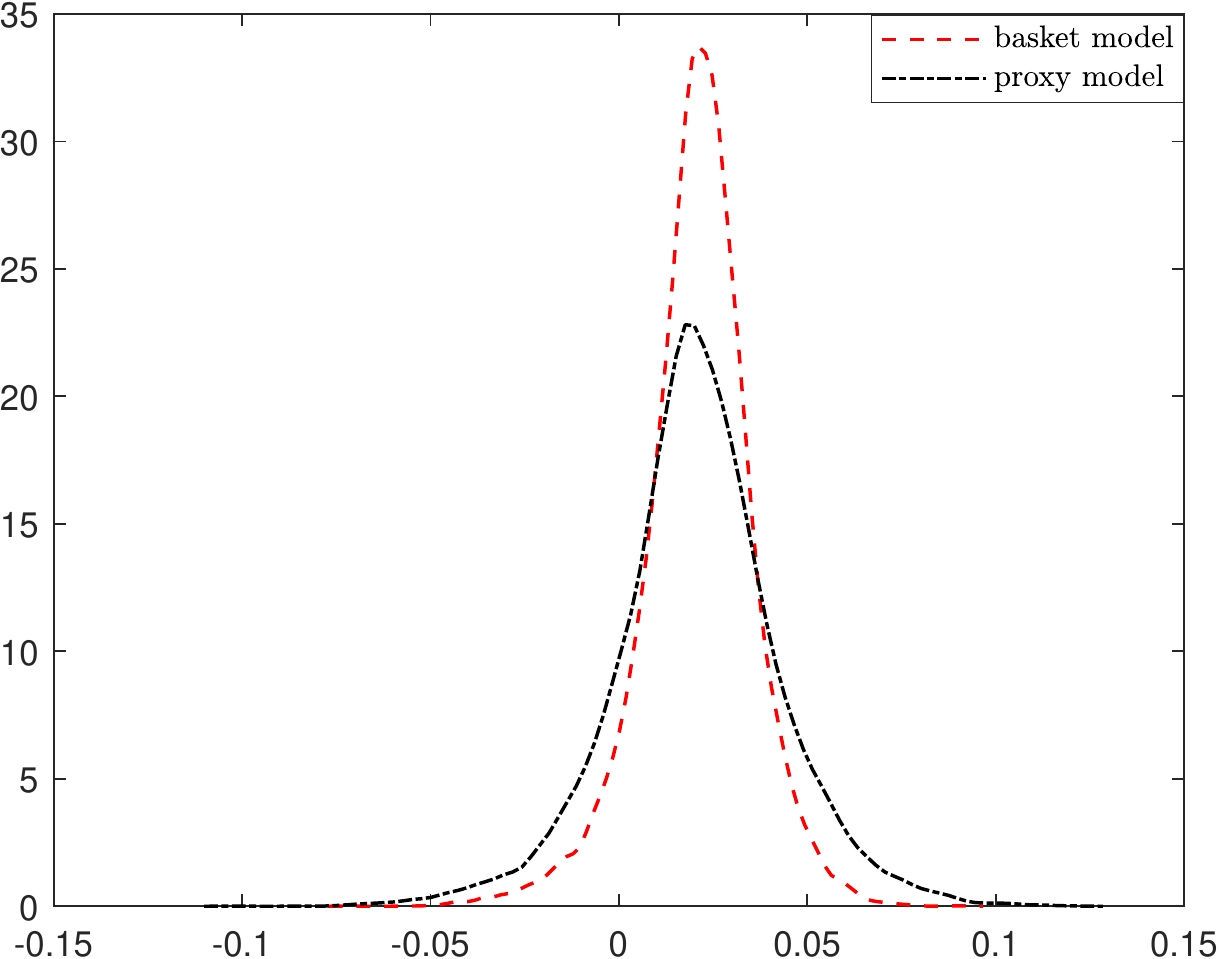}\includegraphics[keepaspectratio=true,scale=0.44]{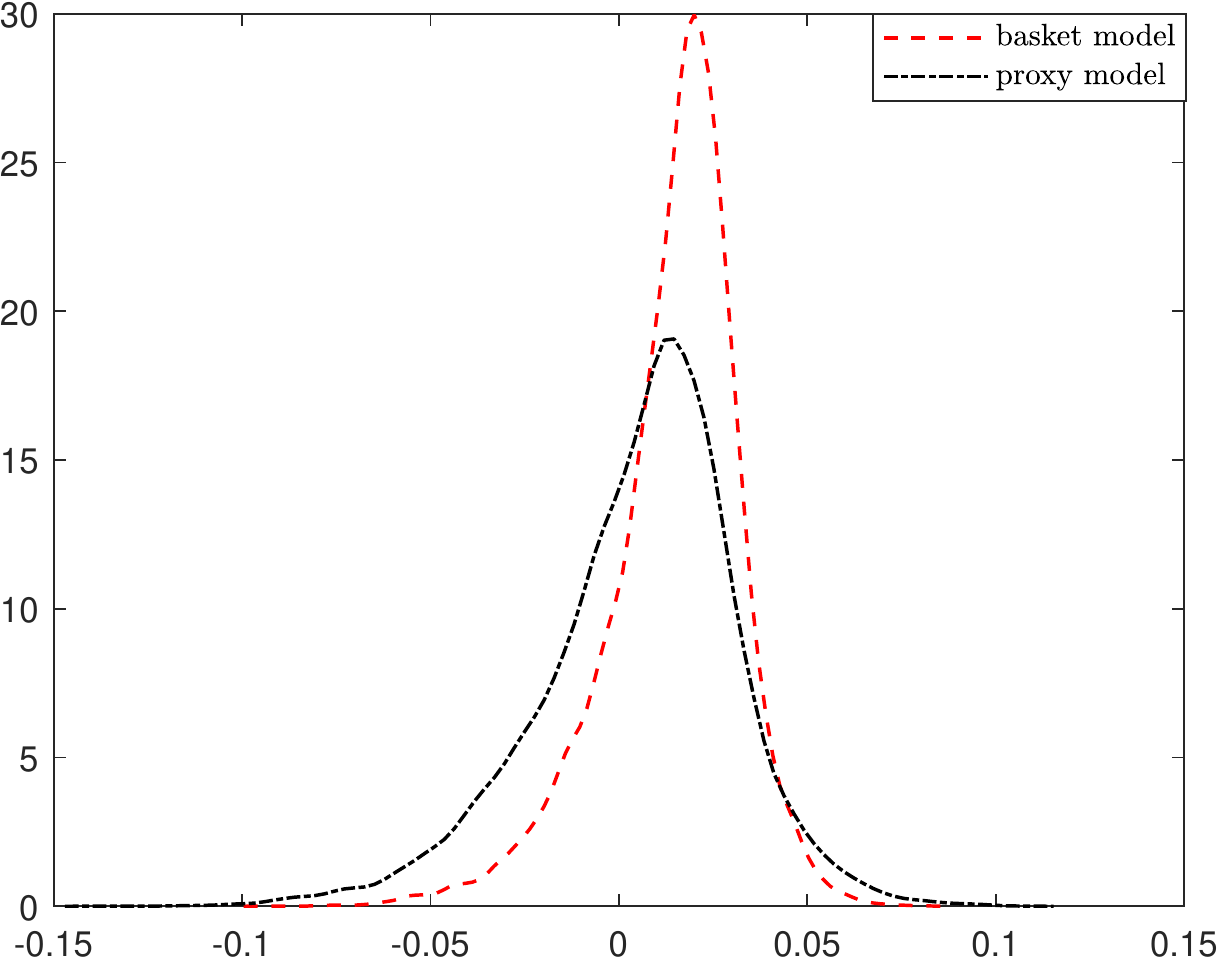}
\caption{Empirical BOF distributions with the proxy model and the original (Basket) model with a basket of bonds, after the downward shock (left),  the upward shock (right) and in the central framework (middle). }\label{BOF_proxy}
\end{figure}
Figure~\ref{BOF_proxy} illustrates  the difference between the original and the proxy models on the Basic Own-Funds distribution in the central setting and after the interest rate shocks. Table~\ref{BOF_table} indicates the corresponding BOF mean values, and Table~\ref{SCR_table} the associated SCR values. We observe two effects. In all cases, we observe that the BOF distribution has less variance and is more peaked in the original model. This is expected: the basket of bonds allows a good cash-flow matching between the nominal value of the expiring bonds and the surrendering policyholders. The second effect  concerns the interest rate shocks. In both cases, the original model performs much better than the proxy model. This is again due to the cash-flow matching. These shocks induce large latent gains or losses on bonds: with the proxy model, the insurer is forced to realize a part of it since he pays policyholders by selling a fraction of its portfolio. For the central setting and the shock on equity, there is no longer important  latent gains or losses on bonds, and the mean BOF values are roughly the same between both models. Note that this is clear for the upward shock since realizing losses reduce the yearly P\&L, but the interpretation for the downward shock is less obvious. Besides, on this second effect, we notice that the upward shock is even more expensive  than the downward shock for the proxy model: from Table~\ref{BOF_table}, the difference with the original basket model is equal to $0.0036$ for the downward shock, versus $0.0092$ for the upward shock. This difference is due to the massive rate of surrenders in the upward shock (see Figure~\ref{Int_sh_0.02}), which is caused by the increase of the  competitor rate~$r_t$. Thus, these increased surrenders have again to be paid by selling a greater fraction of portfolio, leading to realize even more losses. This effect has already been noticed in the literature, see~\cite{KuBeGr}.

\begin{table}[H]
\centering  
\begin{tabular}{|l|c|c|}
  \hline 
  & Basket & Proxy \\
  \hline
  Central &    0.0208 [0.0206,0.02010] & 0.0207 [0.0203,0.0210]\\
  Equity shock &0.0136 [0.0134, 0.0139] & 0.0134 [0.0130,
   0.0137]\\
  Downward shock &  0.0130  [0.0128, 0.0133] & 0.0094  [ 0.0091,
    0.0096]\\
  Upward shock &  0.0145 $[0.0142,0.0147]$ &   0.0053 $[  0.0049,0.0056]$\\
  \hline
\end{tabular}
\caption{Mean value of the BOF in the original and the proxy models with 95\% confidence interval, under the central and shocked settings. }\label{BOF_table}
\end{table}

Let us now comment quickly the different SCR values in Table~\ref{SCR_table}. An obvious remark is that the standard formula that relies on mean values is not sensitive to the BOF distributions and does not reward if they are more peaked with less variance. This is a clear weakness of the standard formula. Thus, the first effect described just above has no impact on the SCR, and for example the values of $SCR_{eq}$ is the same on both models. The second effect has instead some impact on the interest rate modules of the SCR, leading to some improvement of $SCR_{mkt}$. Note that the improvement is nonetheless tamed by the fact that in the aggregated formula~\eqref{def_SCR_mkt}, we use $\varepsilon=0$ for the proxy model and $\varepsilon=1/2$ for the original model with a basket of bonds.

\begin{table}[H]
\centering  
\begin{tabular}{|l|c|r|}
  \hline 
  & Basket & Proxy \\
  \hline
  $SCR_{eq}$ &0.0072  & 0.0073\\
  $SCR_{down}$ &  0.0078 & 0.0113 \\
  $SCR_{up}$ &  0.0063 &   0.0154 \\
  $SCR_{mkt}$& 0.0119 & 0.0170\\
  \hline
\end{tabular}
\caption{Different values of the SCR modules. }\label{SCR_table}
\end{table}

\subsection{Analysis in a low interest rate framework}

In this paragraph, we investigate our model in a framework where interest rates are low (around 0.5\%) to be closer to the current interest rates. This was the rate observed for the 10Y bonds issued by France (OAT) at the beginning of 2019.  As explained in Paragraph~\ref{IR_module}, the multiplicative shocks are no longer relevant in this context and we have applied the last recommendation of the EIOPA given in Table~\ref{table_EIOPA2}. The model parameters are specified in Tables~\ref{Model_param2} and~\ref{LM_param2}. Note that we have considered here a higher structural surrender rate of 10\%, and therefore a smaller value of~$n$, the maximal maturity of the basket of bonds. Again, we have taken a constant allocation in equity $w^s$ that is such that the $SCR_{eq}$ and $SCR_{int}$ are approximately of the same order. 

\begin{table}[H]
\centering  
\begin{tabular}{|l|r|}
  \hline
  Stock model & Short-rate model  \\
  \hline
  $S_0=1$ & $r_0=\theta=0.005$  \\
  $\sigma_S=0.1$ & $\sigma_r=0.01$  \\
   $\gamma=0$ & $k=0.2$  \\
  \hline
\end{tabular} 
\caption{Market-model parameters in the low yield framework.}\label{Model_param2}
\end{table}

\begin{table}[H]
\centering  
\begin{tabular}{|l|l|}
  \hline
  Management Parameters & Liability Parameters \\
  \hline
  Allocation in stock $w^s=0.08$ &Lapse triggering threshold $\beta=-0.01$ \\
  Allocation in bond $w^b=0.92$ & Massive lapse triggering threshold $\alpha=-0.05$ \\
  Participation rate $\pi_{pr}=0.9$ & Maximum lapse dynamic lapse rate $DSR_{max}=0.3$  \\
  Minimum guaranteed rate $r^G=0$ &Static lapse rate $\underline{p}=0.1$ \\
  Competitor rate $r^{comp}_t=r_t$& \\
  Smoothing coefficient of the PSR: $\bar{\rho}=0.5$ &  \\
  Bond portfolio maximal maturity $n=10$ &  \\ 
  \hline
\end{tabular} 
\caption{Liability and management parameters in the low yield framework.}\label{LM_param2}
\end{table}
\begin{table}[H]
\centering  
\end{table}

\begin{figure}[h]
\centering
\includegraphics[keepaspectratio=true,scale=0.44]{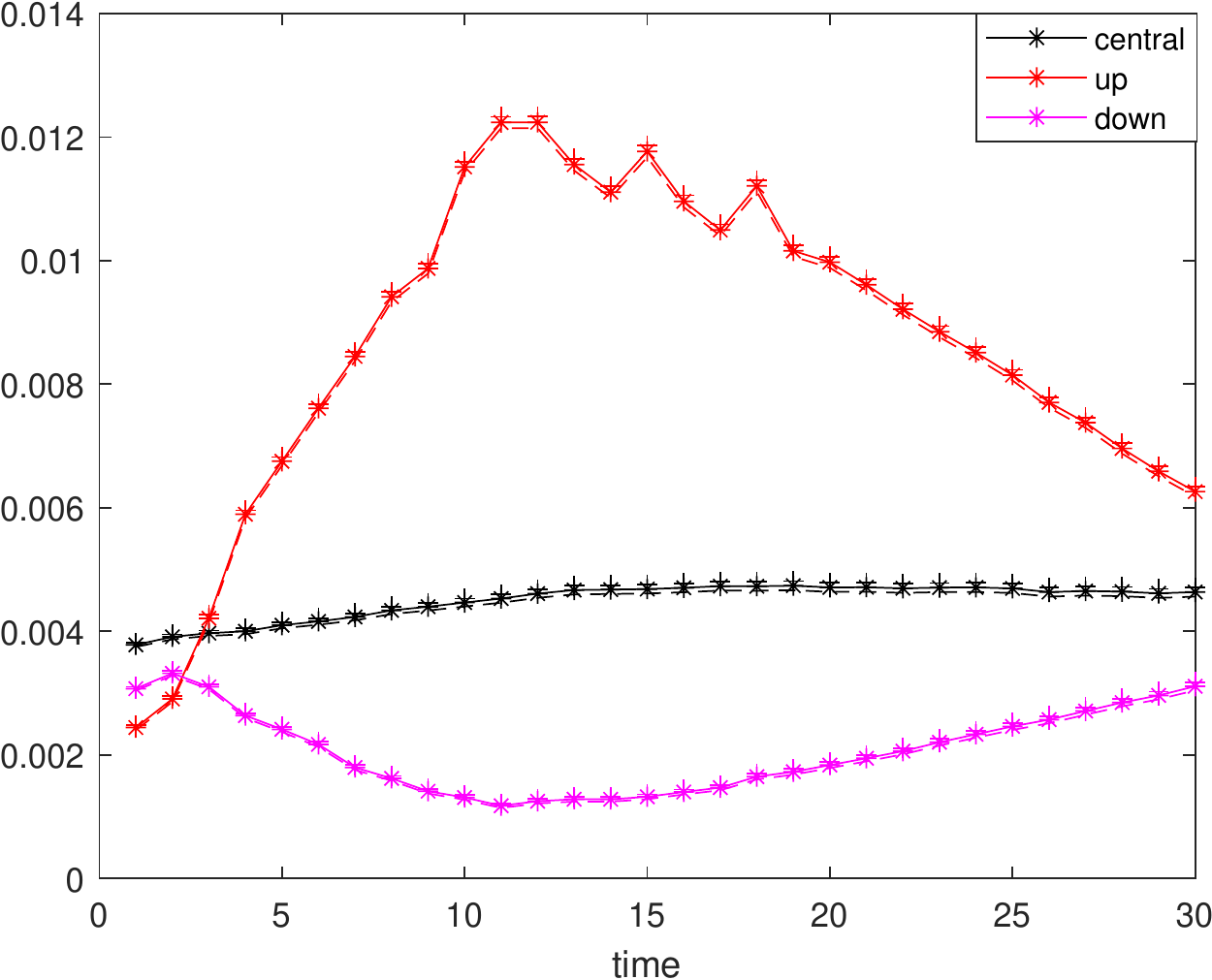}\includegraphics[keepaspectratio=true,scale=0.44]{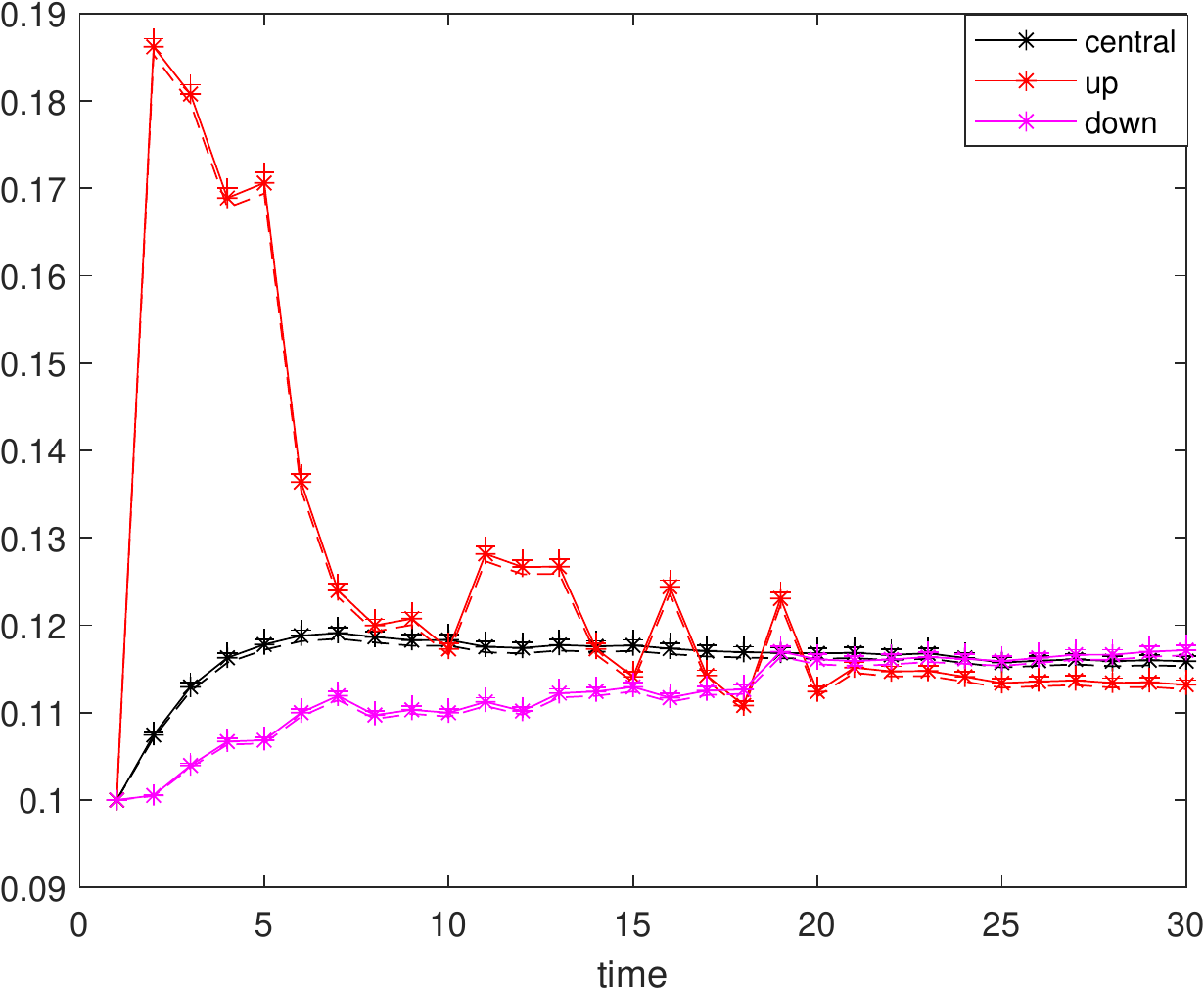}\includegraphics[keepaspectratio=true,scale=0.44]{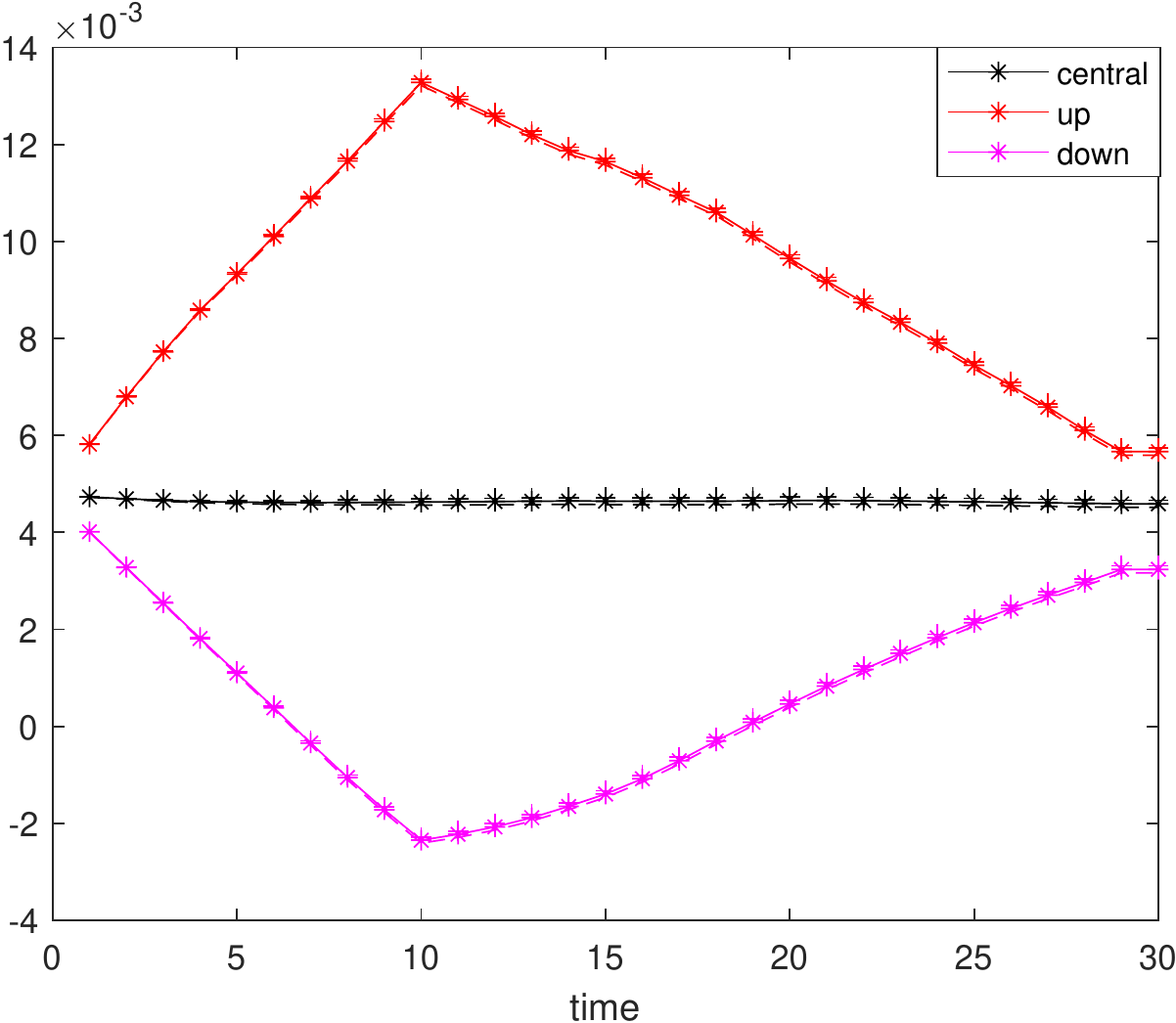}
\caption{Before and after the downward and upward shocks on interest rates. Evolution of the mean crediting rate $\E[r_{ph}(t)]$ (left), of the mean exit rate  $\E[p^e_t]$ (middle) and of the average coupon in the Bond portfolio (right)  in function of the time~$t$.}\label{Int_sh_0.005}
\end{figure}

\begin{figure}[h]
\centering
\includegraphics[keepaspectratio=true,scale=0.55]{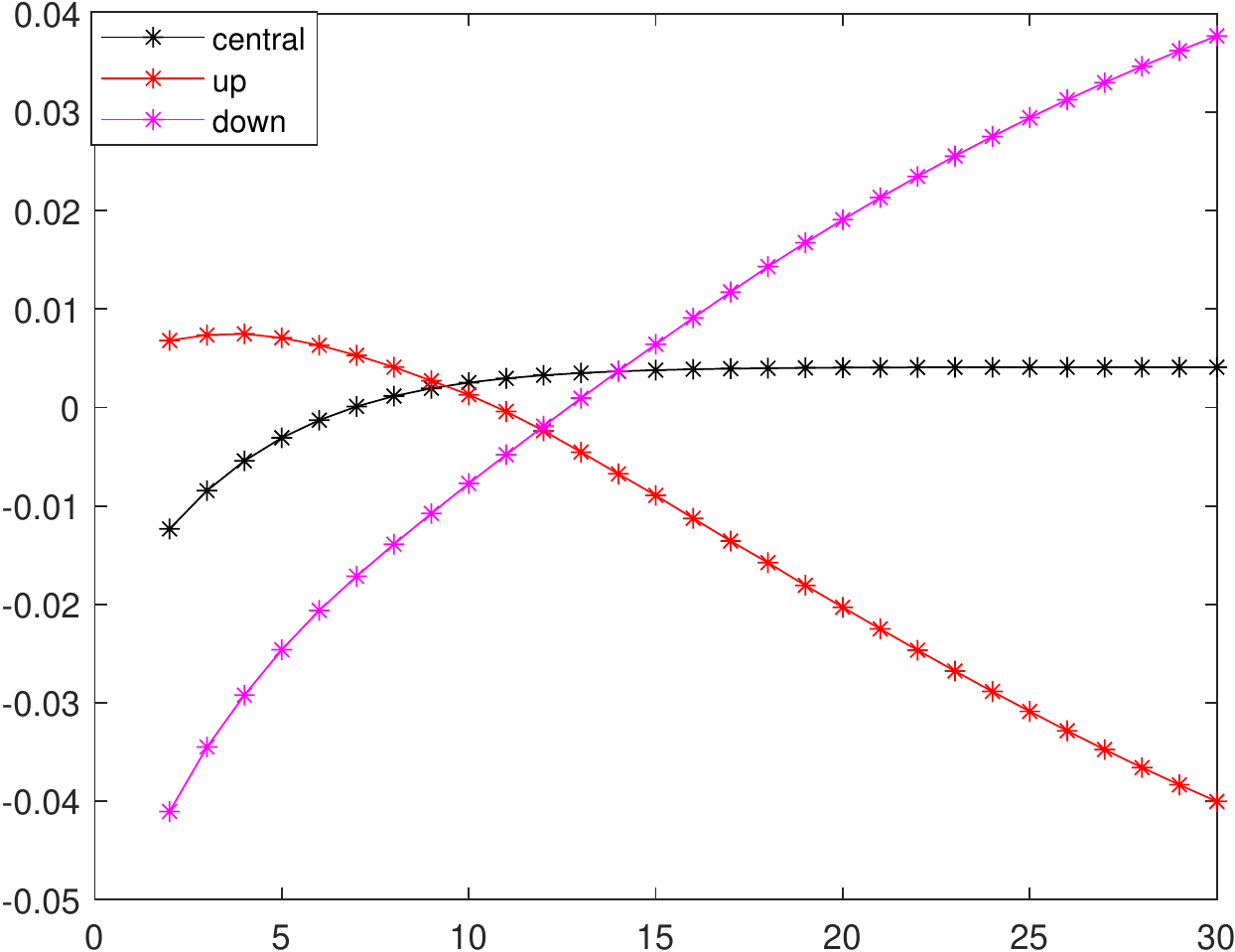} 
\caption{Mean value of the Basic Own-Funds in function of~$n$ (defining the bond combination~\eqref{def_Bbar} in which bonds are invested) in the central framework  and with the upward and downward shocks on interest rates. }\label{BOF_n_0.005} 
\end{figure}

We have plotted in Figure~\ref{Int_sh_0.005} the crediting rate, the surrending proportion and the average coupon rate. The behavior is roughly the same as the one observed in Figure~\ref{Int_sh_0.02} for higher interest rates, and we do not repeat the interpretation. We also observe than on the downward shock, we get negative average coupon values. The small differences between both cases can be explained by the change of method for the shocks. First, we notice on our examples that the additive term in the shocks makes the shocks stronger at the beginning. For example, the spread on the mean exit rate between the upward shock and the central framework is about $0.09$ at year~$2$ instead of $0.05$ in Figure~\ref{Int_sh_0.02} for a $2\%$ interest rate. Of course, the multiplicative shock would have been stronger for a higher interest rate, but a simple calculation made on constant interest rates indicate that the 1Y shock obtained with zero interest rates and Table~\ref{table_EIOPA2} (2.14\%) is almost the same as the one obtained with 3\% interest rates and Table~\ref{table_EIOPA} (2.1\%). Another difference is that the crediting rates change of monotonicity after 10-15 years in the shocked framework. This is mostly due to the fact that the shift functions (left of Figure~\ref{fitted_thetaphi_rbas}) have opposite monotonicity after year 20, while they remain essentially parallel (left of Figure~\ref{fitted_thetaphi}) with the shocks of Table~\ref{table_EIOPA}. Since $n=10$, this has an effect on the coupon rates from year 10 and on the competitor rate from year 20. 
Another interesting plot is the calculation of the BOF in function of~$n$, the maximum maturity of the basket of bonds, which is displayed in Figure~\ref{BOF_n_0.005}. The behavior is very similar to the one observed on the left of Figure~\ref{BOF_n_0.02}. Nonetheless, we see here that the $SCR_{up}$ and $SCR_{down}$ crosses around $n=12$, making this choice optimal for the minimization of the $SCR_{int}$ and even $SCR_{mkt}$ since we have $SCR_{up}>SCR_{down}$ for this choice. Thus, contrary to the previous case, the best choice of $n$ to minimize $SCR_{int}$ is not $1/\underline{p}$. More suprisingly, it does not also satisfy $n\le 1/\underline{p}$, as one should take to have the nominal values of expiring bonds greater than the value of the surrending contracts. This shows anyway that our model can be a useful tool to determine the investment in different bond maturities. 

\vspace{1cm}

{\bf Acknowledgments. } This research benefited from the Joint Research Initiative ``Numerical methods for the ALM'' of AXA Research Fund. A. A. has also benefited from the support of the ``Chaire Risques Financiers'', Fondation du Risque. We thank Vincent Jarlaud and the team ALM of AXA France for useful discussions and remarks.

\bibliographystyle{plain}
\bibliography{biblio}

\end{document}